\documentclass{article}

\usepackage{graphicx}
\usepackage{natbib}
\usepackage{epsfig}
\usepackage{psfrag}
\usepackage{fancyhdr}

\usepackage{color}
\definecolor{gray}{gray}{.5}

\usepackage{bbding}

\usepackage{amsmath,amssymb}
\usepackage{amsbsy}

\newcommand\Ha{\mbox{\textit{Ha}}}  
\newcommand\Gr{\mbox{\textit{G}}}  
\newcommand\Rey{\mbox{\textit{Re}}}  

\begin{document}
\title{\bf Direct Numerical Simulations of Low-$Rm$ 
MHD turbulence based on the least dissipative modes}

\author{ALBAN POTH\'ERAT and VITALI DYMKOU\\
%
Applied Mathematics Research Centre,\\ 
Coventry University, Priory street Coventry CV1 5FB, United Kingdom}


\date{February the 8$^{th}$, 2010}


\maketitle

\begin{abstract}
We present a new spectral method for the Direct Numerical Simulation of 
Magnetohydrodynamic turbulence at low Magnetic Reynolds number. The originality 
of our approach is that instead of using traditional bases of functions, 
it relies on the basis of eigenmodes of the dissipation 
operator, 
which represents viscous and Joule dissipation. We apply this idea to the simple
case of a periodic domain in the three directions of space, with an homogeneous 
magnetic field in the $\mathbf e_z$ direction. The basis is then
 still as subset of the Fourier space, but ordered by growing linear decay rate
 $|\lambda|$ (\emph{i.e} according to the \emph{least dissipative modes}). We 
show that because the  lines of constant energy tend to 
 follow those of  constant $|\lambda|$ in the Fourier space, the 
 scaling for the the smallest scales $|\lambda^{\rm max}|$ in a forced flow can be expressed using 
 this single parameter, as a function of the Reynolds number as $\sqrt{|\lambda^{\rm max}|}/(2\pi k_f)\simeq 0.5\Rey^{1/2}$, where $k_f$ is the forcing wavelength,
 or as a function of the Grashof number $\Gr_f$, which gives a 
 non-dimensional measure of the  forcing, 
as $|\lambda^{\rm max}|^{1/2}/(2\pi k_f)\simeq 0.47\Gr_f^{0.20}$. 
This scaling is also found consistent with heuristic scalings derived by  
 \cite{Alemany1979} and \cite{Potherat2003} for interaction parameter 
$S\gtrsim 1$, and which we are able to numerically quantify as 
$k_\perp^{\rm max}/k_f\simeq 0.5  \Rey^{1/2}$ and 
$k_z^{\rm max}/k_f\simeq 0.8k_f  \Rey/Ha$. 
 Finally, we show that the set of least dissipative modes gives a relevant
 prediction for the scale of the first three-dimensional structure to 
 appear in a forced, initially two-dimensional turbulent flow. This completes 
our numerical 
demonstration  that the least dissipative modes can be used to simulate 
both two- and three-dimensional low-Rm MHD flows. 

\end{abstract}


\section{Introduction}

Turbulence can be described as a flow where a large number of
different patterns evolve in complex interaction with one another. The
knowledge of how much energy each of them carries at a given time then provides 
a reasonably simple statistical representation of the flow. Our purpose is
to apply this very idea to turbulence in liquid metal flows
subjected to an homogeneous external magnetic field,
 by tailoring existing spectral methods to this particular problem.\\
Although simple, these ideas express quite closely the
phenomenology behind  \cite{k41}'s famous theory of homogeneous isotropic
turbulence. Here, the patterns are isotropic vortices sorted in three 
categories, according 
to their size $l_k$ (or wavelength $k$): the large scales where energy is
injected in the flow through some unspecified forcing, the inertial range, 
where mid size vortices pass on energy to smaller scales and the smallest 
scales of size $k_\kappa\sim Re^{3/4}$ where kinematic energy is dissipated by 
viscous friction ($Re=UL/\nu$ stands for the Reynolds number built on velocity  
$U$ and length $L$, that are typical of the large scales, as well as the fluid
kinematic viscosity $\nu$). This early picture has 
been a lot further refined since then, to account for more complex effects
such as intermittency (see \cite{frisch95} or \cite{Davidson2004}
for an overview).\\
The description of the flow in terms of patterns is also well reflected
in the more mathematical spectral approach of turbulence, in which the
solution is sought as a decomposition over the elements $\mathbf u_i$ of a 
basis that spans the functional space it evolves in:
\begin{equation}
\mathbf u=\sum_i c_i(t) \mathbf u_i(\mathbf x).
\label{OrthDecomp}
\end{equation}
The spatial dependence ($\mathbf x$) representing the flow patterns is carried 
by $\mathbf u_i$ while the time dependence (t) appears in the coefficients of 
the expansion $c_i$ only, so when (\ref{OrthDecomp}) is injected into the set 
of Partial Differential Equations that governs the problem, the latter reduces 
to a simpler system of Ordinary Differential Equations (see \cite{canuto06_1}
for a detailed account of spectral methods in fluid mechanics). Apart from 
clear advantages in terms of simplicity and precision, spectral methods can also 
be tailored to the physical reality they describe by choosing 
a basis $(\mathbf u_i)$ that represents realistic flow patterns. 
This basis 
can be obtained from the set of eigenvectors and adjoint eigenvectors of the 
operator derived from the linear part of the motion equations,
with the boundary conditions of the problem. In
incompressible homogeneous turbulence in a spatially periodic domain, the
corresponding operator is the self-adjoint Stokes operator. Its eigenvectors 
are Fourier functions (\cite{constantin85_jfm}), which are
classically related  to vortices of wave-vector $\mathbf k$. 
When the flow is isotropic, vortices of all shapes
are present in statistically equal number, so they are only sorted according
to their size $\|\mathbf k\|$, which facilitates the direct comparison with 
 Kolmogorov's phenomenology.\\

The picture is quite different for turbulence in liquid metals, where the
application of a strong magnetic field $\mathbf B$ breaks isotropy. The fluid
motion induces eddy currents that produce strong Joule dissipation and  
interact with the magnetic field  to yield the Lorentz force.  When the 
magnetic Reynolds number $Rm$ is small, as in most experiments at the 
laboratory scale, the magnetic field induced in turn by these currents can 
be neglected so the total magnetic field is externally imposed and not altered 
by the fluid motion. In the frame of this so-called Low $Rm$ approximation 
(see \cite{Roberts1967}), the Lorentz force mainly damps velocity variations
along the magnetic field lines so vortices tend to be elongated in this
direction, resulting in a strongly anisotropic flow. This effect is 
counteracted by
inertial effects that tend to break up long vortices and promote isotropy in
the flow. Just how isotropic the flow is, is determined by the ratio
between the Lorentz force and inertia, expressed by the interaction parameter
$S=\sigma B^2 L/(\rho U)$, where $\sigma$ and $\rho$ are the fluid's electric
conductivity and density. For large $S$, in a three-dimensional
cubic periodic domain, when all vortices extend from one boundary to the other,
the flow is perfectly two-dimensional, so a transition exists between
two- and three-dimensional turbulence. These effects were pointed out
in the 1960's (\cite{moffatt67}) while \cite{Sommeria1982} analysed the conditions
for a channel flow perpendicular to the magnetic field $\mathbf B$ to be
quasi two-dimensional. More recently, \cite{dav97} explained how vortices evolve
using the conservation of angular momentum.\\
  Spectral methods have been numerically implemented to study this type of flow in three-dimensional periodic domains in several important pieces of work, 
starting with \cite{schumann76} who showed that the free decay of initially 
isotropic turbulence under the influence of an homogeneous magnetic field in a 
three-dimensional periodic box at high $S$ could lead to a two-dimensional 
state. \cite{Zikanov1998} found that  initially isotropic 
MHD flows held steady on average by application of a forcing localised in a 
spherical shell of the Fourier space exhibited intermittent shifts between two 
and three-dimensional states for $S\sim 1$. Intermittency was also observed by 
\cite{thess07_jfm} in both forced and decaying MHD flows in a tri-axial 
ellipsoid.
Most of these studies, however, have used the basis derived
from the Stokes operator, and analysed the flow in terms of the modulus of the 
structure's wavevector $k$, when
clearly, anisotropy imposes that vortices of same $k$ but oriented along or
across the magnetic field should undergo very different levels of Joule
dissipation and eventually carry very different levels of energy. Also,
since no clear MHD equivalent to the Kolmogorov laws had been derived at 
the time, Kolmogorov laws themselves were used to impose a global cutoff
frequency on $k$ when once again, the resolution required to resolve the flow
completely would be expected to decrease when spanning directions from across
to along the magnetic field direction. Therefore, determining a more 
"MHD-suitable" basis, and obtaining  MHD equivalent to the Kolmogorov laws 
for the dissipative scales in both two- and three-dimensional MHD forced 
turbulence are the precise questions we wish to address in this work, by going 
back to the initial idea of using a basis of
functions that imitates flow patterns as closely as possible. We  focus
our attention on the configuration of a cubic domain, periodic in the three
spatial directions, with an homogeneous magnetic field in the $z$ direction.
Although physically not realistic, these assumptions offer a simple but 
still
meaningful test case for the application of our ideas, keeping in mind that
results more directly comparable to experiments will have to come out of a
configuration where  boundaries that intercept the magnetic field lines at 
least, will be physical walls.\\
        In the frame of the low $Rm$ approximation, the Lorentz force appears
as a linear term in the Navier-Stokes equation so the linear part of the latter
 is in fact the sum of the Stokes operator and that related to the Lorentz
force (see \cite{Roberts1967}). We have previously solved the spectral problem 
for this operator (\cite{Potherat2003}), shown that it was
self-adjoint and that its sequence of eigenfunctions (the \emph{least 
dissipative modes}) was able to finely
mimic the anisotropic properties of MHD turbulence.  We also showed that
this sequence of modes achieved an upper bound for the attractor dimension of
the system that was consistent with estimates obtained heuristically for the 
size of the smallest scales. It is
worth mentioning that the spectral analysis of the same operator, but in the
case
where the boundaries orthogonal to $\mathbf B$ are physical walls leads to a
sequence of eigenfunctions that exhibit the correct Hartmann boundary layer
profile in the vicinity of these walls (see \cite{Potherat2006}, and \cite{moreau90} 
for a review of the theory of these layers). In the present
work, we will therefore numerically implement our previously found basis 
 in order to extract the relevant modes and determine the MHD 
equivalent of the Kolmogorov scales. In
section \ref{sec:principle}, we first recall and complement the properties of
the linear part of the Navier-Stokes equation found in \cite{Potherat2003}. We 
then implement this basis in an existing spectral code and determine some 
Kolmogorov-like laws for the small scales in three-dimensional MHD flows which 
should serve as a criterion to resolve the flow completely in section 
\ref{sec:modeDetect}. Since an essential property of MHD turbulence is that it 
can be two-dimensional or three-dimensional, we 
devote section \ref{sec:trans} to testing whether DNS based on the least 
dissipative modes can reproduce this feature. This leads us to find out 
the lengthscale of vortices in which three-dimensionality first appears when 
the intensity of the forcing is increased in an initially two-dimensional flow.
\section{Principle of DNS based on the least dissipative modes}
\label{sec:principle}
\subsection{Problem formulation}
\label{sec:problem}
We consider an incompressible, conducting fluid (density $\rho$, electrical 
conductivity $\sigma$ and kinematic viscosity $\nu$) in a three-dimensional
periodic cube $\Omega$ of size $L_0$ under imposed homogeneous and steady 
magnetic field $B{\bf e}_z$. In the frame of the low-$Rm$ approximation, the 
governing equations can be reduced to the closed system made of momentum and 
mass conservation, which involve the flow velocity ${\bf u}({\bf x},t)$ and 
pressure $p({\bf x},t)$ only (see \cite{Roberts1967} and \cite{Sommeria1982}). 
A third equation deduced from electric current conservation and the Ohm's law 
can be used to reconstruct the electric potential and the electric current 
\textit{a posteriori}.
We shall, however, only need here the equations for ${\bf u}({\bf x},t)$ and 
$p({\bf x},t)$. These can be written in non-dimensional form by choosing 
reference length $L$, time $L^2/\nu$, velocity $\nu/L$, pressure 
$\rho\nu^2/L^2$ and a dimensionless external force $||{\bf f}||/L^{3/2}$,
where $\|\cdot\|=(\int|\cdot|^2d\Omega)^{1/2}$ is the usual norm in $L_2(\Omega)$ space. The Navier Stokes equations are then written:
\begin{eqnarray} \label{NSLorForNonDimForm}
\begin{aligned}
&\frac{\partial}{\partial t}{\bf u}({\bf x},t)+({\bf u}\cdot
\nabla){\bf u} + \nabla p=\nabla^2 {\bf u}-\Ha^2\nabla^{-2}\frac{\partial^2 {\bf u}}{\partial z^2}+\Gr{\bf
f}({\bf x},t),\\[2mm]
 &\nabla\cdot{\bf u}=0,
\end{aligned}
\end{eqnarray}
where $\Ha  = LB\sqrt{ \frac{\sigma}{\rho \nu}}$ is the Hartmann 
number while 
$\Gr=\frac{L^{3/2}}{\nu^2}||{\bf f}||$ is the Grashof number, which represents 
the forcing normalised by viscous forces (as in \cite{Doering1995}). 
Consequently, the solution of (\ref{NSLorForNonDimForm}) is defined by the only 
two relevant control parameters $\Ha$ and $\Gr$ in (\ref{NSLorForNonDimForm}). 
The choice of $L$ is not straightforward as it is not imposed by the geometry. 
It is noteworthy that if it is set to 
$L=\frac{1}{B}\sqrt{\frac{\rho\nu}{\sigma}}$,
then the governing equations depend on the single dimensionless parameter 
$\Gr/\Ha^3$. This reference length however ignores
the dynamics of the large scales present in the flow. One would instead 
expect a better suited reference length to follow the forcing scale to some 
extent. Since, however, the latter is not specified at this stage,
 we shall choose $L=L_0$, as it 
represents \emph{de facto} the largest achievable scale in our problem, 
and denote $\Ha_0$, the Hartmann number built on $L_0$. It is 
worth stressing that we shall not try to minimise or ignore the effect of 
the boundaries where 
periodic conditions are applied. In particular, we shall also analyse  
two-dimensional flows where structures extend across the whole domain in the 
$z$ direction. Although clearly not experimentally achievable, this 
configuration has often been used as an interesting toy-model for the 
study of the transition between two-dimensional and three-dimensional flows 
(\cite{nakauchi92_pf,Zikanov1998,thess07_jfm}). Therefore, 
contrarily to  many previous studies of turbulence where periodic domains are 
used to represent a small volume taken out of an homogeneous flow, and where 
structures of the size of the domain should therefore be avoided, the 
conditions under which structures extend over the full domain along $z$ will 
be of interest in this work. For this reason, the 
length $L_0$ will be a meaningful parameter of the problem, wherever 
such two-dimensional vortices are considered (in section \ref{sec:trans}).\\

Two further non-dimensional numbers can be defined that are 
traditionally used in MHD turbulence: the usual Reynolds number 
$\Rey = \frac{UL_{\rm int}}{\nu}$, with integral length scale 
\begin{eqnarray}
L_{\rm int}=\frac{\pi}{2{||\bf u||}^2}\int\limits_{0}^{\infty}{\bf||
k||}^{-1}E(k)dk
\end{eqnarray}
gives a measure of the intensity of turbulence (Here, $\mathbf k$ is 
the three-dimensional wavevector that appears in the Fourier transform of 
$\mathbf u$, $E(k)$ is the spectral power density of all wavevectors of 
norm $k$ and $U=(\int Edk)^{1/2}$ is a reference velocity).  Also, the magnetic interaction 
parameter $S=\sigma B^2 L_0/(\rho U)$  
represents the ratio of the Lorentz force to inertia. In freely 
decaying turbulence where boundaries are ignored, taking $U$ as a reference 
velocity from the initial 
velocity field and $L_{\rm int}$ as a reference length, $S$ becomes the only 
non-dimensional parameter that governs the 
problems. In our case however, only $G$ and $\Ha_0$ are known \textit{a priori}.
In this sense, they are the control parameters for this problem.\\
The problem is fully defined by the addition of periodic boundary conditions
\begin{eqnarray} \label{BC}
\begin{aligned}
{\bf u}(x,y,z,t)&={\bf u}(x+a,y,z,t)\\[-1mm]
                 &={\bf u}(x,y+b,z,t)\\[-1mm]
                 &={\bf u}(x,y,z+c,t),\quad a,b,c\in \mathbb{Z}
\end{aligned}
\end{eqnarray}
and of the initial condition
\begin{eqnarray} \label{IC}
\begin{aligned}
{\bf u}(x,y,z,0)={\bf u_{\rm i}}(x,y,z).
\end{aligned}
\end{eqnarray}
These, together with the mass conservation, which simply implies that $\bf u$ 
is a solenoidal vector field, are taken into account by specifying that the 
solution $\bf u$ is sought in the functional space $V^2$, a solenoidal 
subspace of Hilbert space $H^2$. Since the spectral method we wish to 
implement is derived from the spectral properties of governing equations, 
these ought to be written in abstract form, with help of the Helmholtz 
decomposition:
\begin{eqnarray}\label{NSLorForOperForm}
\begin{aligned}
\frac{\partial}{\partial t}{\bf u} &= D_{\Ha_0}{\bf u}+B({\bf u},{\bf u})+\Gr{\bf f},\\[2mm]
{\bf u}|_{t=0} &= u_{\rm i}.
\end{aligned}
\end{eqnarray}
Details of the mathematical framework can be found in \cite{dyp09}. The 
advantage of this form is that it gathers the linear part of the equations 
into a single operator that operates in $V^2$ onto itself:
\begin{eqnarray} \label{LinOper}
\begin{aligned}
D_{\Ha }=P\bigg (\nabla^{2}-\Ha^2\nabla^{-2}\frac{\partial^2}{\partial z^2}\bigg) \quad :\quad
V^2\rightarrow V^2.
\end{aligned}
\end{eqnarray}
$P$ denotes the orthogonal projection onto the subspace of solenoidal 
fields, and nonlinear terms are represented by the bilinear operator 
$ B({\bf u},{\bf u})=P ({\bf u}\cdot \nabla){\bf u}$.\\
In the absence of magnetic field, $\Ha_0=0$ and the system reduces to the 
usual Navier-Stokes equation. Periodic boundary conditions then ensure that the 
eigenfunctions of the Stokes operator form a basis of 
$V^2$ (\cite{Temam2001}). They can thus be used for the 
spectral decomposition in order to reduce the problem to 
a simpler system of ordinary differential equations. For $\Ha_0\neq0$, the 
physical relevance of 
the linear part can be seen by noticing that 
the Lorentz force only appears in $D_{\Ha_0}$. The spectral properties of this 
operator are therefore expected to express the mode-selecting dissipation that 
results from its action on the flow. This makes the set of eigenfunctions of 
$D_{\Ha_0}$ a good 
candidate for the choice of the basis of modes required in the solution's 
expansion (\ref{OrthDecomp}). We have previously found these in 
(\cite{Potherat2003}) and shown that they constituted a basis of $V^2$, so we 
shall now summarise and extend these results derived from the spectral 
characteristics of the dissipation operator $D_{\Ha}$.
\subsection{Spectral properties of the $D_{Ha}$ operator for any given $Ha$}
\begin{figure}
\centering
\begin{tabular}{ccc}
 &
 \begin{minipage}[b]{.43\linewidth}
   \psfrag{x_kp}{$k_\perp$}
   \psfrag{y_kz}{$k_z$}
   \psfrag{k2}{$k_z$}
   \psfrag{kz}{$k_\perp$}
   \psfrag{kz}{$k_{z}^{\rm max}$}
   \psfrag{kp}{$k_\perp^{\rm max}$}
   \centering\epsfig{figure=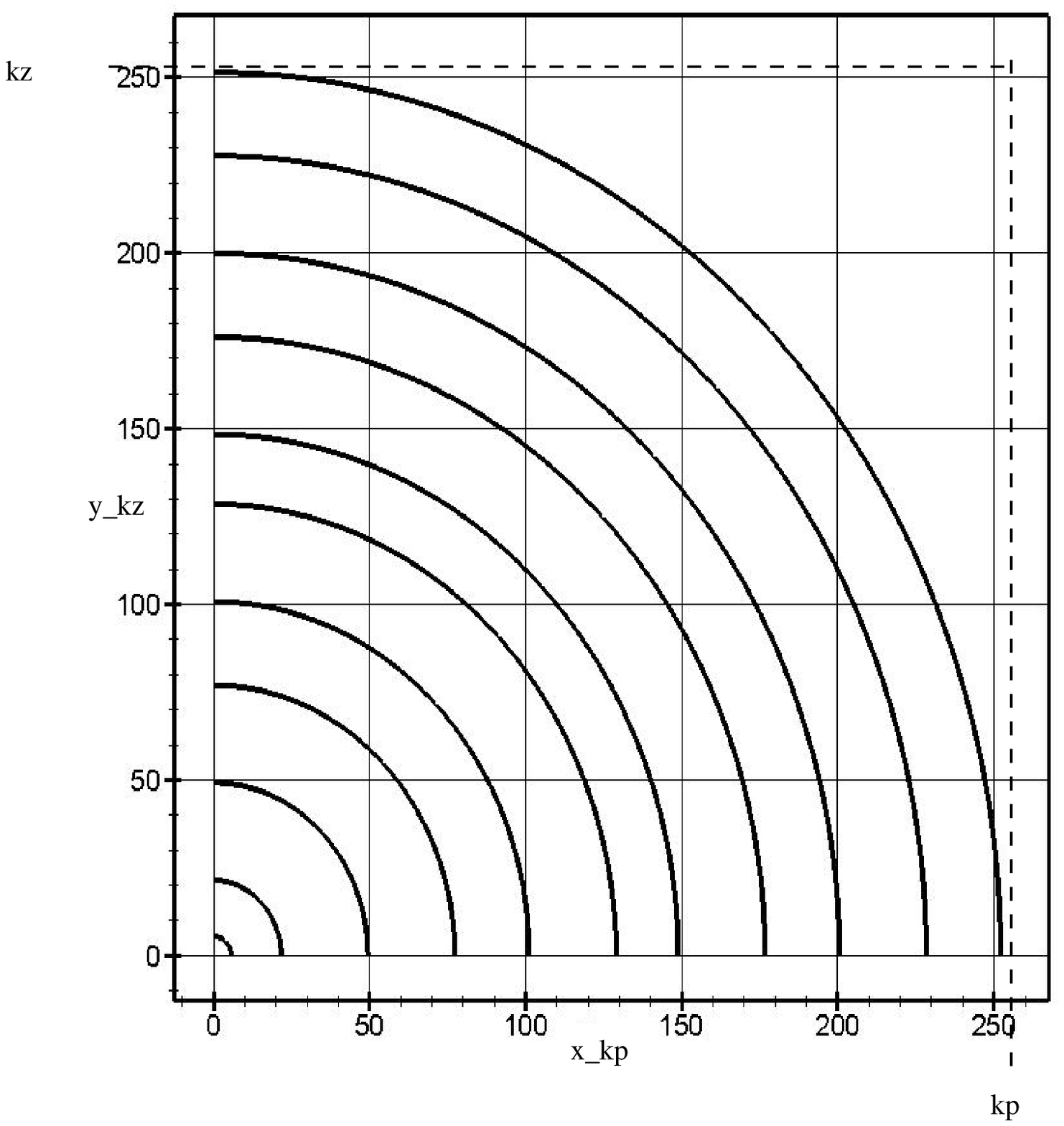,width=\linewidth}
 \end{minipage}
 & 
 \begin{minipage}[b]{.43\linewidth}
   \psfrag{x_kp}{$k_\perp$}
   \psfrag{y_kz}{$k_z$}
   \psfrag{k2}{$k_z$}
   \psfrag{kz}{$k_\perp$}
   \psfrag{kz}{$k_{z}^{\rm max}$}
   \psfrag{kp}{$k_\perp^{\rm max}$}
   \centering\epsfig{figure=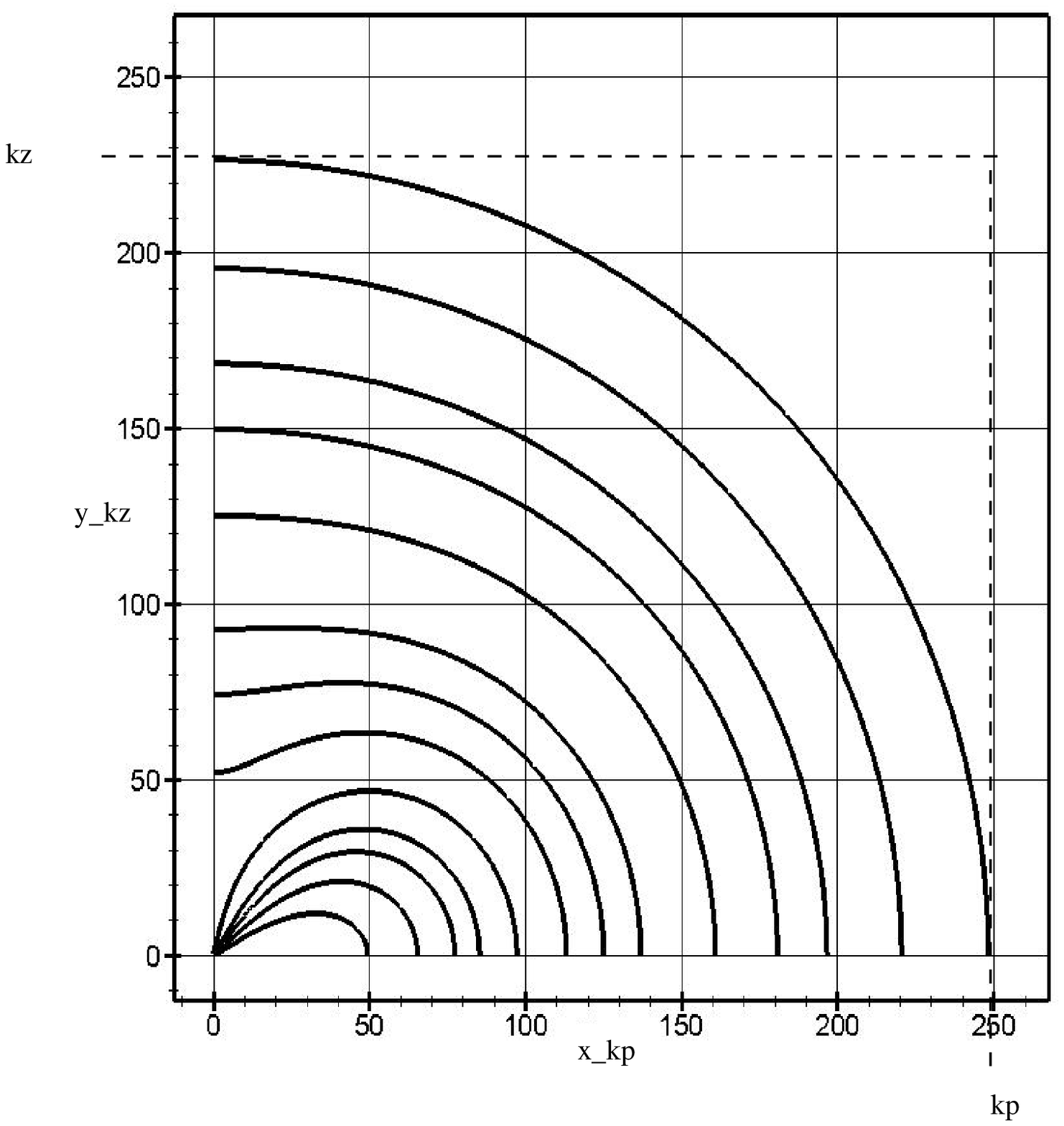,width=\linewidth}
 \end{minipage}
\\
  &
  (a) $\Ha =0$
  &
  (b) $\Ha =630$
 \\
 &
 \begin{minipage}[b]{.43\linewidth}
   \psfrag{x_kp}{$k_\perp$}
   \psfrag{y_kz}{$k_z$}
   \psfrag{k2}{$k_z$}
   \psfrag{kz}{$k_\perp$}
   \psfrag{kz}{$k_{z}^{\rm max}$}
   \psfrag{kp}{$k_\perp^{\rm max}$}
   \centering\epsfig{figure=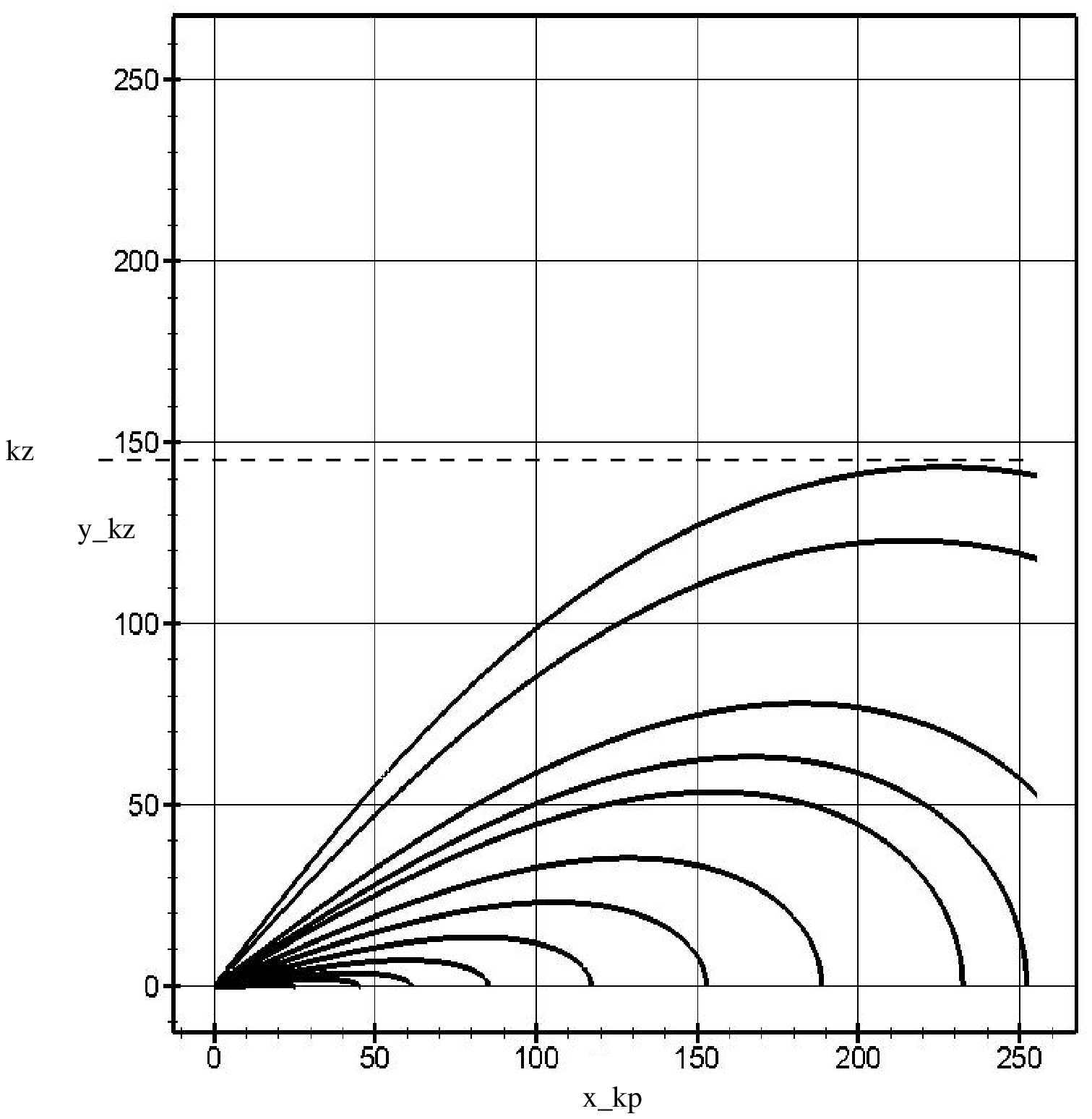,width=\linewidth}
 \end{minipage}
 &
 \begin{minipage}[b]{.43\linewidth}
   \psfrag{x_kp}{$k_\perp$}
   \psfrag{y_kz}{$k_z$}
   \psfrag{k2}{$k_z$}
   \psfrag{kz}{$k_\perp$}
   \psfrag{kz}{$k_{z}^{\rm max}$}
   \psfrag{kp}{$k_\perp^{\rm max}$}
   \centering\epsfig{figure=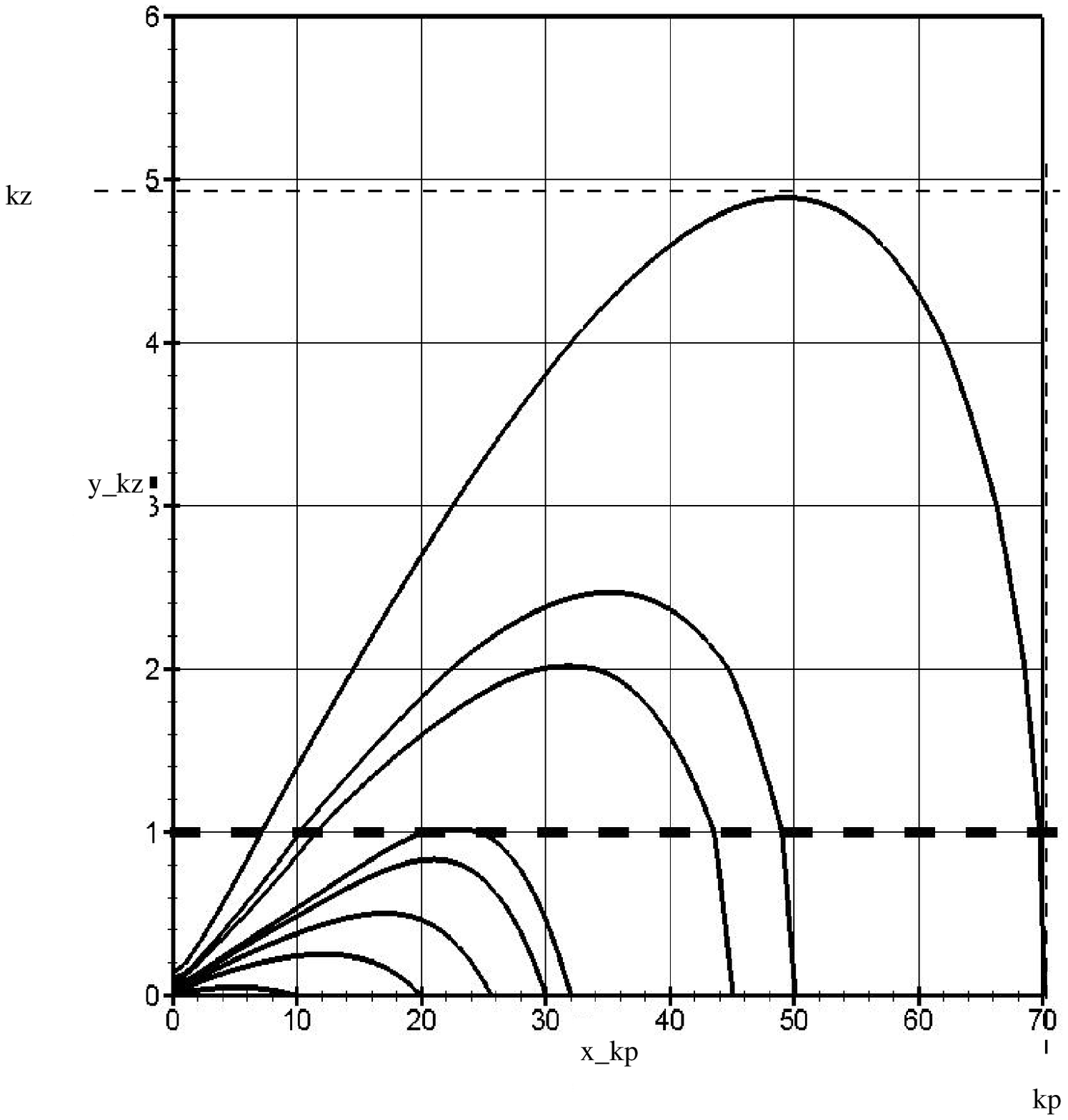,width=\linewidth}
 \end{minipage}
 \\
 
 &
 (c) $\Ha =3140$
 &
 (d) $\Ha =3140$ 
 \\
 \end{tabular}
\caption{Iso-$\lambda$ curves for different values of $\Ha$. Note that all 
families of curves (except those for $\Ha=0$) can be scaled down 
to a single family in the $(k_\perp/Ha,{ k}_z/\Ha)$} plane where 
$k_\perp=\sqrt{k_x^2+k_y^2}$. Values of 
$k_\perp^{\rm max}$ and  $k_z^{\rm max}$ are marked on arbitrary iso-$\lambda$ curves to illustrate how they are related to $\lambda^{\rm max}$.
\label{IsoLambda}
\end{figure}

$D_{\Ha}$ is a linear operator. The boundary conditions are accounted for in the 
definition of the domain 
of the operator, defined as $D(A)=V^2(\Omega)$. Since $\Omega$ is 
bounded, the natural injection of $V^2$ into $L_2(\Omega)$ is compact, thus 
$D_{\Ha }$, as an operator in $L_2(\Omega)$, is compact (\cite{Temam2001}). 
Also, this operator is 
self-adjoint and therefore possesses a 
discrete set of eigenvalues $(\lambda_\mathbf k)$ and eigenfunctions 
${\bf v}^{\mathbf k}$ that form an orthonormal basis of the 
$L_2(\Omega)$   
space. We have shown in \cite{Potherat2003} that the eigenfunctions 
${\bf v}^{\bf k}=(v^{\bf k}_i)_{i\in\{x,y,z\}}$ are a subset of the usual 
Fourier space: 
\begin{eqnarray} \label{eigenfunction}
\begin{aligned}
v^{\bf k}_i=V_ie^{{j2\pi\bf k\cdot x}},
\end{aligned}
\end{eqnarray}
with wavenumbers ${\bf k}=(k_x,k_y,k_y)\in \mathbb{Z}^3$, constants 
$V_i\in\mathbb{C}$ and where $j$ is the imaginary unit. The corresponding 
eigenvalues are
\begin{eqnarray} \label{eigenvalue}
\begin{aligned}
\lambda_{\bf k}=
-4\pi^2(k_x^2+k_y^2+k_z^2)-\Ha ^2\frac{k_z^2}{k_x^2+k_y^2+k_z^2}.
\label{eq:lambda}
\end{aligned}
\end{eqnarray}
We denote the set of all eigenvalues (\ref{eigenvalue}) by 
$\sigma_\infty(D_{\Ha})$. Since $\lambda_{\bf k}$ represents the linear 
decay rate of mode ${\bf v}_{\bf k}$ by $D_{Ha}$, and $\lambda_{\bf k}<0$, 
$(\lambda_{\bf k})$ and $v_{\bf k}$ can be arranged by growing dissipation. 
This singles out $\lambda_{\bf k}$ as a spectral parameter that naturally 
reflects the effects of the Lorentz force. From the definition 
(\ref{eigenvalue}), we see that for $\Ha=0$, 
$|\lambda_{\bf k}|/(2\pi)^2$ reduces to the square length 
$k^2=||{\bf k}||^2$ of the wave vector ${\bf k}$ which is the usual spectral parameter in non-MHD 
isotropic turbulence (see Figure\ref{IsoLambda}(a)). In the MHD case,  
different values of the magnetic field $B$ or of 
the reference length $L$ that enter the definition of $\Ha$ yield  
different sets of eigenvalues (see 
Figure\ref{IsoLambda}(a)-(d)). Such dependency is absent in the usual Fourier 
basis ordered by growing ${\|\bf k\\||}$. The main novelty 
introduced by using this basis thus doesn't reside in the elements of the 
basis themselves but rather in the fact that they are ordered by growing 
values of $|\lambda_{\mathbf k}|$ instead of by growing $k$. This earns these 
modes their denomination of \emph{least dissipative}. Furthermore, we 
previously showed 
(\cite{Potherat2003}) that the set of least dissipative modes required to 
describe the flow possessed the anisotropy 
properties predicted heuristically for such MHD flows. 
In the light of (\ref{eigenvalue}), the sequence 
$(-\lambda_{\bf k})^{1/2}/(2\pi)$ therefore appears as an anisotropic 
generalisation of the usual $k$-sequence, and the spectral decomposition 
(\ref{OrthDecomp}) of $\bf u$ can now be rewritten as
\begin{eqnarray}\label{EigenfunctionDecomposition}
{\bf u}({\bf x},t)=\sum\limits_{|\lambda_{\bf k}|<|\lambda^{\rm max}|}c_{\lambda_{\bf k}}(t){\bf v}_{\lambda_{\bf k}}({\bf x}),
\end{eqnarray}
where $c_{\bf k}(t)$ are the expansion coefficients, ${\bf v}_{\lambda_{\bf k}}({\bf x})$ are the eigenvectors of $D_{\Ha}$ for 
eigenvalue $\lambda_{\bf k}$  and $\lambda^{\rm max}$ 
defines the maximum resolution required to resolve the flow completely.

\subsection{Choice of the set of least dissipative modes}
At this point, we still lack two parameters to be able to choose the set of 
modes to fully resolve a given flow, defined by the values of $\Gr$ 
(or $\Rey$) and $\Ha_0$. Firstly, the 'shape' of the set of modes is 
determined by the value of $\Ha$ only. We have however defined $\Ha_0$ using 
the domain size $L_0$, as a reference length. Clearly, for $\Ha$ to reflect the 
actual physics of the flow, another reference length $L$ should be found 
that accounts for the forcing scale in one way or another. 
Secondly, the number of modes $N$ required to fully resolve the flow or, 
equivalently, the largest value of $|\lambda|$, $|\lambda^{\rm max}|$ in
(\ref{EigenfunctionDecomposition}) must be determined in such a way that 
the flow is fully represented by its projection onto the set of $N$ least 
dissipative modes defined by $|\lambda|<|\lambda(N)|=|\lambda^{\rm max}|$. For 
this, the global attractor of the motion equations has to be entirely included 
in the functional subspace spanned by the $N$ least dissipative modes. 
Consequently, if 
$d_M$ is the dimension of this attractor, or equivalently the number of 
degrees of freedom of the flow, we must have 
$|\lambda^{\rm max}|\geq|\lambda(d_M)|$. Unfortunately, it is difficult 
to obtain a precise estimate for $d_M$. Its physical interpretation, however, 
can be easily understood: in both the non-MHD and the MHD case, the reason why 
$d_M$ is finite is that viscous 
dissipation introduces a cutoff at the small scales, beyond which 
flow structures carry a vanishingly small amount of energy. 
\cite{constantin85_jfm} give an elegant illustration 
of the physical meaning of these mathematical concepts. This cutoff 
wavelength can be estimated 
heuristically, which, in turn leads to scalings for $N$.
The most famous example is that of the three-dimensional non-MHD case, where 
the heuristic Kolmogorov scale $k^{\rm max} = k_\kappa\simeq 
C_\kappa\Rey^{3/4}$ ($=|\lambda^{\rm max}|^{1/2}/{2\pi}$ in our 
notations, and where $C_\kappa>1$, \cite{k41} ) gives an estimate that is 
precise enough to be used as a criterion to fix the number of determining 
modes as $N\simeq C_\kappa^3\Rey^{9/4}$ in a Fourier-based DNS. In 
two-dimensional turbulence, 
 a precise estimate for the attractor dimension (\cite{Doering1995}) and a 
heuristic scaling for the size of the smallest, or \emph{Kraichnan scales}, 
(\cite{Kraichnan1967,ohkitani89}) coincide precisely with $k^{\rm max} = 
k_k\simeq \Gr^{1/3} (1+\log \Gr)^{1/6}$ where $k^{\rm max}=|\lambda^{\rm max}|^{1/2}/{2\pi}$ .\\
In the MHD case, viscous dissipation still determines the cutoff scale, 
even though Joule dissipation extracts energy at all scales. 
 \cite{Alemany1979} and \cite{Potherat2003}  used this idea, further 
assumed that the anisotropy $k_\perp/k_z$ was scale--independent and that 
inertia balanced the Lorentz force at all scales to derive some heuristic
scalings for the cutoff value $\lambda^{\rm max}$ and $N$, when 
$S\gtrsim1$:
\begin{eqnarray}
 N                & \simeq C_0\frac{\Rey ^{2}}{\Ha }, \\
\frac{\sqrt{|\lambda^{\rm max}|}}{2\pi k_f} & \simeq C_\lambda \Rey ^{1/2}.
\label{eq:heur_scal_lambda}
\end{eqnarray}
We have here expressed $\lambda^{\rm max}$ with respect to the largest 
forcing scale in the problem $L_f=L_0/k_f$ 
to reflect the fact 
that for spatially periodic domains, the forcing scale is a relevant large 
scale 
that determines the small scales while the size of the computational domain 
isn't.  
Since the set of least dissipative modes is a subset of that of Fourier modes, 
these scalings can be more classically expressed in terms of the smallest 
scales across (subscript $\perp$) and along the magnetic field by virtue of 
the properties of (\ref{eq:lambda}):
\begin{eqnarray}
\begin{aligned}
 \frac{k_z^{\rm max}}{k_f}      & \simeq \pi k_f C_\lambda^2 \frac{\Rey }{\Ha },    
\qquad
 \frac{k_{\perp}^{\rm max}}{k_f} & \simeq C_\lambda \Rey ^{1/2}.
\label{eq:sscale_mhd}
\end{aligned}
\end{eqnarray}
We have been able to partly confirm these scalings by finding an upper bound 
for the attractor dimension (\cite{Potherat2003}).  
$C_0$ or $C_\lambda$  however remain to be evaluated, so no practical 
criterion currently exists 
for the number of determining modes in flows where a magnetic field is present.
The next section is therefore
devoted to searching numerically the values of $L$ and $\lambda^{\rm max}$.  
In particular, we shall estimate the lowest values of $C_\lambda$  
for which the flow is fully resolved for $S\gtrsim1$. When $S>>1$, the 
flow becomes two-dimensional so the set of least dissipative modes becomes the 
two-dimensional isotropic set of Fourier modes defined by  $N\gtrsim k_k^2$ 
(or $\lambda^{\rm max}=(2\pi k_k)^2$). When 
$S<<1$, the effects of the Lorentz force become small and the set of least 
dissipative modes differs little from that of the usual three-dimensional 
isotropic set of Fourier modes  $N\simeq C_\kappa^3\Rey^{9/4}$ (or $\lambda^{\rm max}=(2\pi k_\kappa)^2$).\\
At this point, it is important to notice that a flow described by the set of 
least dissipative modes with $\lambda^{\rm max}$ determined by the rules above is 
resolved exactly, without any approximation, as all energy and dissipation 
containing modes are contained in the attractor. In particular, a clear 
distinction should be made between solving the equations by projection on the 
full set of least dissipative modes, which is a type of Direct Numerical 
Simulation,  and approaches such as Large Eddy Simulations where part of the 
spectrum is modelled and not resolved. Both approaches could even be combined 
to achieve important reductions in computational cost.

\section{Determination of the exact set of modes required to resolve the flow
for $S\gtrsim 1$}
\label{sec:modeDetect}

\subsection{Numerical system and procedure}\label{sec:NumProc}

We base our DNS on the eigenfunctions of the dissipation operator. 
Since these are a subset of the usual Fourier modes, we use the
code developed by
\cite{Knaepen2004} and \cite{vorobev05} where the problem formulated in
section \ref{sec:problem} was implemented and fully tested. It relies on
 traditional spectral methods based on a Fourier 
decomposition, with Fast Fourier Transform and a fourth-order low-storage 
time-integration Runge-Kutta scheme (see \cite{Rogallo1981} and \cite{Williamson1980}). 
The alias error resulting from the bilinear products is removed by 
phase-shifting method (\cite{Rogallo1981,patterson71}), which
allows us to retain all of the Fourier modes but
requires eight evaluations during each time step. 
We adapt this code to 
our needs of performing calculations using set of modes that satisfy 
$|\lambda_{\bf k}|<|\lambda^{\rm max}|$, simply by setting unneeded modes to 
zero when required.
In all calculations presented in the whole of section \ref{sec:modeDetect}, 
initial velocities are set to zero ($\mathbf u(t=0)=0$). The flow is driven
by two distinct types of constant forcing $\mathbf f$ in 
(\ref{NSLorForOperForm}), that respectively favour two-dimensional and 
three-dimensional structures. The two-dimensional forcing is applied to  
Fourier modes with wavevectors
${\bf k_f}=(k_{fx},k_{fy},k_{fz})\in\{(6,6,0),(7,7,0),(9,9,0)$\}
\begin{eqnarray} \label{eq:2dforce}
\begin{aligned}
{\bf f}_{2D}({\bf x},t)=\sum\limits_{\bf
k_f}\bigg(\sin(k_{fx}{2\pi}x)\cos(k_{fy}{2\pi}y){\bf
e}_x+\cos(k_{fx}{2\pi}x)\sin(k_{fy}{2\pi}y){\bf
e}_y\bigg),
\end{aligned}
\end{eqnarray}
and tends to generate a flow with no velocity component nor
velocity variations in the $z$-direction. Since the numerical algorithm would 
not otherwise allow the solution of the problem to be three-dimensional at all, we add a 
small constant force of amplitude $\varepsilon= 10^{-3}$ (relative to 
$\mathbf f$) in each ball $||{\bf k-k_f}||<2$. There are several other reasons 
for this choice: firstly, the forcing has to be a 
combination of the set of modes used for the expansion. In this regard, a 
practically $z$-independent forcing can be used to simulate both 
two-dimensional flows (for which the effect of the small three-dimensional 
component of the forcing falls within the numerical error) and 
three-dimensional flows. 
The second reason is that this type of constant weakly three-dimensional 
forcing strongly resembles that 
obtained in liquid metal experiments by injecting electric current though 
metallic electrodes embedded in insulating Hartmann walls (\cite{Sommeria1986},
\cite{Sommeria1988}, \cite{Delannoy1999}). 
Our most recent experiments on electrically driven channel flows under 
transverse magnetic fields  (\cite{kpa09,kp10_prl}) have indeed confirmed 
the previous theoretical prediction that in such experiments, even for high 
values of $Ha$, inertia induced some slight velocity variations along the 
magnetic field lines, so that three-dimensional vortex 
instabilities such as 
those analysed by \cite{thess07_jfm} do not occur in strictly two-dimensional, 
or even strictly quasi two-dimensional flows, but rather is some weakly 
three-dimensional flow (\cite{psm00}), which our  
weakly three-dimensional forcing  imitates.\\
Finally, \cite{vorobev05} have suggested that the two or three-dimensional 
nature of the forcing had no noticeable influence on the anisotropy of 
intermediate and small scales. This is supported by the properties of the least 
dissipative modes, as they imply that the small scales are determined by $\Gr$, 
which only carries the intensity and the scale of the forcing, and $Ha$ 
(\cite{Potherat2003}). 
To check this point further, we have performed a series of  
computations in the same conditions as those described above, but with a 
three-dimensional forcing. The latter was chosen of the ABC type 
(\cite{mininni06}) so as to act on the three components of the velocity, and  
expressed as:
\begin{eqnarray}
{\bf f}_{3D}({\bf x},t)=
(\cos(k_{fy}y)+1.1\sin(k_{fz}z))\mathbf e_x
+(1.1\cos(k_{fz}z)+0.9\sin(k_{fx}x))\mathbf e_y \nonumber \\
+(0.9\cos(k_{fx}x)+\sin(k_{fy}y))\mathbf e_z \qquad \text{with }\mathbf k_f=(6,6,6).
\end{eqnarray}
All calculated cases are summarised in table \ref{table:simul1}.
\subsection{Determination of the length scale $L_{\rm opt}$}
\label{sec:lopt}
\begin{figure}
\centering
\begin{tabular}{ccc}
 \begin{minipage}[b]{.4\linewidth}
\centering
\psfrag{y}{$k_z$}
\psfrag{x}{$k_\perp$}
\includegraphics[width=\linewidth]{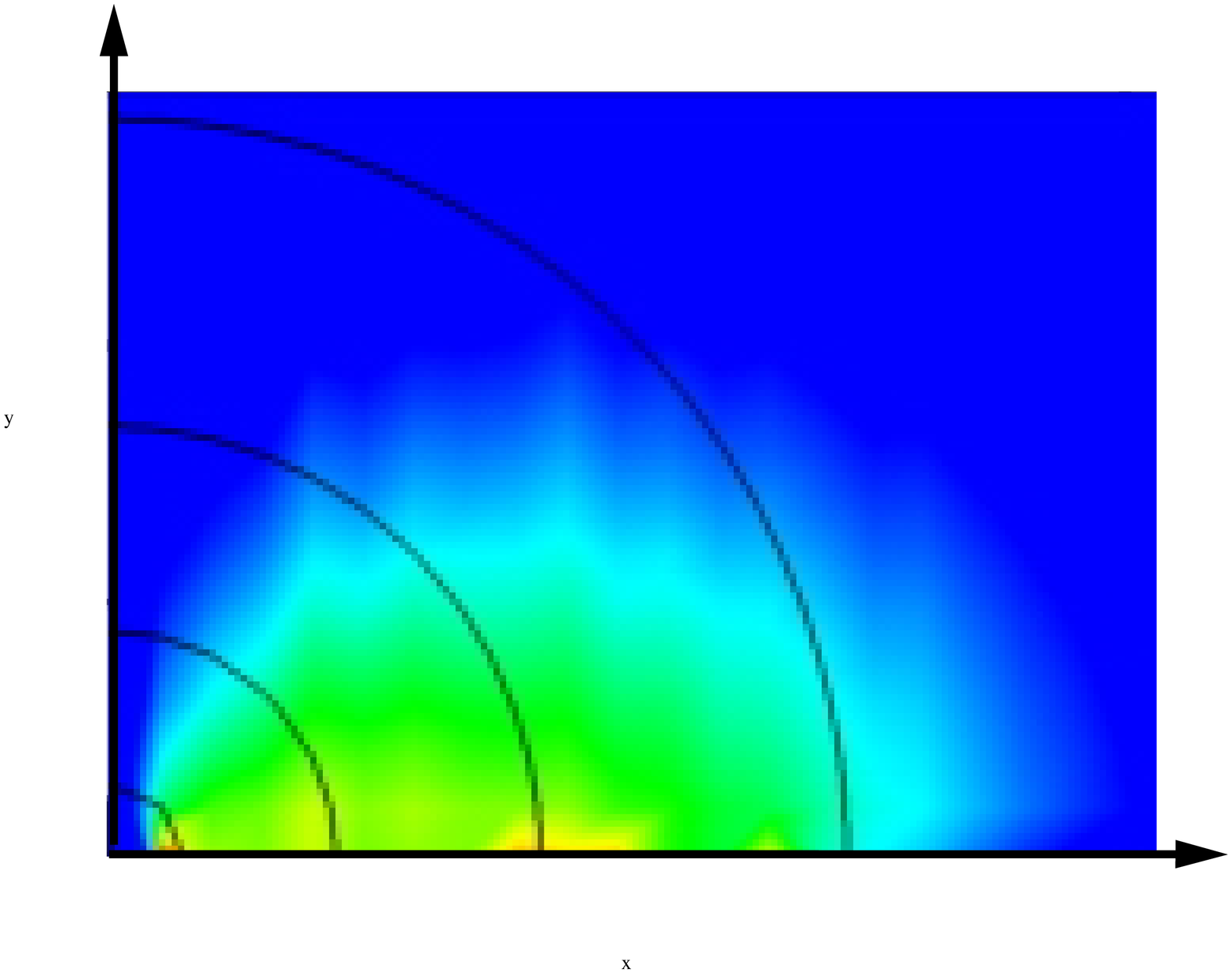}
 \end{minipage}
 &
 \begin{minipage}[b]{.4\linewidth}
\psfrag{y}{$k_z$}
\psfrag{x}{$k_\perp$}
   \centering\epsfig{figure=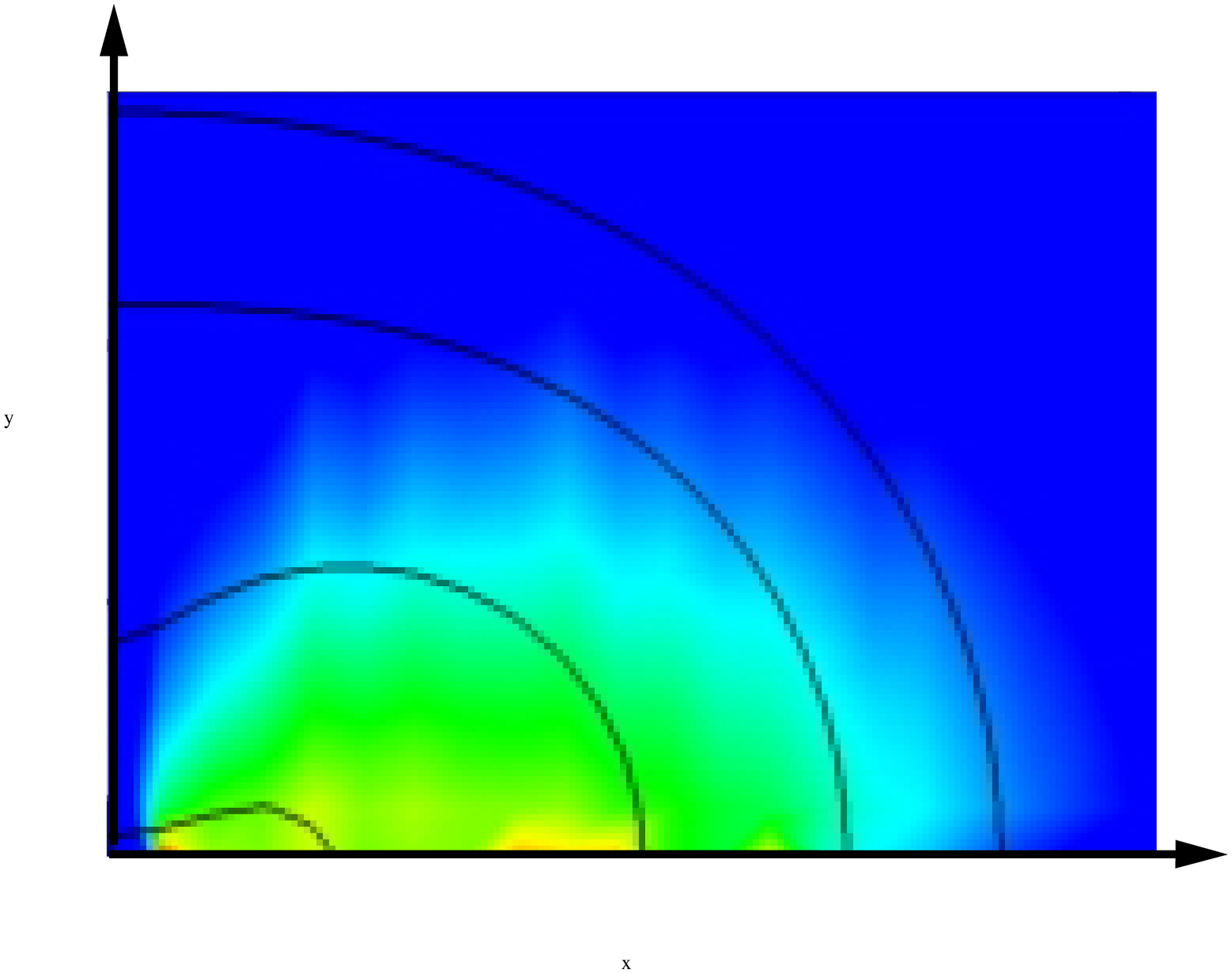,width=\linewidth}
 \end{minipage}
 \\
  (a) $\frac{L}{L_0}=0$
  &
  (b) $\frac{L}{L_0}=\frac{1}{10}2\pi$
 \\
 \begin{minipage}[b]{.4\linewidth}
\psfrag{y}{$k_z$}
\psfrag{x}{$k_\perp$}
   \centering\epsfig{figure=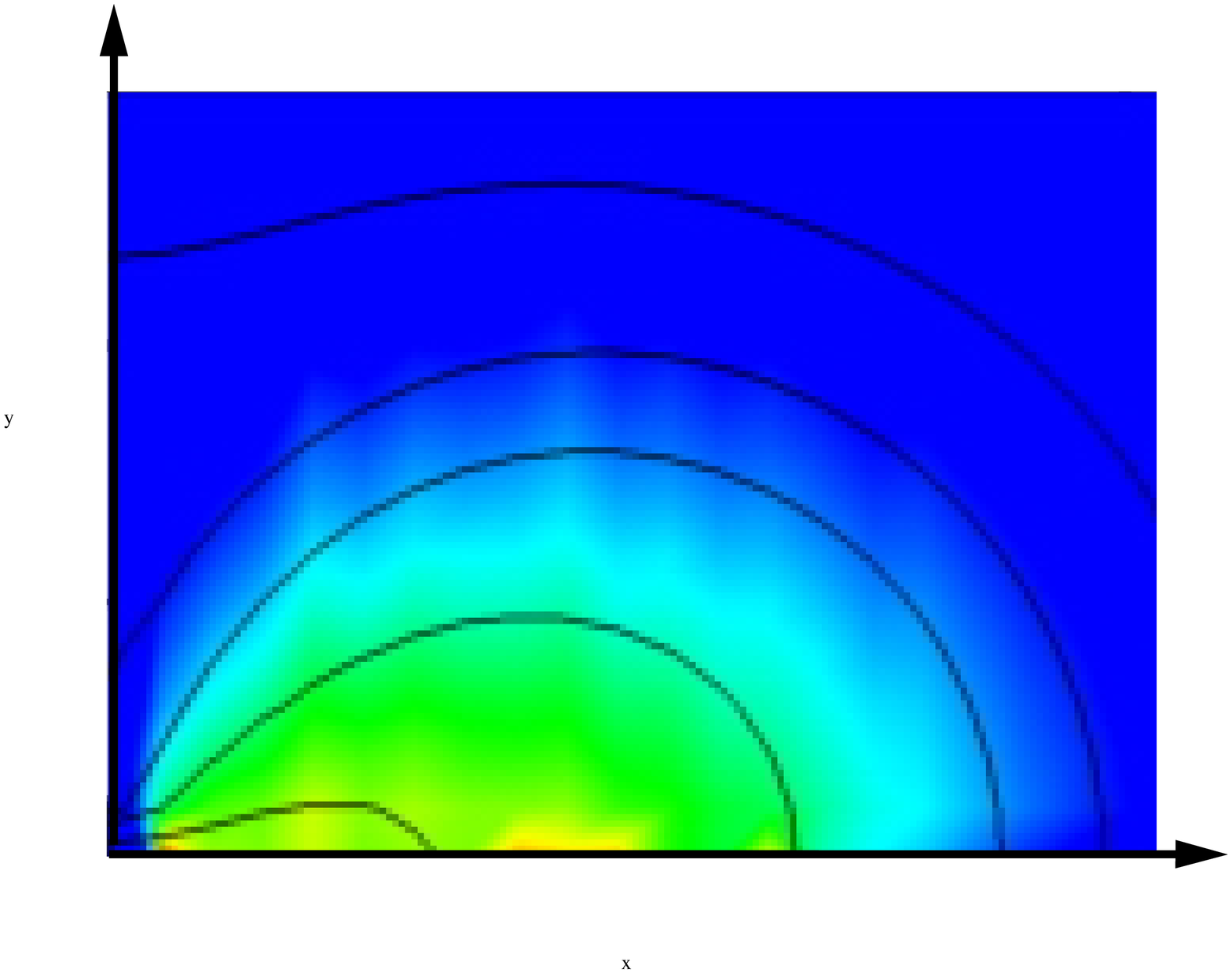,width=\linewidth}
 \end{minipage}
 &
 \begin{minipage}[b]{.4\linewidth}
\psfrag{y}{$k_z$}
\psfrag{x}{$k_\perp$}
   \centering\epsfig{figure=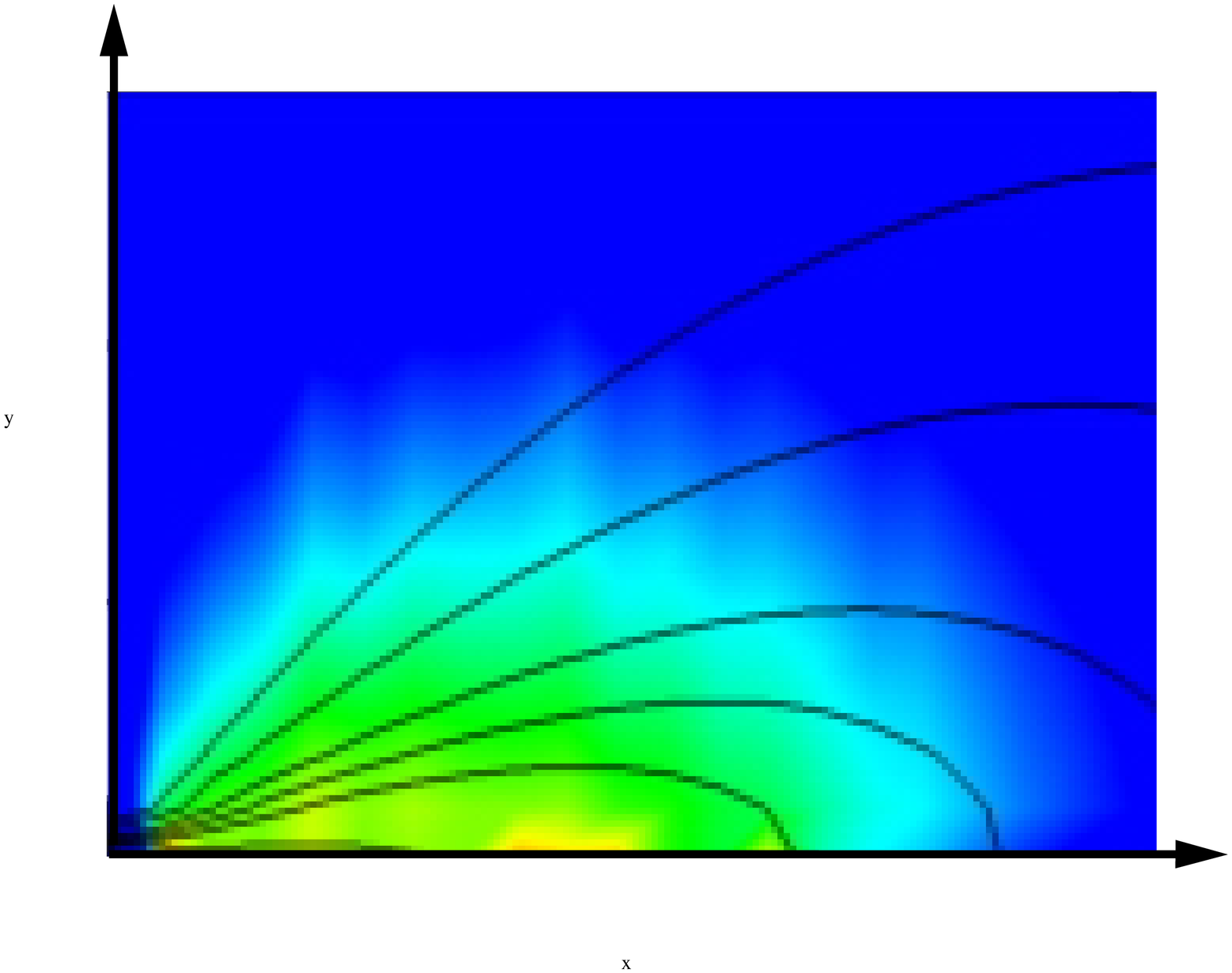,width=\linewidth}
 \end{minipage}
 \\
 (c) $\frac{L}{L_0}=\frac{3}{10}2\pi,\quad {\rm optimal}$
 &
 (d) $\frac{L}{L_0}=2\pi$
 \\
 \end{tabular}
 \caption{Contours of spectral density of energy $E(k_\perp,k_z)$ (colours) 
with iso-$k$ and iso-$\lambda$ curves (solid lines) for several values of $L$ 
at $\Ha_0=80$ and $\Gr=2.94\times10^7$.}
\label{lambda-shells} 

\end{figure}
We first address the problem of choosing the best suited reference length 
$L$ that enters the definition of the Hartmann number $\Ha$,  for a flow at 
given $\Ha_0$ and $\Gr$ (or $\Rey$). This problem appears only 
in three-dimensional flows as in two-dimensional flows, the least dissipative 
modes reduce to the isotropic set of two-dimensional Fourier modes.
At this point, one should remember that the 
choice of the basis is arbitrary and should not have any impact on the final 
solution, as long as its elements can be combined to obtain all the energy 
and dissipation--carrying modes. In the particular case of a basis of least 
dissipative Fourier modes, this gives us the freedom to leave $L$ as a 
free parameter \emph{a priori}, and to fix it so as to obtain a 
basis that contains the least possible non-energetic, non-dissipative modes,
that are superfluous for the description of the solution. How this can be 
done can be understood by analogy with 
the non-MHD case where the flow is expected to be isotropic in regions of the 
Fourier space located far enough from the forcing. There, the energy of a given 
mode $\bf k$ is expected to depend on $\|{\bf k}\|$ only. Similarly, for 
the spectral parameter $\lambda$ to be physically relevant to the MHD case we 
would expect each eigenmode of $D_{\Ha }$ of eigenvalue $\lambda$ located 
far enough from the forced modes $\mathbf k_f$ to carry 
approximately the same amount of energy. The erratic nature of turbulent flows,
however,  
makes it impossible to satisfy this condition exactly, so we shall instead look 
for the optimal value $L_{\rm opt}$ of $L$ that minimises the functional:
\begin{eqnarray}\label{L-optim-Criteria}
\begin{aligned}
\Sigma_{E_\lambda}(L)=\sum\limits_{v\in\sigma(D_{\Ha})}\sum\limits_{{\bf k}:\lambda({\bf k})=v}\left |\frac{ E_{\lambda} ({\bf k})}{ E_{\lambda} }-1 \right|, 
\end{aligned}
\end{eqnarray}
where $\sigma(D_{\Ha})$ refers to the finite set of eigenvalues of $D_{Ha}$ 
for the numerical resolution considered, $E_{\lambda}$ denotes the energy 
summed over all modes of eigenvalue $\lambda$
and 
$E_{\lambda} ({\bf k})$ is the spectral energy density at point $\bf k$ of the iso-$\lambda$ 
surface. $\Sigma_{E_\lambda}(L)$ gives one possible overall measure of how 
strongly $E$ varies over shells shaped according to the iso-$\lambda$ surfaces 
in the Fourier space. 
\begin{figure}
\centering
\begin{tabular}{cc}
 \begin{minipage}[b]{.47\linewidth}
\centering\epsfig{figure=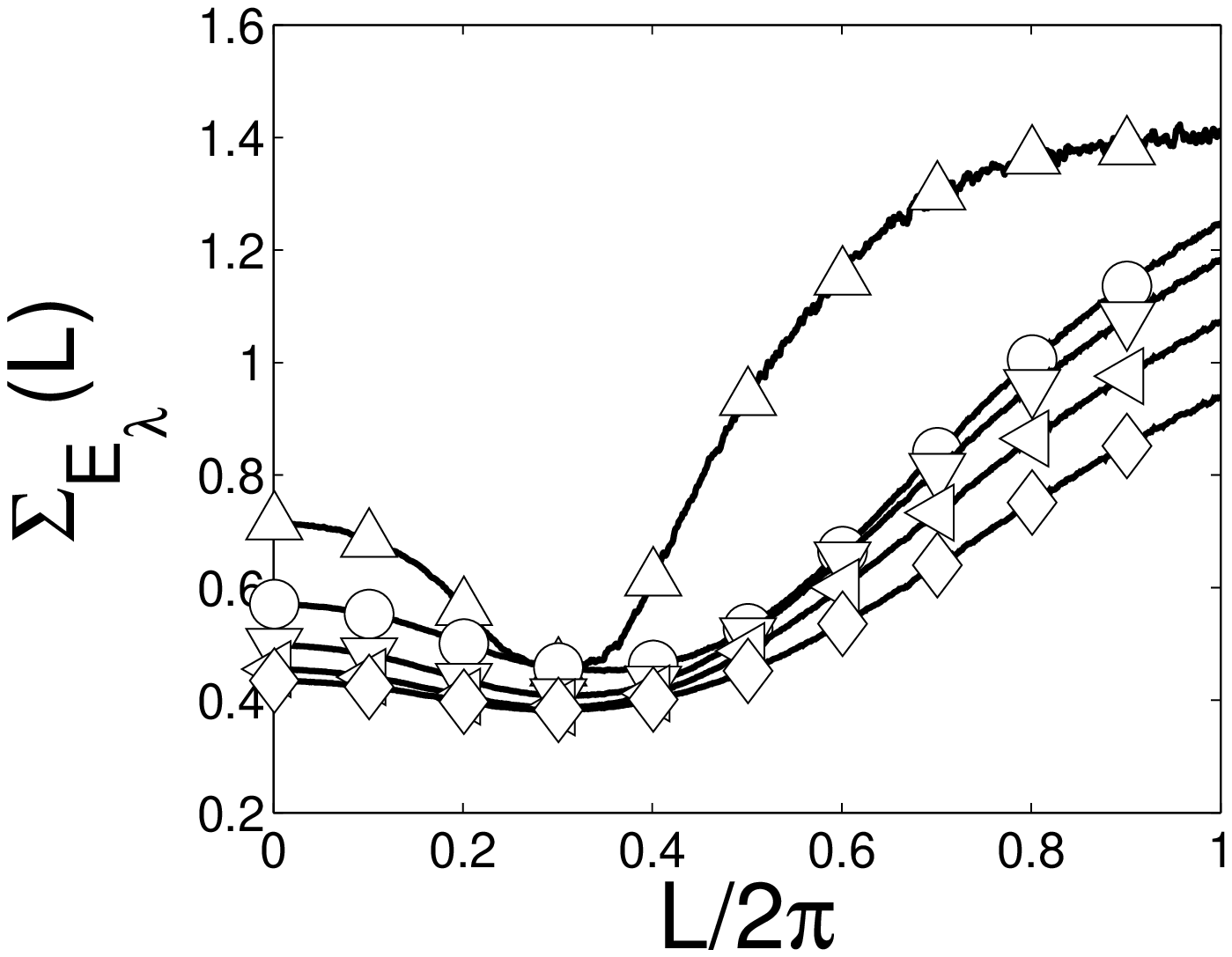,width=6cm,height=6cm}

\end{minipage}
&
\begin{minipage}[b]{.47\linewidth}
\centering\epsfig{figure=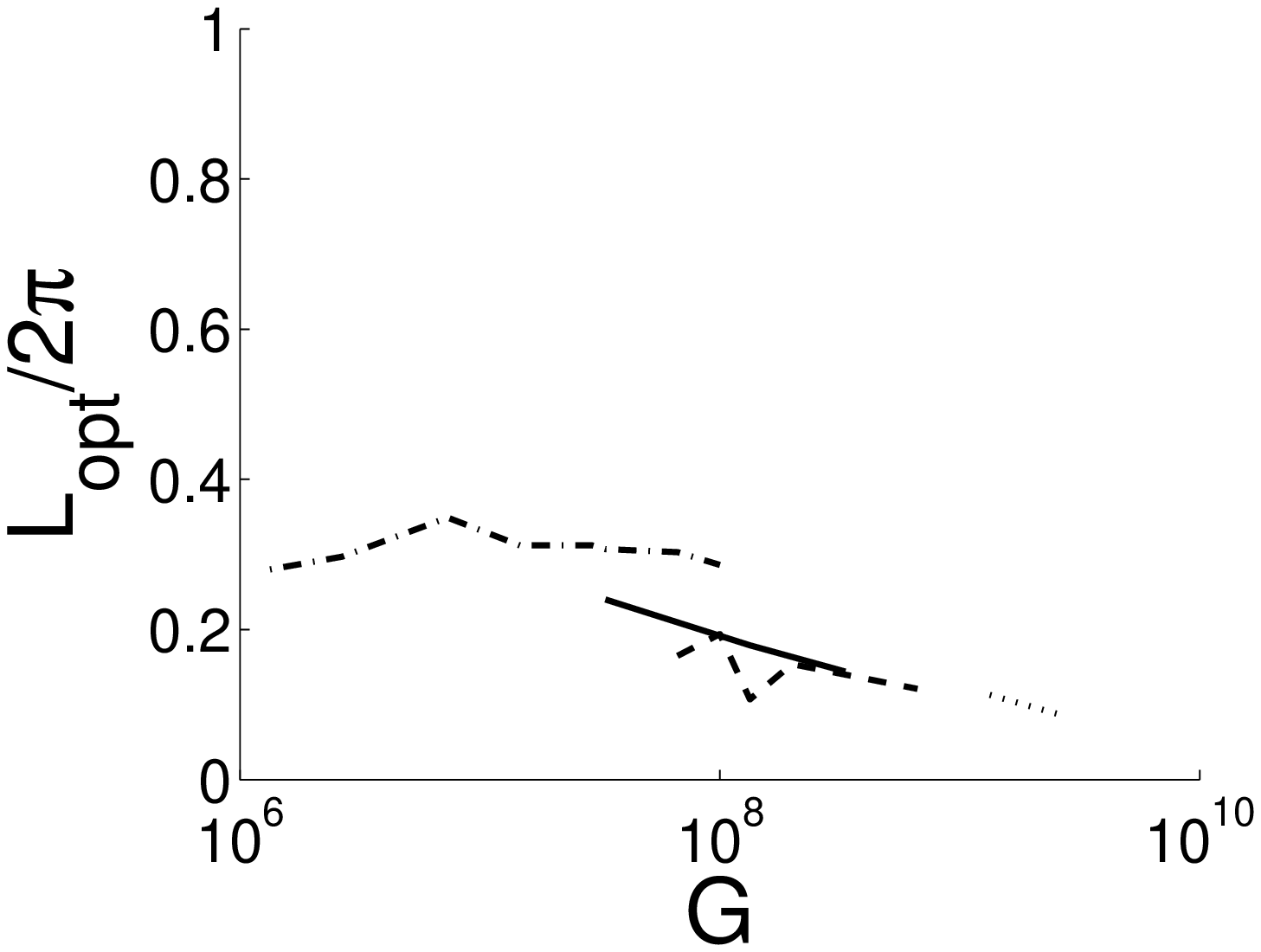,width=6cm,height=6cm}
\end{minipage}
\\
 \end{tabular}

\caption{\textit{Left:}Variations of $\Sigma_{E_\lambda}(L)$ for the 
cases listed in table \ref{table:simul1}. The minima indicate 
$L=L_{\rm opt}$. Symbols are those from table \ref{table:simul1}.
\textit{Right:} variations of $L_{\rm opt}$ with $\Ha_0$ and $\Gr$
for $\Ha_0=80$ and 2D forcing (dash-dot), $\Ha_0=400$ and 2D forcing (dashed)
, $\Ha=1000$ and 2D forcing (dotted) and $\Ha=400$ and 3D forcing (solid).} 
\label{minimize}
\end{figure}
In practice, we start from a "traditional" DNS resolved up to the Kolmogorov 
scale (these cases are gathered in table \ref{table:simul1}), and therefore 
over-resolved in the MHD case (on the basis that the attractor dimension 
decreases monotonically when $\Ha$ increases (\cite{Potherat2003}). This yields a reference solution from which 
$E(k_\perp,k_z)$ can be extracted. We then calculate the minimum of 
functional $\Sigma_{E_\lambda}(L)$ 
numerically (the variations of $\Sigma_{E_\lambda}(L)$ are shown on figure 
\ref{minimize}, left). This is illustrated on a typical example for $\Ha_0=80$ and 
$G=2.94\times10^7$ on figure \ref{lambda-shells} where the sets of 
iso-$\lambda$ curves 
are plotted for several values of $L$ along with the contours of 
$E(k_\perp,k_z)$. One sees that the iso-$\lambda$ curves corresponding to 
$L_{\rm opt}/L_0=0.3\times 2\pi$  on Figure \ref{lambda-shells}(c) follow the 
energy distribution well, as opposed to iso-$\bf k$ lines, shown on Figure 
\ref{lambda-shells}(a) that cross many different levels of energy. This 
shows that the basis of the least dissipative modes does carry the 
morphology of the energy distribution quite realistically, provided 
we choose $L\simeq L_{\rm opt}$.
 It can be seen from the variations of $L_{\rm opt}$ with $\Gr$ for $\Ha_0\in\{80,400\}$ on figure \ref{minimize}, that it depends
 little on either $G$ or $\Ha_0$,  around $0.3L_0\times2\pi$ for $\Ha_0=80$ and $0.2L_0\times2\pi$ 
for $\Ha_0=400$. The fact that it still varies
a little with $\Ha_0$ 
is certainly due in part to the "non--universality" introduced by the forcing, 
as the energy distribution clearly departs from the 
iso-$\lambda$ lines in the vicinity of the forced modes. At a given $\Rey$, or 
$\Gr$, the 
influence of these modes increases with $\Ha_0$, as for higher $Ha_0$, the 
energy tends to stay closer to the $(k_x,k_y)$ plane, which brings the smallest 
scales closer to the forced modes $\mathbf k_f$. 
For the purpose of performing DNS based on the least dissipative modes, a  
precise determination of $L_{\rm opt}$ is however not necessary 
as energy and dissipation spectra $E(\lambda)$ and $D(\lambda)$ obtained with 
$L$ departing by around $\pm 30\%$ from $L_{\rm opt}$, using 
only modes in the region $|\lambda|<|\lambda^{\rm max}|$ (where $\lambda^{\rm max}$  was fixed according to (\ref{L-optim-Criteria}) derived in the next 
section) yielded no significant discrepancy with those obtained from 
calculations based on $L_{\rm opt}$ exactly. This robustness 
also confirms that as long as  the iso-$\lambda$ curves 
follow the contours of energy well enough in the vicinity of the small scales, 
then the set of Fourier modes
 determined by $\lambda^{\rm max}$ 
contains very few non--relevant modes. 
Also, since $\Ha_{\rm opt}= BL_{\rm opt}\sqrt{\sigma/{(\rho\nu)}}$ gives the 
most physically relevant measure of the Lorentz force, we shall now prefer 
it to $\Ha_0$ to express the laws for the small scales
(\ref{eq:heur_scal_lambda}) and (\ref{eq:sscale_mhd}).
\subsection{Scaling laws for $\lambda^{\rm max}$}
\label{sec:scaling}

\begin{figure}
\begin{center}
\begin{tabular}{cccc}
 $ f_{2D} $,
&$ f_{2D}$,
&$ f_{2D}$,
&$ f_{3D}$,
\\
 $\Ha_0=80$,
&$\Ha_0=400$,
&$\Ha_0=1000$,
&$\Ha_0=400$,
\\
  $G=7.34\times10^7$
& $G=10^8$
& $G=2.67\times10^9$
& $G=1.2\times10^{10}$
\\
\psfrag{x}{$k_\perp$}
\psfrag{y}{$k_z$}
\includegraphics[width=2cm,height=2cm]{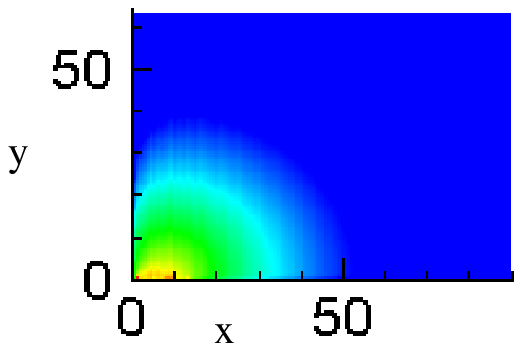}
&
\psfrag{x}{$k_\perp$}
\psfrag{y}{$k_z$}
\includegraphics[width=3cm,height=2.5cm]{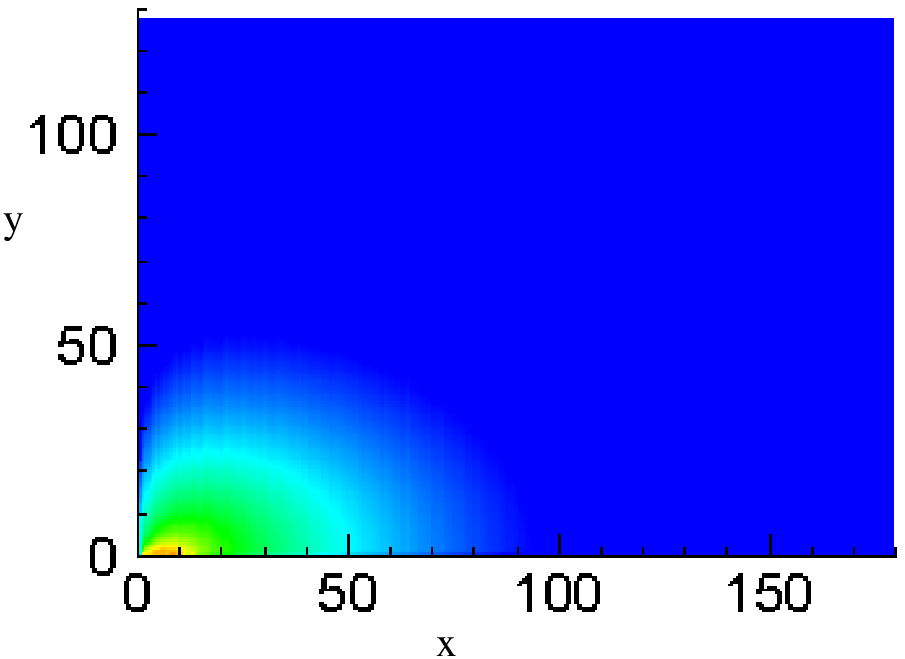}
&
\psfrag{x}{$k_\perp$}
\psfrag{y}{$k_z$}
\includegraphics[width=3.5cm,height=2.5cm]{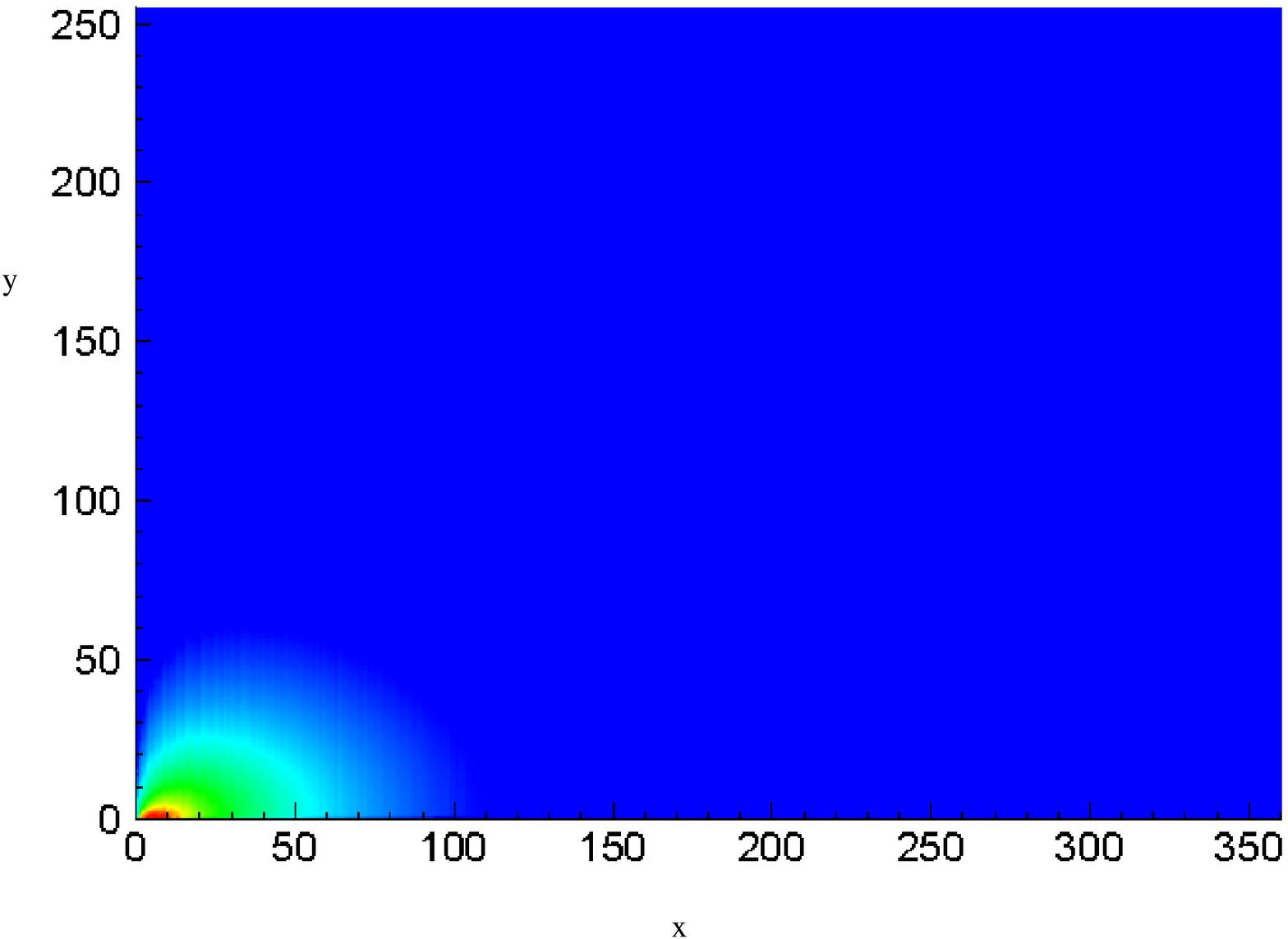}
&
\psfrag{x}{$k_\perp$}
\psfrag{y}{$k_z$}
\includegraphics[width=3cm,height=2.5cm]{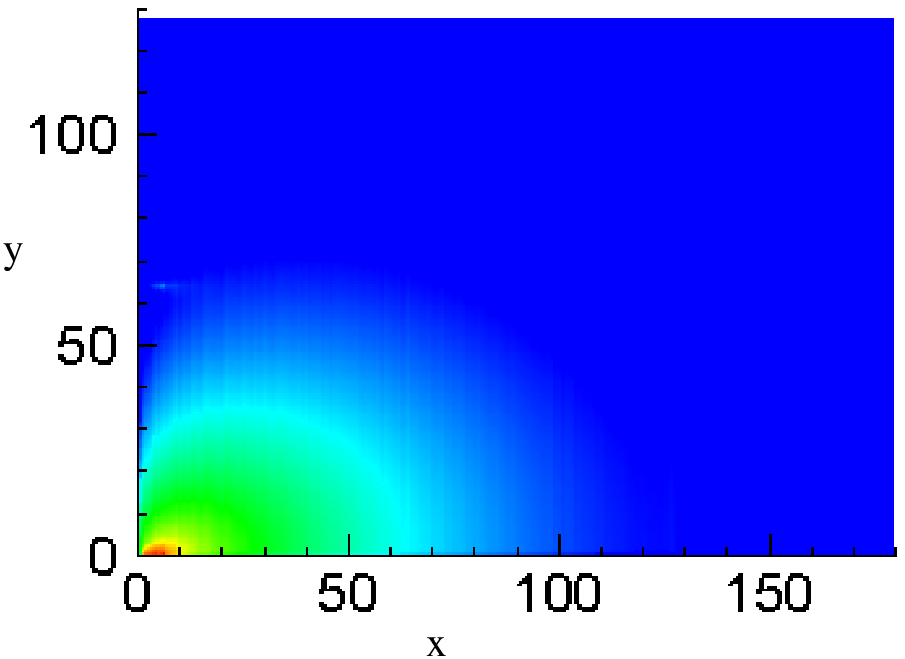}
\\
\psfrag{x}{$k_\perp$}
\psfrag{y}{$k_z$}
$k\leq64 (128^3)$&$k\leq128 (256^3)$ & $k\leq256(512^3)$ & $k\leq128(256^3)$
\\
\psfrag{x}{$k_\perp$}
\psfrag{y}{$k_z$}
\includegraphics[width=2cm,height=2cm]{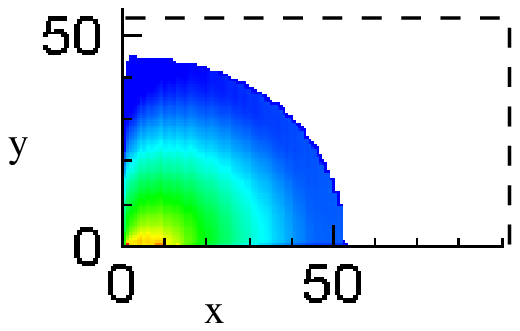}
&
\psfrag{x}{$k_\perp$}
\psfrag{y}{$k_z$}
\includegraphics[width=3cm,height=2.5cm]{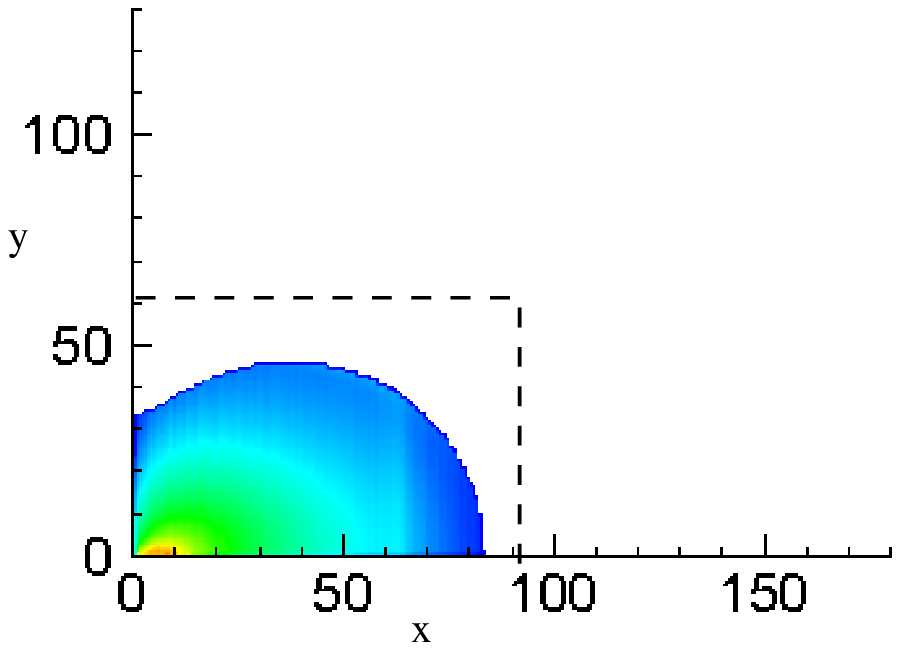}
&
\psfrag{x}{$k_\perp$}
\psfrag{y}{$k_z$}
\includegraphics[width=3.5cm,height=2.5cm]{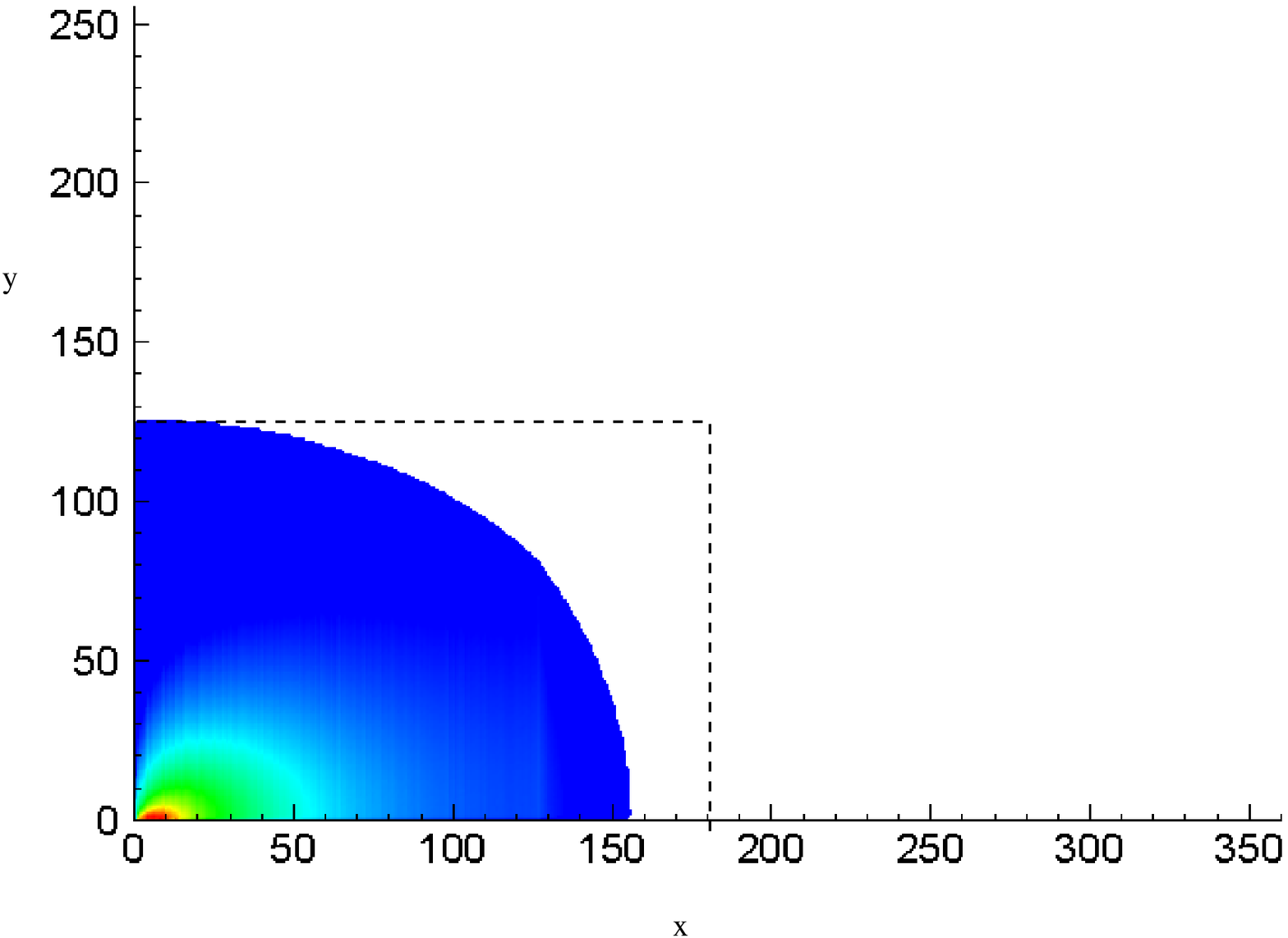}
&
\psfrag{x}{$k_\perp$}
\psfrag{y}{$k_z$}
\includegraphics[width=3cm,height=2.5cm]{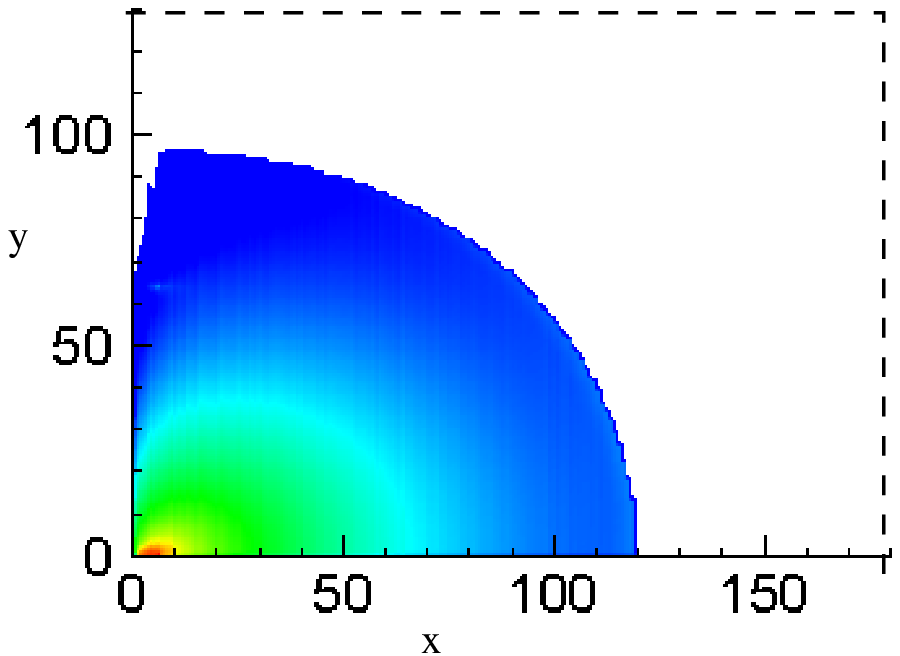}
\\
$C_\lambda=0.59$	&$C_\lambda=0.59$	&$C_\lambda=0.59$	&$C_\lambda=0.48$
\\
$\frac{|\lambda|^{1/2}}{2\pi}\leq53 (128^3)$ 
&
$\frac{|\lambda|^{1/2}}{2\pi}\leq 84 (128^3)$ 
&
$\frac{|\lambda|^{1/2}}{2\pi}\leq 152 (256^3)$
&
$\frac{|\lambda|^{1/2}}{2\pi}\leq 120 (256^3)$
\\
\psfrag{x}{$k_\perp$}
\psfrag{y}{$k_z$}
\includegraphics[width=2cm,height=2cm]{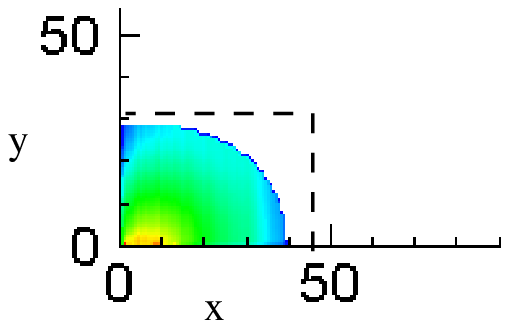}
&
\psfrag{x}{$k_\perp$}
\psfrag{y}{$k_z$}
\includegraphics[width=3cm,height=2.5cm]{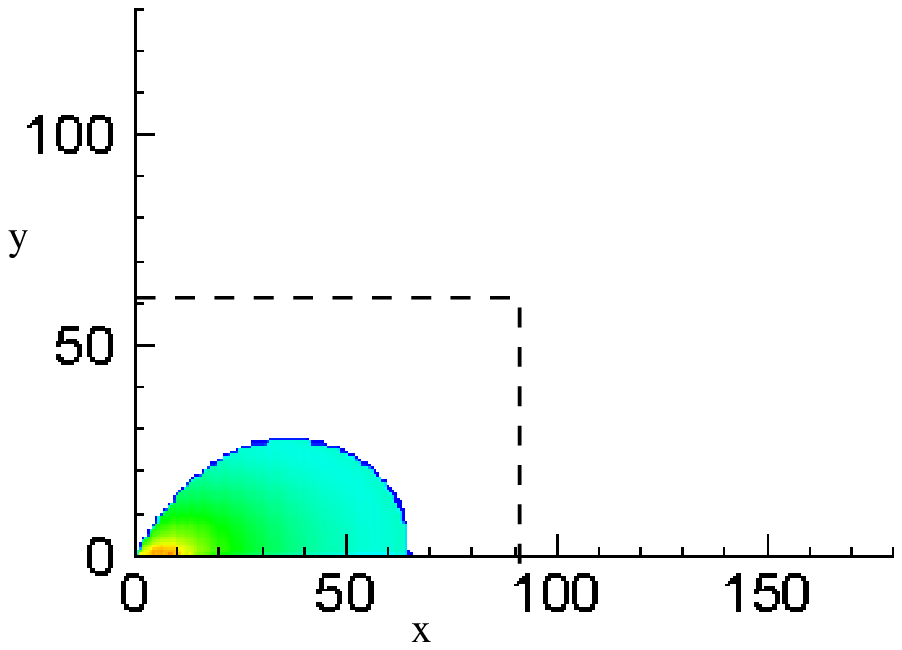}
&
\psfrag{x}{$k_\perp$}
\psfrag{y}{$k_z$}
\includegraphics[width=3.5cm,height=2.5cm]{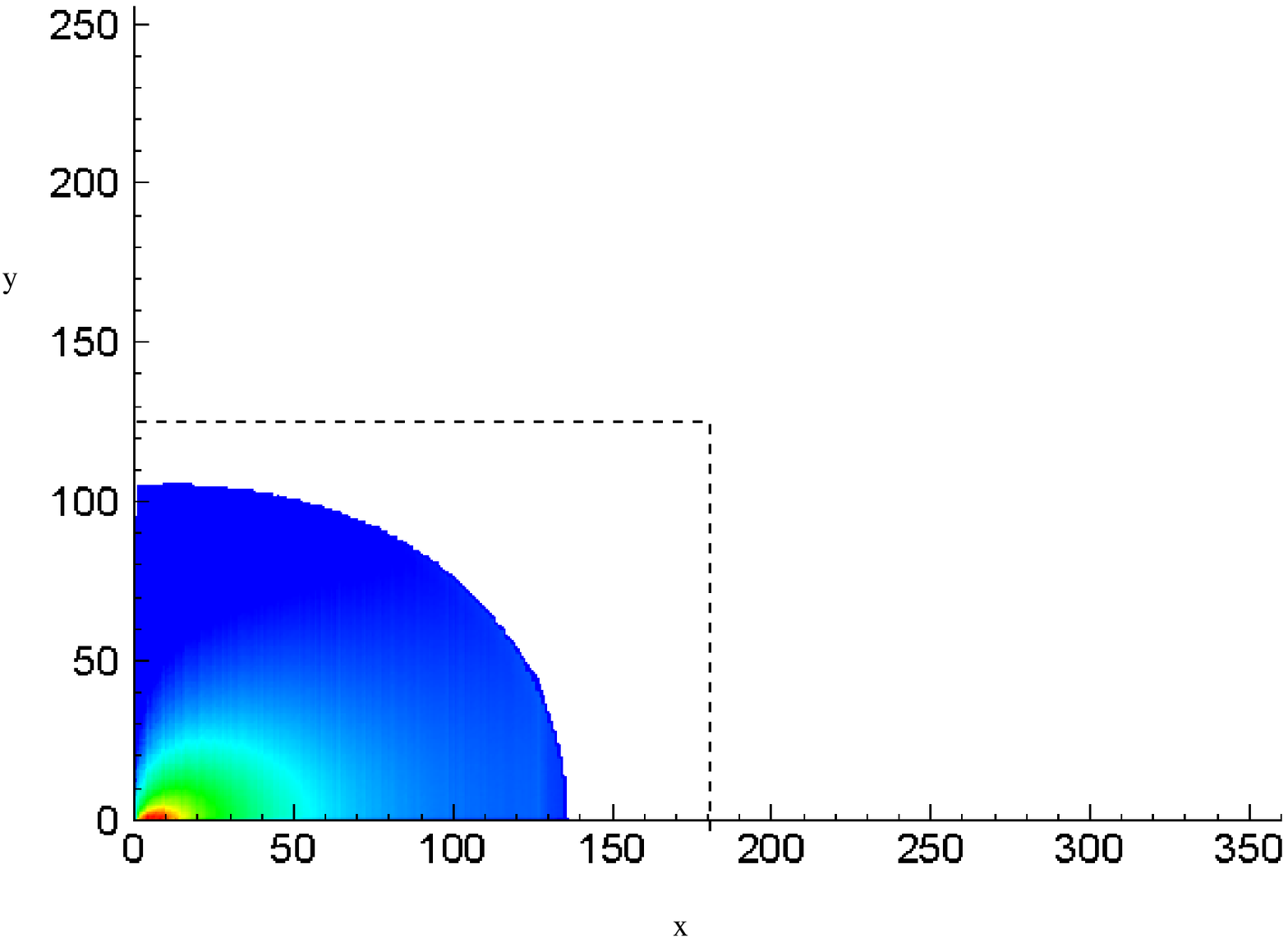}
&
\psfrag{x}{$k_\perp$}
\psfrag{y}{$k_z$}
\includegraphics[width=3cm,height=2.5cm]{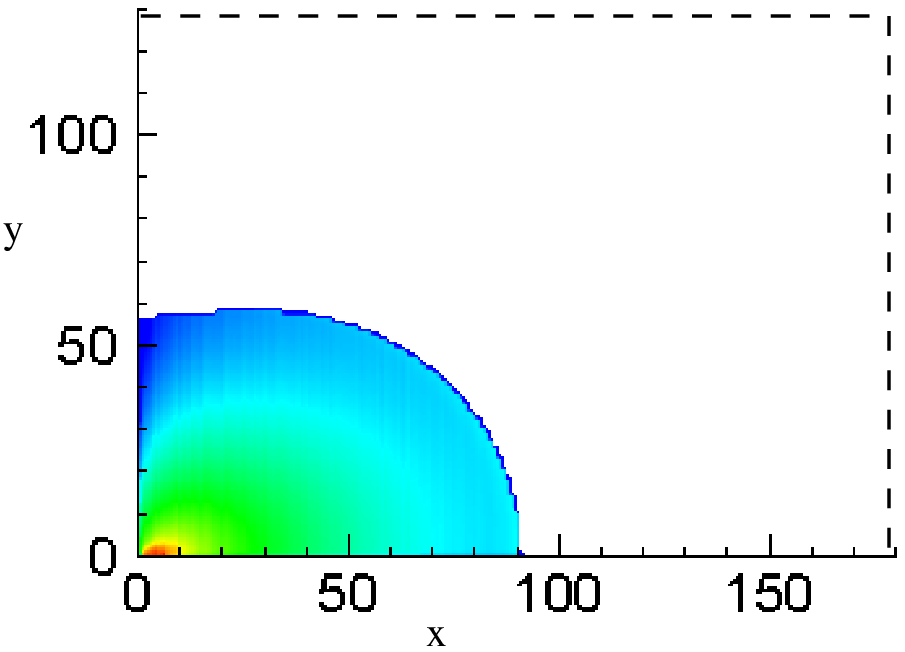}
\\
$C_\lambda=0.47$	&$C_\lambda=0.47$	&$C_\lambda=0.47$	&$C_\lambda=0.38$
\\
$\frac{|\lambda|^{1/2}}{2\pi}\leq 40 (64^3)$  
&
$\frac{|\lambda|^{1/2}}{2\pi}\leq 65 (128^3)$
&
$\frac{|\lambda|^{1/2}}{2\pi}\leq 136 (256^3)$    
&
$\frac{|\lambda|^{1/2}}{2\pi}\leq 91 (256^3)$
\\
\psfrag{x}{$k_\perp$}
\psfrag{y}{$k_z\:$}
\includegraphics[width=2cm,height=2cm]{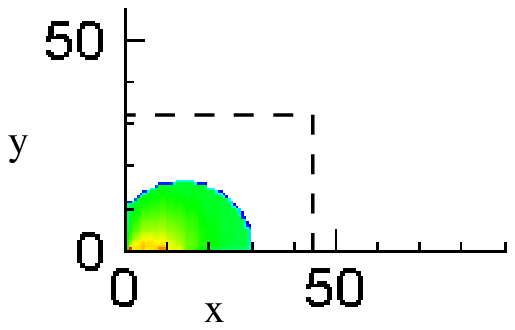}
&
\psfrag{x}{$k_\perp$}
\psfrag{y}{$k_z\quad$}
\includegraphics[width=3cm,height=2.5cm]{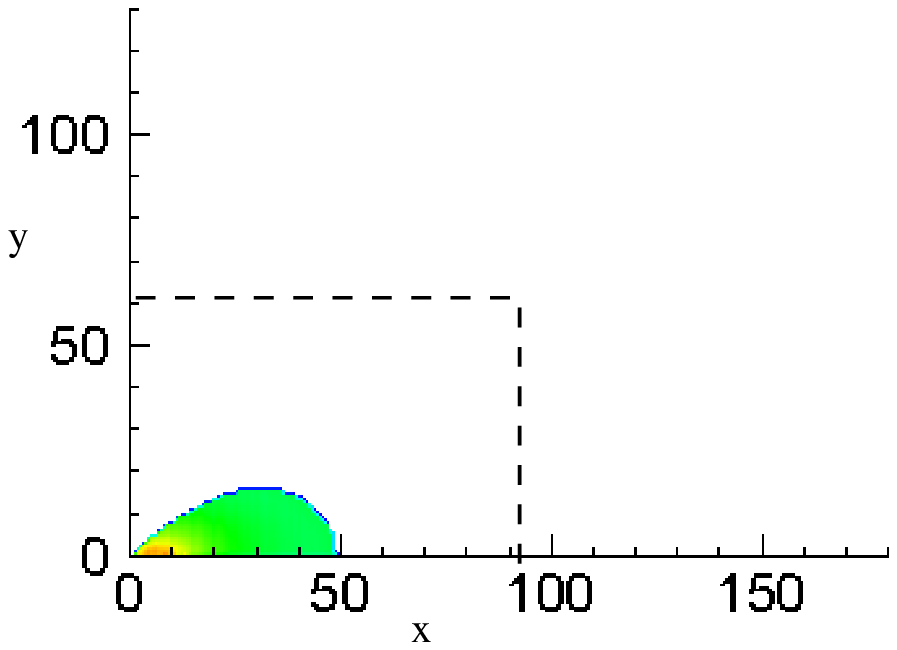}
&
\psfrag{x}{$k_\perp$}
\psfrag{y}{$k_z\:$}
\includegraphics[width=3.5cm,height=2.5cm]{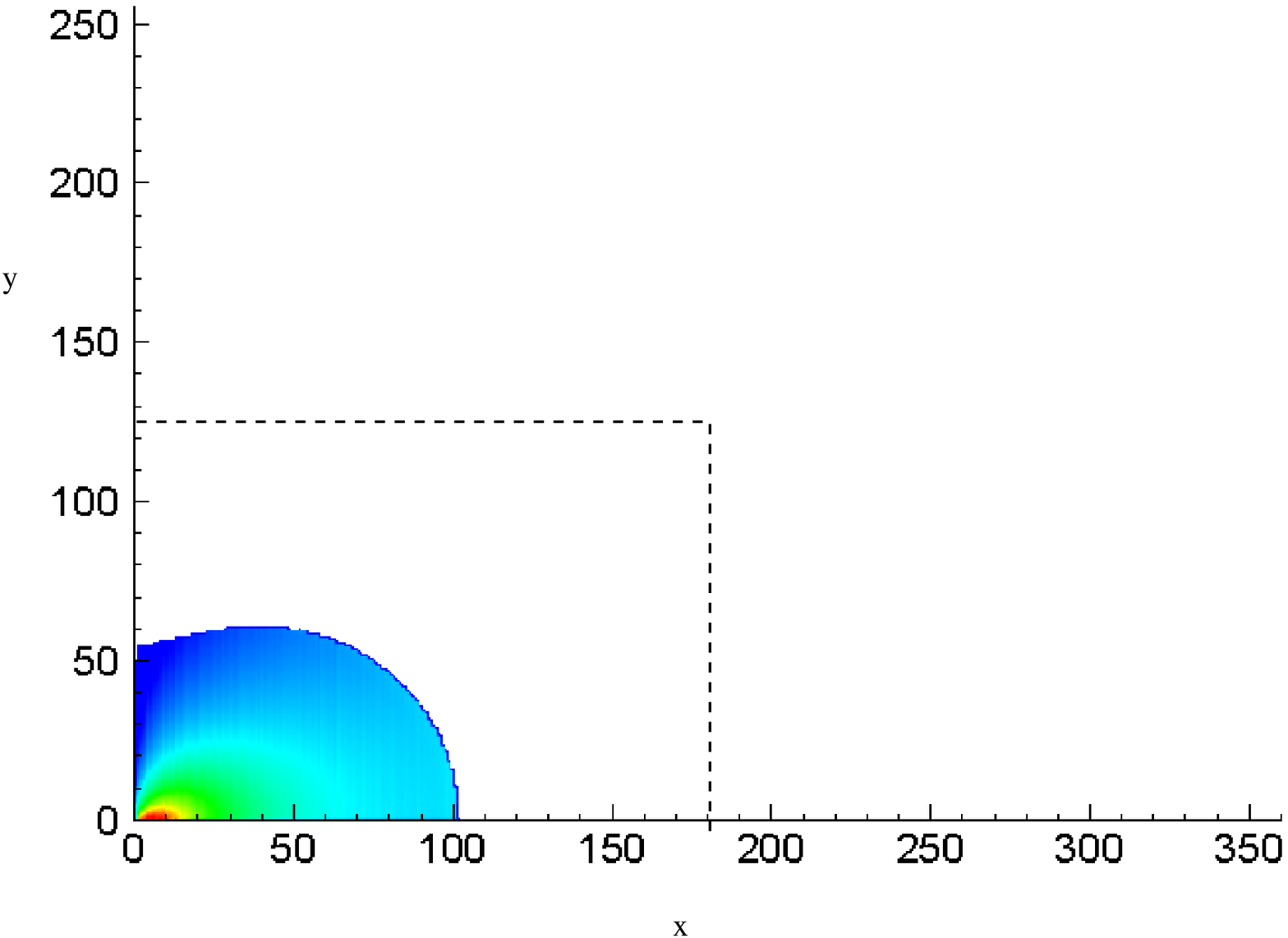}
&
\psfrag{x}{$k_\perp$}
\psfrag{y}{$k_z\:$}
\includegraphics[width=3cm,height=2.5cm]{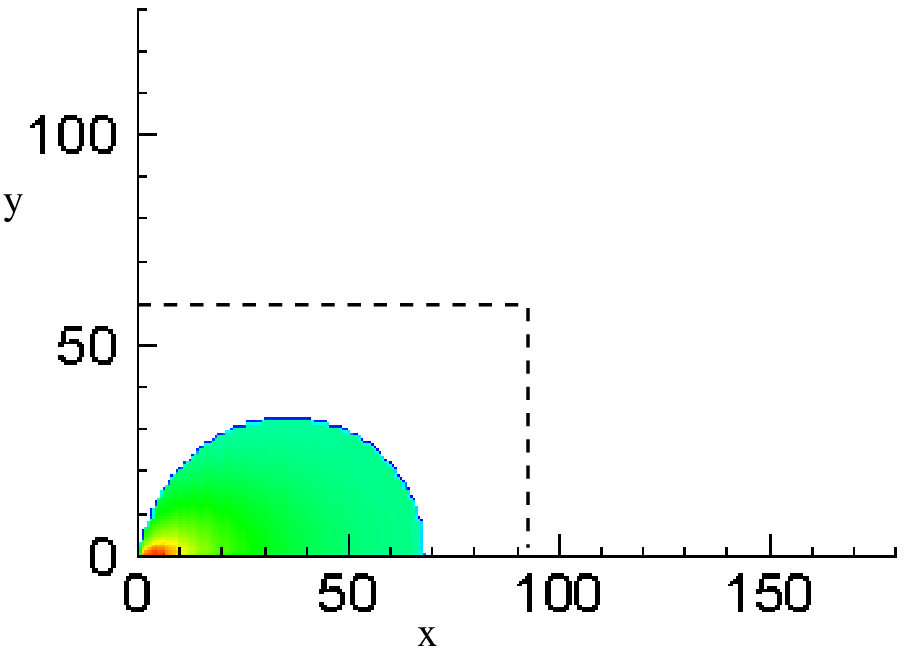}
\\
$C_\lambda=0.35$	&$C_\lambda=0.35$	&$C_\lambda=0.35$	&$C_\lambda=0.29$
\\
$\frac{|\lambda|^{1/2}}{2\pi}\leq 30 (64^3)$
&
$\frac{|\lambda|^{1/2}}{2\pi}\leq 49 (128^3)$ 
&
$\frac{|\lambda|^{1/2}}{2\pi}\leq 102 (256^3)$ 
&
$\frac{|\lambda|^{1/2}}{2\pi}\leq 68 (128^3)$
\\

\end{tabular}
\caption{Logarithmic energy distribution in the $(k_\perp,k_z)$-plane. Blue dots 
correspond to low energy modes. Each column represents flows calculated 
with the same control parameters and the same forcing (indicated at the top), 
but with different resolutions, determined by the value of $C_\lambda$ or, 
equivalently, by  
the spectral domain of resolution defined by $\lambda<\lambda^{\rm cut}$, both
indicated below each graph. 
Calculations from the first line are resolved up to the 
Kolmogorov scale $k_{\kappa}=C_\kappa\Rey ^{3/4}$, with $\Rey=\ 97,\ 245,\ 1140,\ 512$ (see table \ref{table:simul1}).
The dashed lines indicate the embedding resolutions used in our code (in 
brackets). Modes in the white area within this rectangular domain are set to 0.}

\label{AngularSpectra}
\end{center}
\end{figure}

\begin{table}
\begin{center}
{\small
  \begin{tabular}{ l l cc r  l r l r l l c c cc cc cc}
 $\Ha_0$  &&  $\Gr$  &&$n_x\times n_y\times n_z$  && $\Rey$ && $C_\lambda$ && $C_\kappa$ && $\alpha_E(0.5)$ && $\Ha_{\rm opt}/2\pi$ && $S_{\rm opt}$ &&symbol
\\[3pt]
\hline
 &&  && 2D forcing &&  &&  &&  &&  &&  &&
\\
\hline

80  &&  2.67$\times 10^6$  && $064\times064\times064$ &&   70  && 0.55 &&  1.33  && 0.999 &&  23.70 && 8.02  && $\triangle$\\
80  &&  7.34$\times 10^7$  && $128\times128\times128$ &&   97  && 0.81 &&  2.07  && 0.996 &&  27.80 && 7.97  && $\circ$   \\
80  &&  1.47$\times 10^7$  && $128\times128\times128$ &&  159  && 0.63 &&  1.43  && 0.998 &&  24.80 && 3.87 && $\triangledown$ \\
80  &&  2.94$\times 10^7$  && $128\times128\times128$ &&  194  && 0.57 &&  1.23  && 0.996 && 24.90 && 3.20 && $\vartriangleleft$  \\
80  &&  3.34$\times 10^7$  && $128\times128\times128$ &&  195  && 0.57 &&  1.23  && 0.995  && 24.50 && 3.08 && $\square$  \\
80  &&  6.67$\times 10^7$  && $128\times128\times128$ &&  264  && 0.50 &&  0.98  && 0.998 &&  24.20 && 2.21 && $\diamond$\\
400 &&  6.67$\times 10^7$  && $128\times128\times128$ &&  216  && 0.71 && 1.13  && 0.995  &&  65.80 && 20.04 && $\blacktriangle$ \\
400 &&  1.00$\times 10^8$  && $256\times256\times256$ &&  245  && 1.11 && 2.07  && 0.986  &&  77.40 && 24.45 && $\bullet$ \\
400 &&  1.33$\times 10^8$  && $128\times128\times128$ &&  282  && 0.53 && 0.93  && 0.998  &&  42.70 && 6.47 && $\blacktriangledown$\\
400 &&  2.00$\times 10^8$  && $256\times256\times256$ &&  343  && 0.90 && 1.61  && 0.994  &&  61.40 && 10.99 && $\blacktriangleleft$ \\
400 &&  6.67$\times 10^8$  && $512\times512\times256$ &&  575  && 1.28 && 1.09  && 0.986  &&  48.25 && 4.05 && ${\blacksquare}$
\\
1000&&  1.33$\times 10^9$  && $512\times512\times256$ &&  935  && 1.08 &&  1.52 && 0.998 && 112.70 && 13.58 && $\color{gray}\blacktriangle$  \\
1000&&  2.67$\times 10^9$  && $512\times512\times512$ && 1140  && 0.94 &&  1.3  && 0.996 &&  86.00 &&  6.44 && $\color{gray}\bullet$   \\

\hline
 &&  && 3D forcing &&  &&  &&  &&  &&  &&
\\
\hline
400  &&  1.5$\times 108$    && $256\times256\times256$ &&   465  &&
0.71 && 1.28 && 1.000  && 95.00 && 19.41 && $\color{gray}\blacksquare$
\\ 
400  &&  6.0$\times 10^{8}$  && $256\times256\times256$ &&   512  &&
0.62 && 1.19 && 0.999  && 71.00 && 9.84  &&
$\color{gray}\blacktriangledown$\\ 
\hline
\end{tabular}
\caption{ Summary of all cases calculated with initial condition
$\mathbf u(t=0)=0$: Grashof number $\Gr$, embedding spectral resolution $n_x\times n_y\times n_z$, Reynolds number $\Rey$,  resolution ($C_\lambda$ and 
$C_\kappa$), fraction of the total energy $\alpha_{3D}$ contained in modes with 
lower $|\lambda|$ than the value given by (\ref{eq:final_scal_lambda}), 
``optimal`` Hartmann number $Ha_{\rm opt}=Ha_0 (L_{\rm opt}/L_0)$, and 
``optimal`` interaction parameter $S_{\rm opt}=Ha_{\rm opt}^2/(4\pi^2 Re)$.}
\label{table:simul1}
}
\end{center}
\end{table}
Having chosen $L=L_{\rm opt}$, we have fixed a family of modes, indexed by the 
corresponding sequence of values of $\lambda$. We now need to know how many of 
these modes are required to resolve the flow fully, for given values of 
$\Ha_{\rm opt}$ and $\Gr$ (or $\Rey$). A usable estimate for this 
number is obtained through a  value for the numerical constant $C_\lambda$ that 
appears in the scaling law for the smallest scales 
$\lambda^{\rm max}(\Ha_{\rm opt},\Rey)$ 
 (\ref{eq:heur_scal_lambda}). To find it, we select four cases covering 
different values of $Ha$, $\Gr$, two and three-dimensional forcing.
In each case, we first calculate the established state with resolution up to 
the Kolmogorov scale 
$k_\kappa$ (summarised in table 
\ref{table:simul1}). 
Since this case is over-resolved, it serves as a reference for the energy 
and dissipation distribution  in the Fourier space. 
We then recalculate several times the same flow, but resolved up to 
$|\lambda|^{1/2}=|\lambda^{\rm cut}|^{1/2}=C_\lambda\Rey^{1/2}$ with different values 
of $C_\lambda$ and compare the corresponding power and dissipation density spectra 
$E(\lambda)$ and $D(\lambda)$ to those obtained in the reference DNS 
resolved to the Kolmogorov scale. 
Finally, the impact of the reduction in resolution on time dependent-flows is assessed by applying 
the same procedure to the freely decaying flow that follows a shutdown of the 
forcing in the established regime in the reference case, at $t=t_{\rm decay}$.\\  
Figure \ref{AngularSpectra} summarises all calculated 
cases along with resolution and embedding resolution. The latter is of no 
incidence on the solution but gives a measure of the reduction in computational 
cost incurred by using our ``$\lambda$-based`` approach, and this, even though 
the spectral code we are using hasn't been optimised for it.\\

%
%
%
%
%
%
\begin{figure}
\begin{center}
\begin{tabular}{cc}
$f_{2D}$, $\Ha_0=80$, $G=7.34\times10^7$
&$ f_{2D} $, $\Ha_0=400$, $G=10^8$\\
\psfrag{x}{$0.5\log(|\lambda|)$}
\psfrag{y}{$\log(E(|\lambda|^{1/2}))$}
\includegraphics[width=6cm,height=4cm]{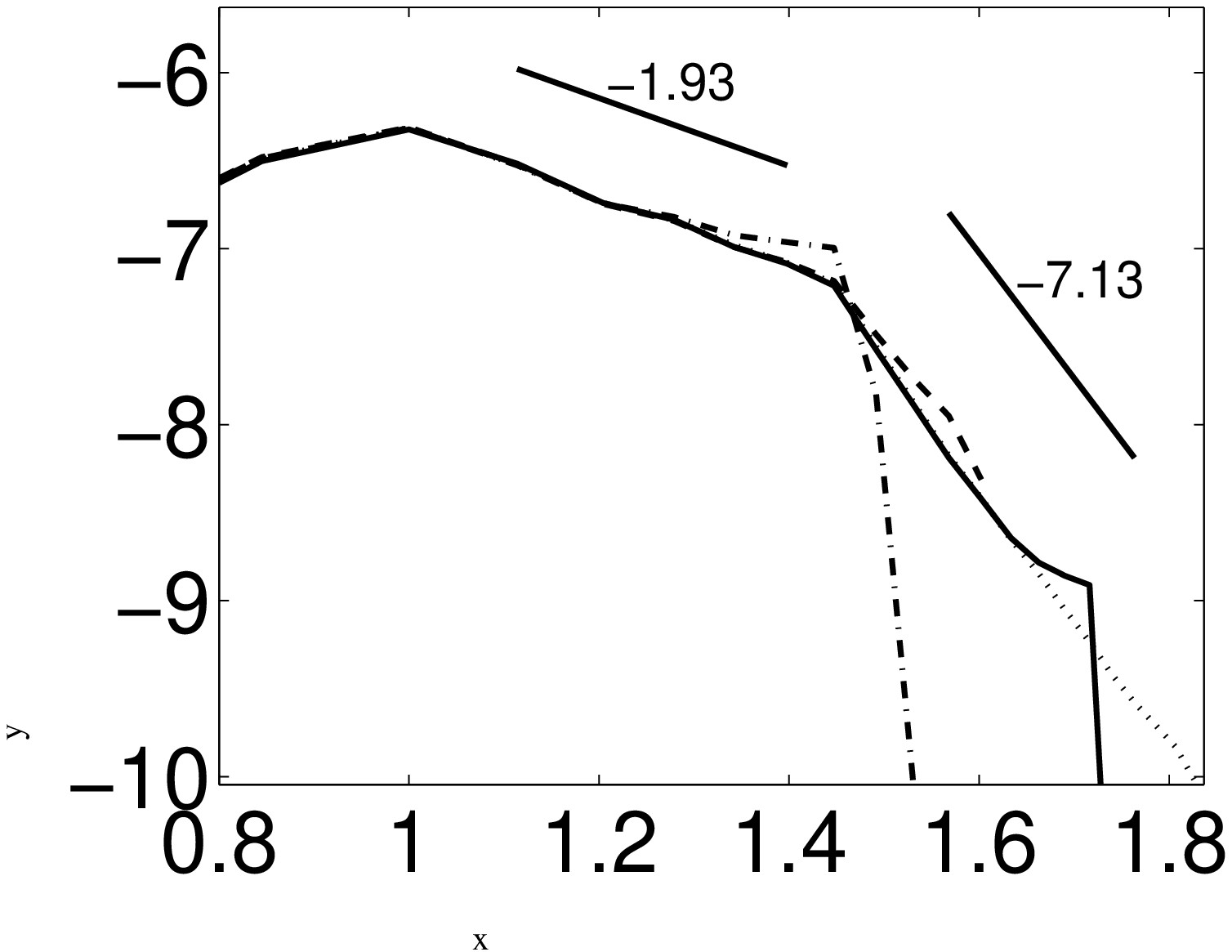}
&
\psfrag{x}{$0.5\log(|\lambda|)$}
\psfrag{y}{$E(|\lambda|^{1/2})$}
\includegraphics[width=6cm,height=4cm]{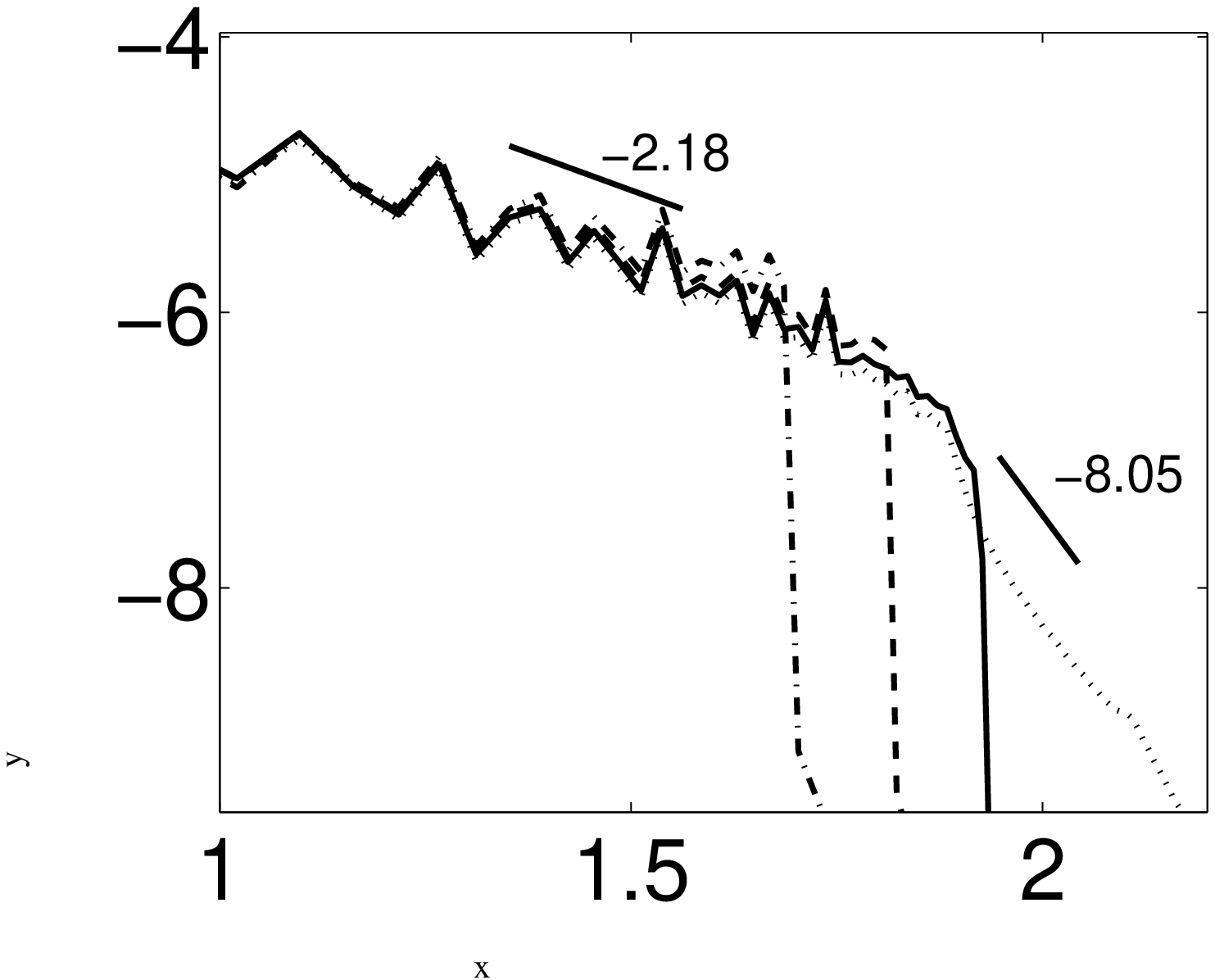}
\\
\psfrag{x}{$0.5\log(|\lambda|)$}
\psfrag{y}{$\log(D(|\lambda|^{1/2}))$}
\includegraphics[width=6cm,height=4cm]{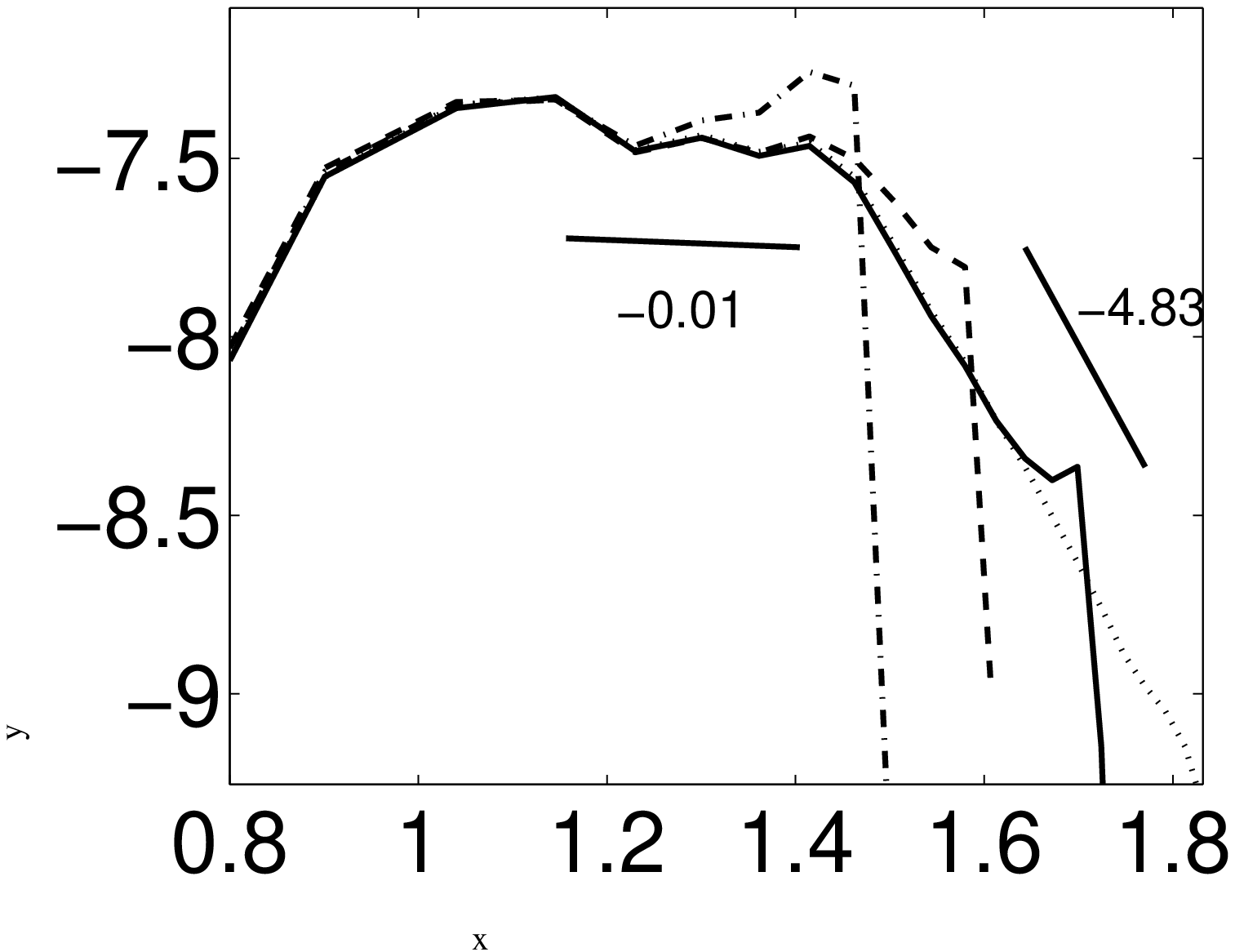}
&
\psfrag{x}{$0.5\log(|\lambda|)$}
\psfrag{y}{$\log(D(|\lambda|^{1/2}))$}
\includegraphics[width=6cm,height=4cm]{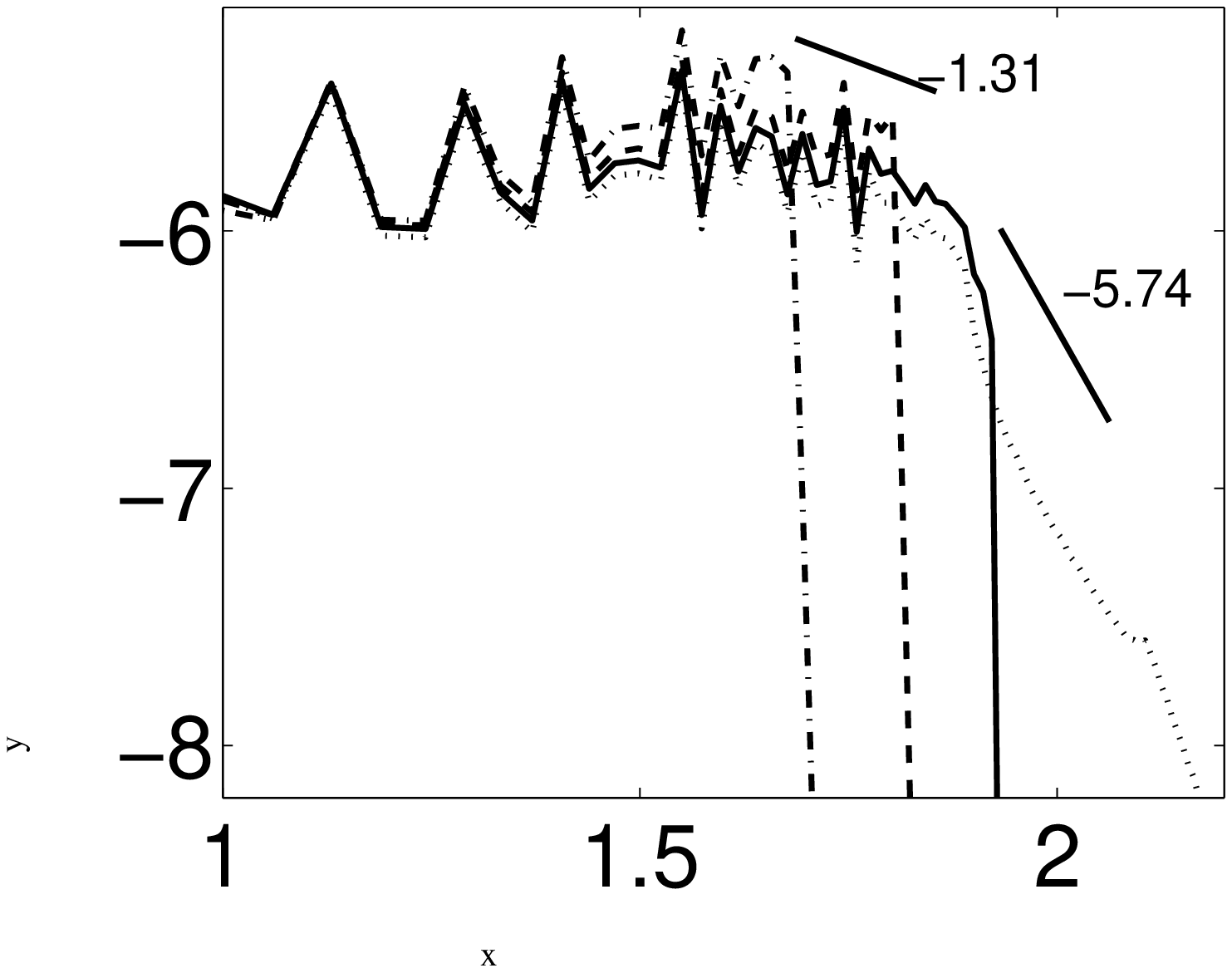}
\\
\psfrag{x}{$\log(k)$}
\psfrag{y}{$\log(E(k))$}
\includegraphics[width=6cm,height=4cm]{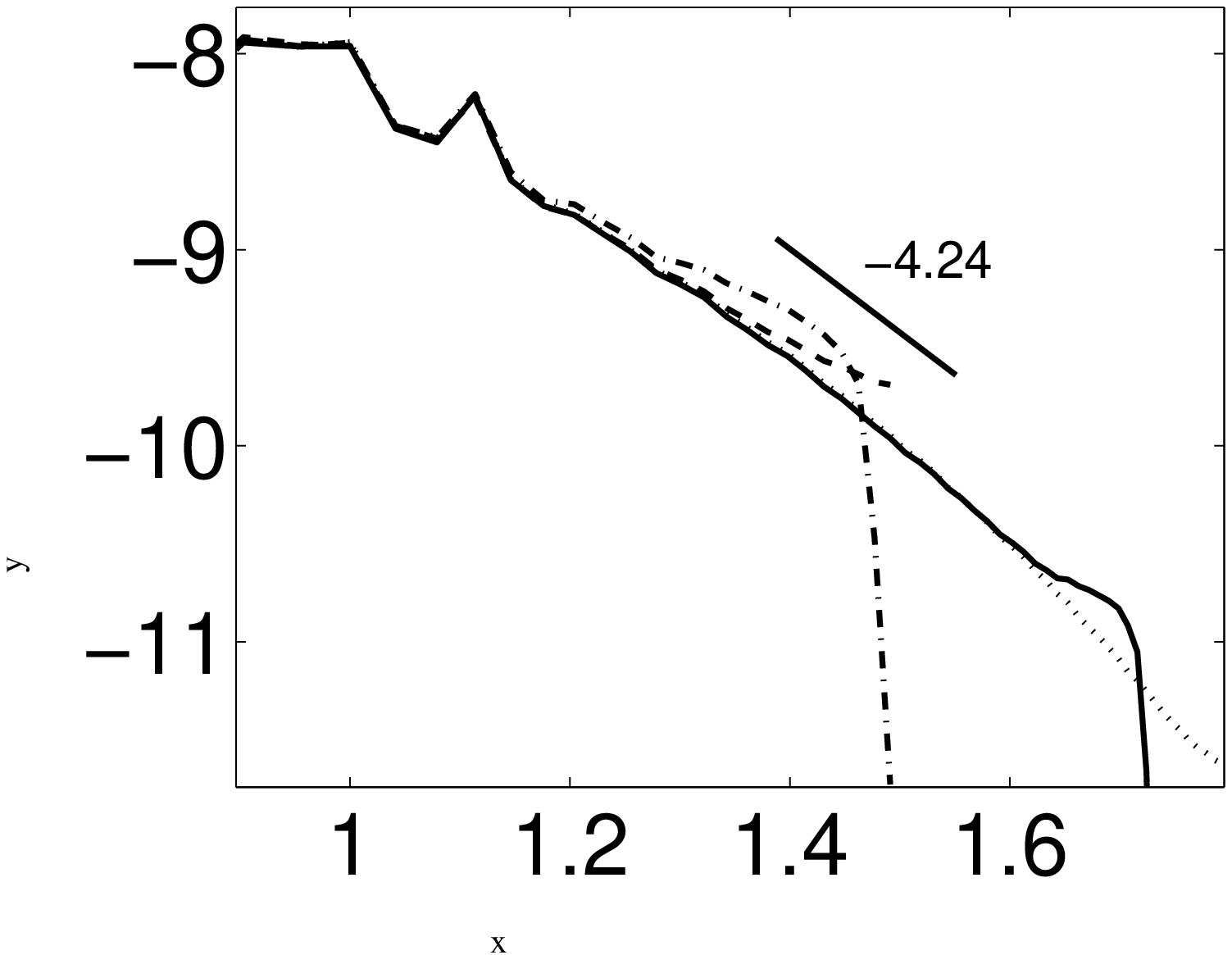}
&
\psfrag{x}{$\log(k)$}
\psfrag{y}{$\log(E(k))$}
\includegraphics[width=6cm,height=4cm]{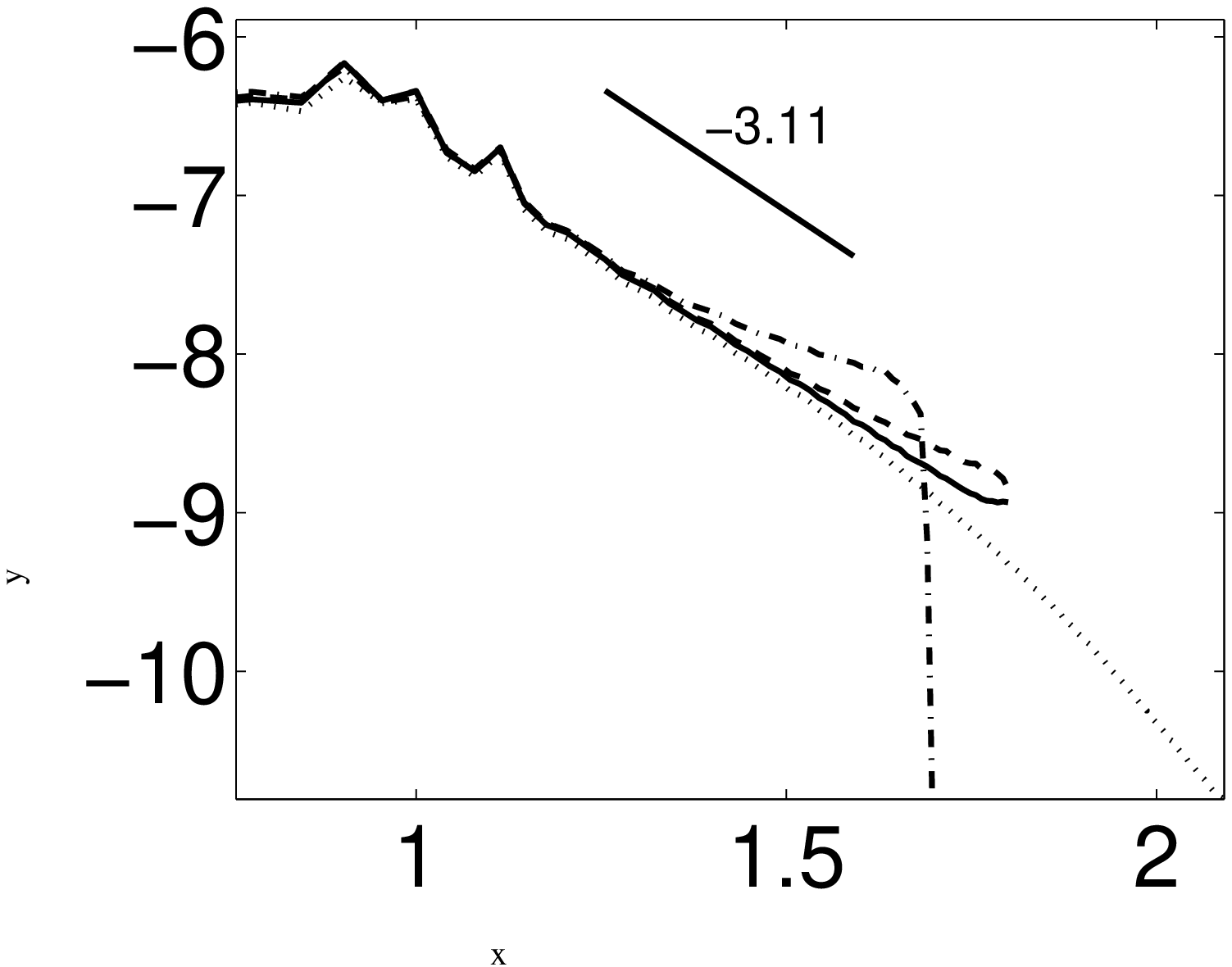}
\\
\psfrag{x}{$tS_0$}
\psfrag{y}{$E(t)/E(t_{\rm decay})$}
\includegraphics[width=6cm,height=4.5cm]{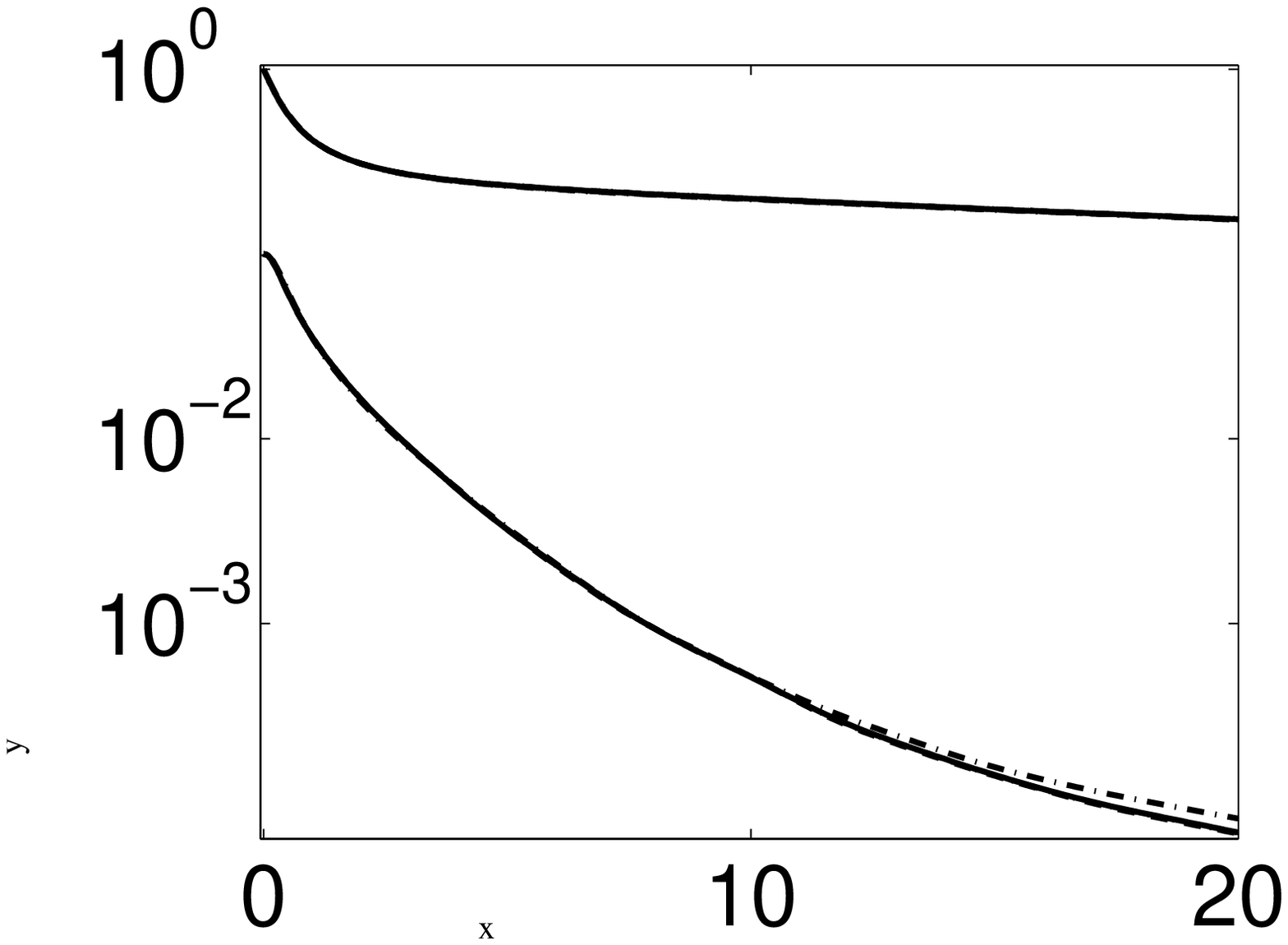}
&
\psfrag{x}{$tS_0$}
\psfrag{y}{$E(t)/E(t_{\rm decay})$}
\includegraphics[width=6cm,height=4.5cm]{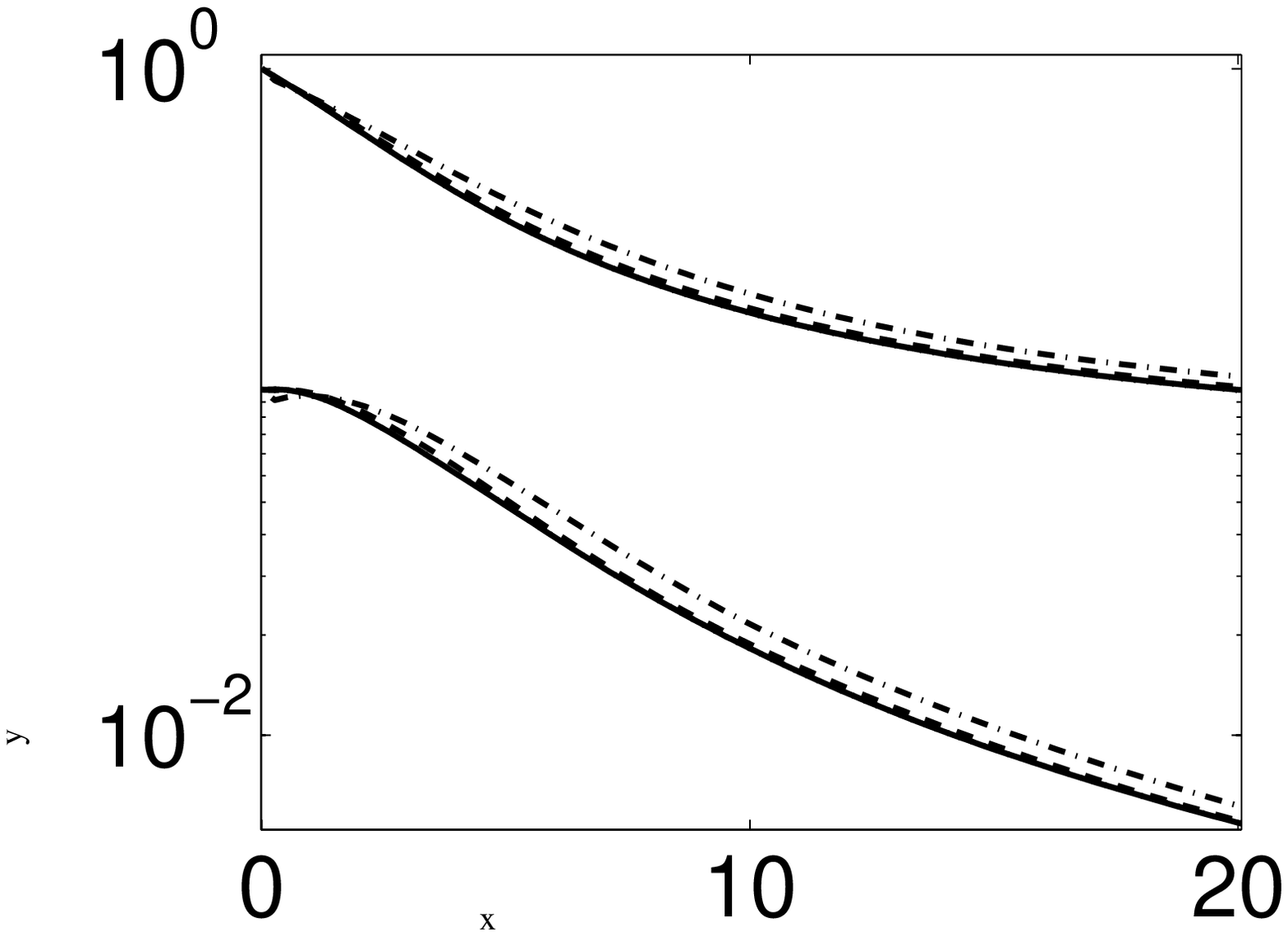}
\\
\end{tabular}

\caption{
{\it Top:} energy density spectrum in $\lambda$-shells (statistically steady flow), {\it  middle-top:} dissipation density spectrum in $\lambda$-shells 
(statistically steady), {\it  middle-bottom:} energy density spectrum in $k$-shells (statistically steady), {\it bottom:} evolution of the total kinetic energy 
of freely decaying
flows normalised by the total energy at the time when the forcing was shut
down $t_{\rm decay}$. For given $\Ha_0$ and $\Gr$, initial conditions are taken
from the same statistically steady reference flow resolved up to the Kolmogorov scale for all values of $\lambda^{\rm cut}$.
Each column presents data from the corresponding cases from figure 
\ref{AngularSpectra}, from which  different resolutions are represented by the 
following curves: {\it figure \ref{LambdaEnergy} left, right and figure 
\ref{LambdaEnergy2} left:} $C_\lambda=0.35$ (dash-dot) , $C_\lambda=0.47$ 
(dash), $C_\lambda=0.59$ (solid), {\it figure \ref{LambdaEnergy2} right:}  $C_\lambda=0.29$ (dash-dot) , $C_\lambda=0.38$ 
(dash),  $C_\lambda=0.48$ (solid). Dotted lines correspond to the reference 
case resolved up to the Kolmogorov scales $C_\kappa \Rey^{3/4}$.
}
\label{LambdaEnergy}
\end{center}
\end{figure}
\begin{figure}
\begin{center}
\begin{tabular}{cc}
$ f_{2D}$, $\Ha_0=1000$, $G=2.7\times10^9$
& $f_{3D}$, $\Ha_0=400$,  $G=1.2\times10^{10}$ 
\\
\psfrag{x}{$0.5\log(|\lambda|)$}
\psfrag{y}{$\log(E(|\lambda|^{1/2}))$}
\includegraphics[width=6cm,height=4cm]{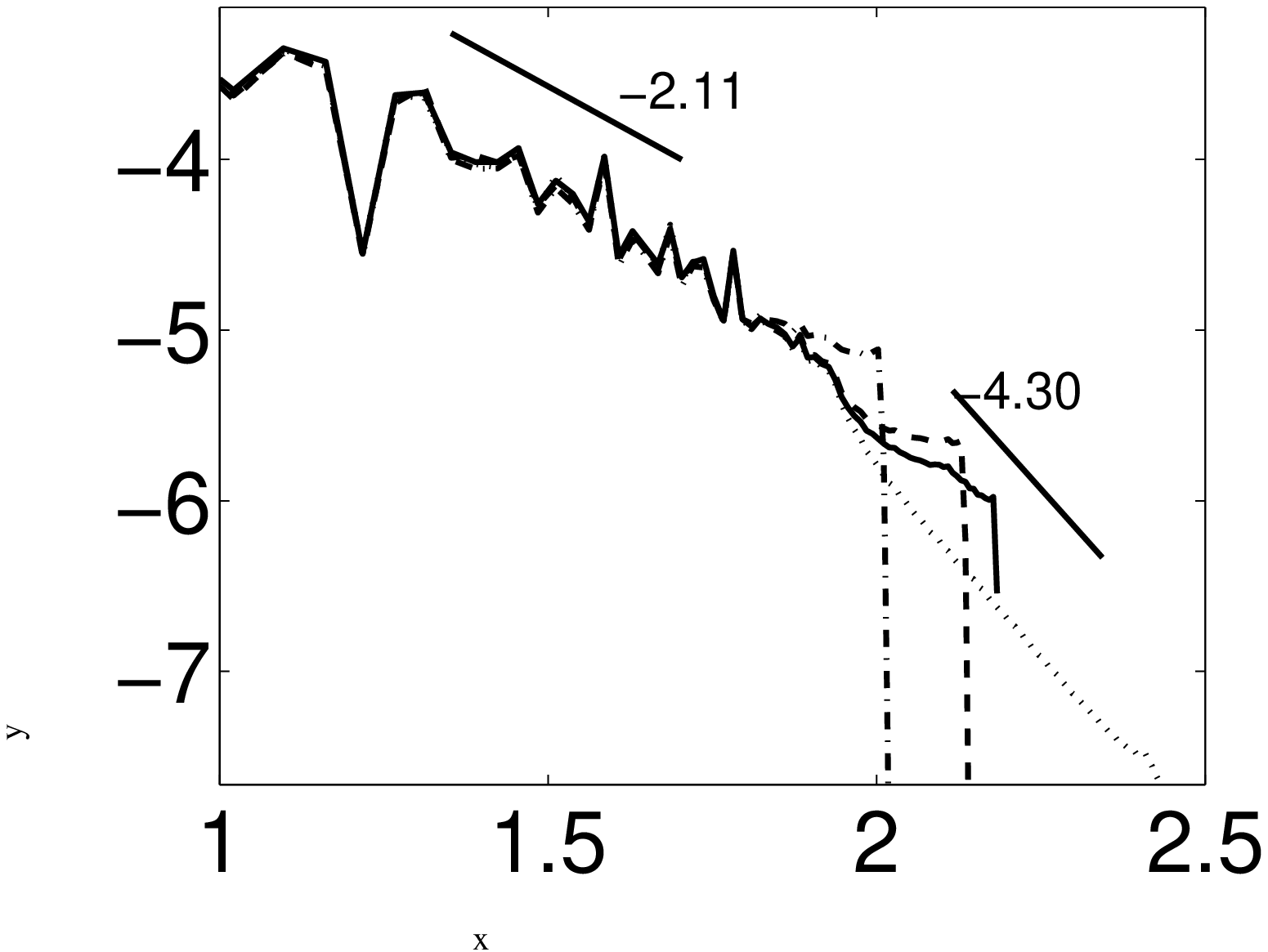}
&
\psfrag{x}{$0.5\log(|\lambda|)$}
\psfrag{y}{$\log(E(|\lambda|^{1/2}))$}
\includegraphics[width=6cm,height=4cm]{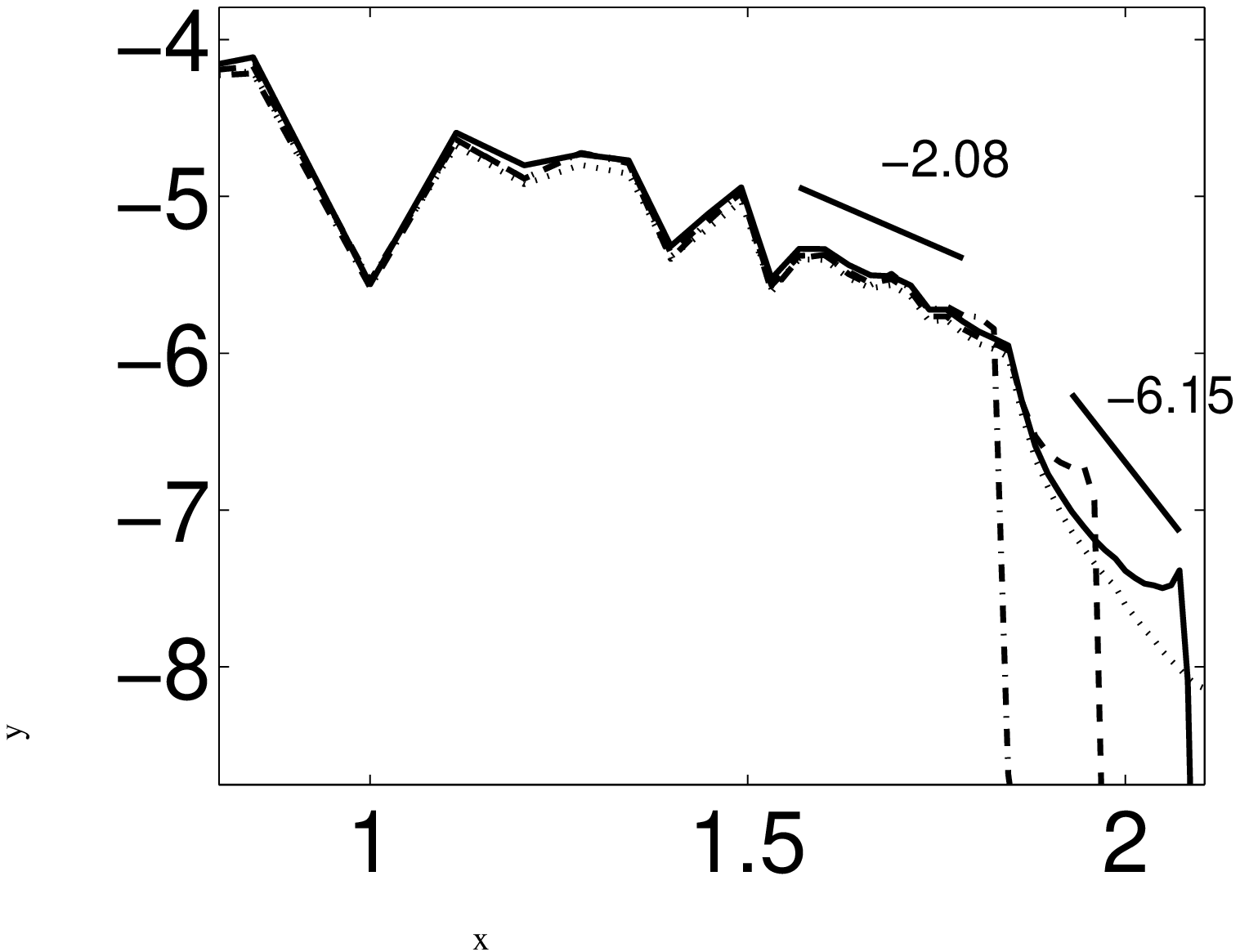}
\\
\psfrag{x}{$0.5\log(|\lambda|)$}
\psfrag{y}{$\log(D(|\lambda|^{1/2}))$}
\includegraphics[width=6cm,height=4cm]{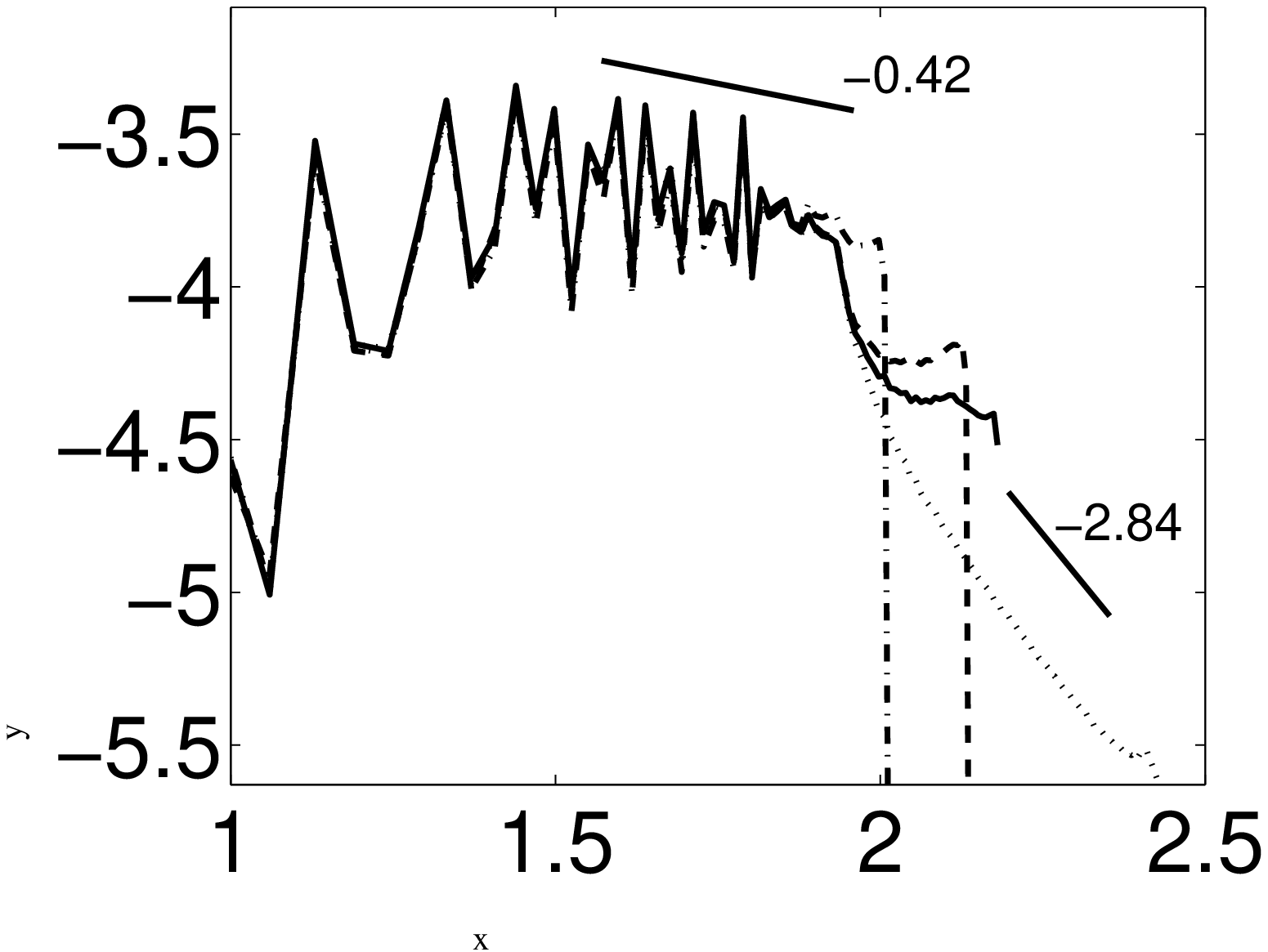}
&
\psfrag{x}{$0.5\log(|\lambda|)$}
\psfrag{y}{$\log(D(|\lambda|^{1/2}))$}
\includegraphics[width=6cm,height=4cm]{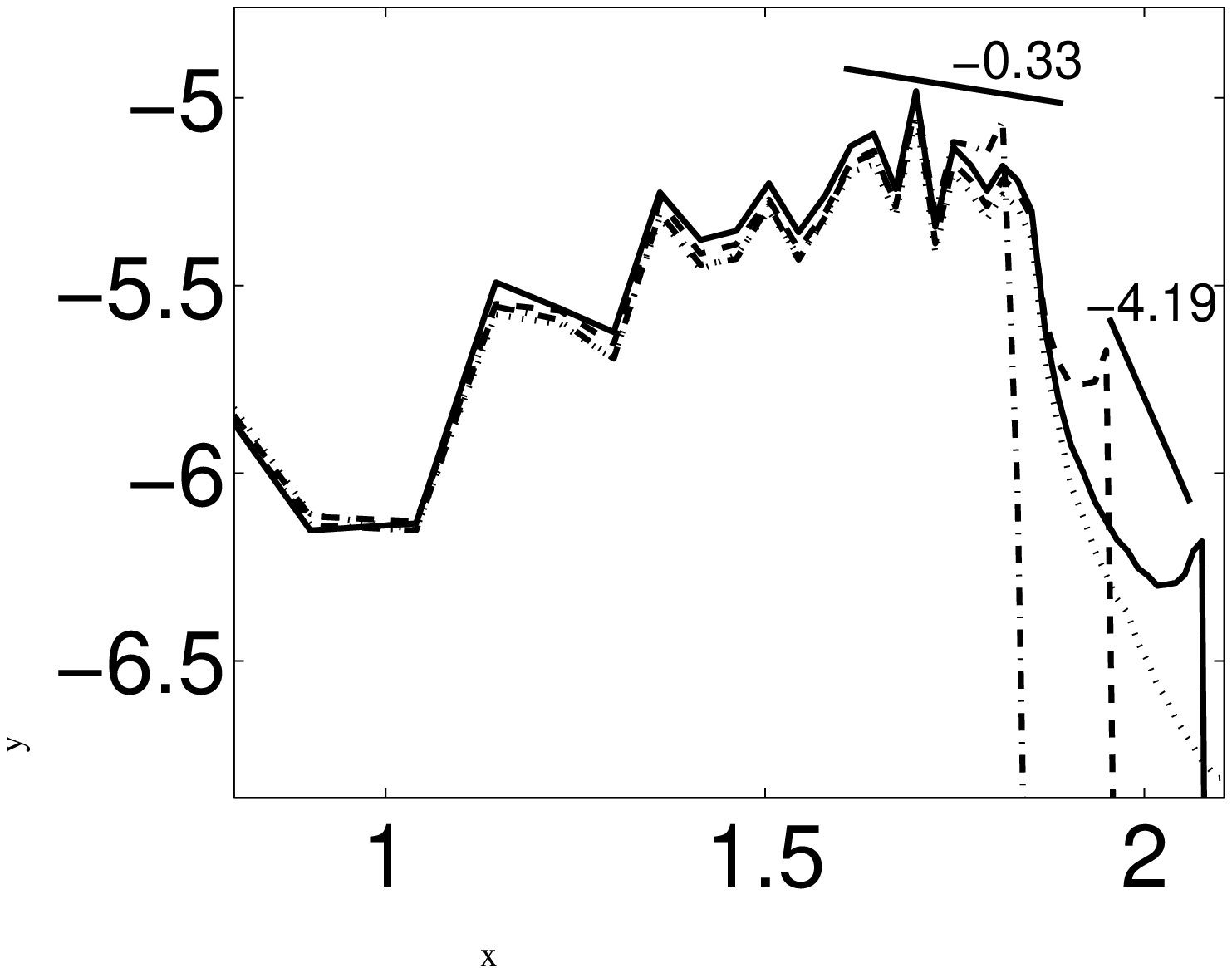}
\\
\psfrag{x}{$\log(k)$}
\psfrag{y}{$\log(E(k))$}
\includegraphics[width=6cm,height=4cm]{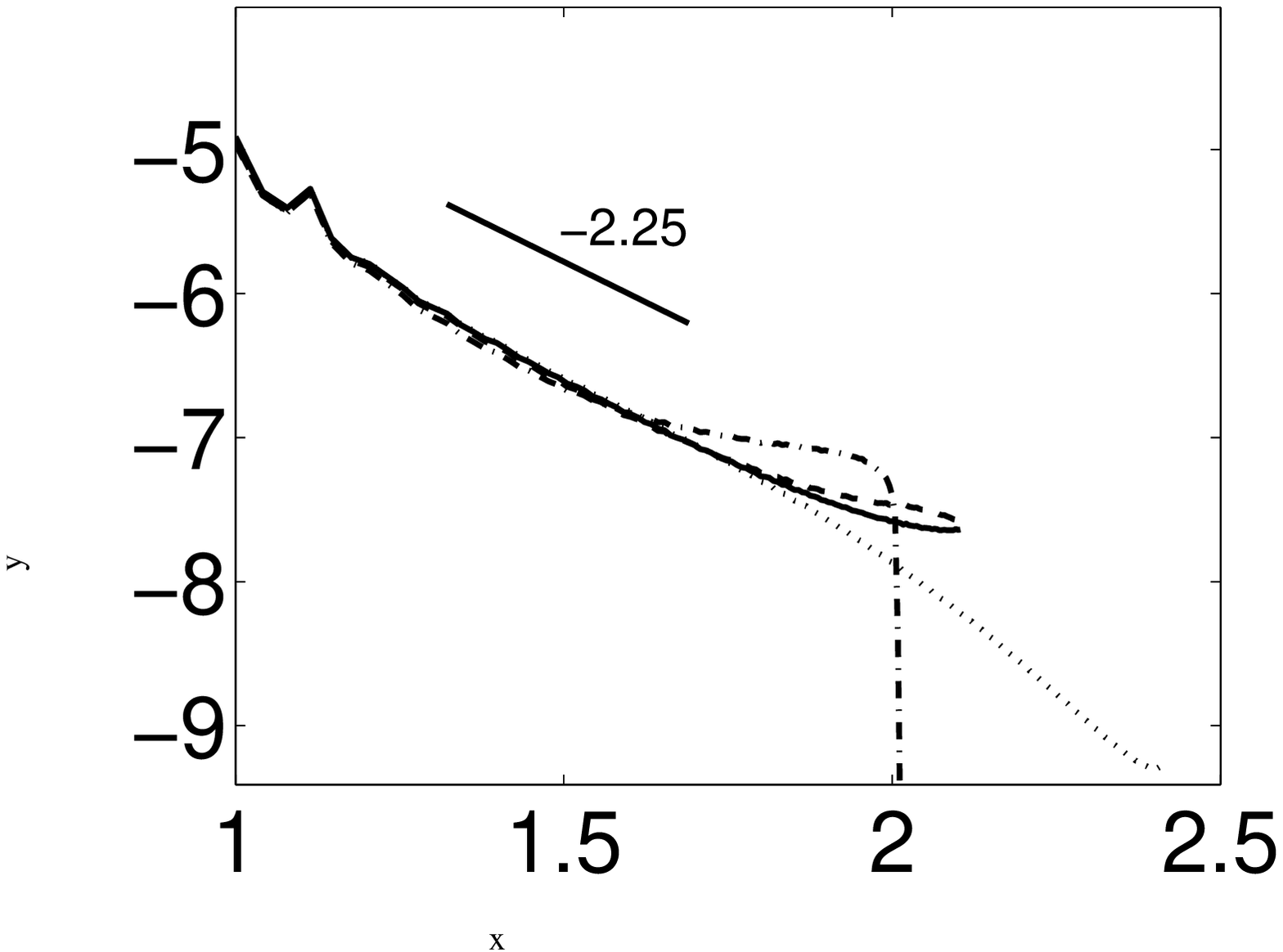}
&
\psfrag{x}{$\log(k)$}
\psfrag{y}{$\log(E(k))$}
\includegraphics[width=6cm,height=4cm]{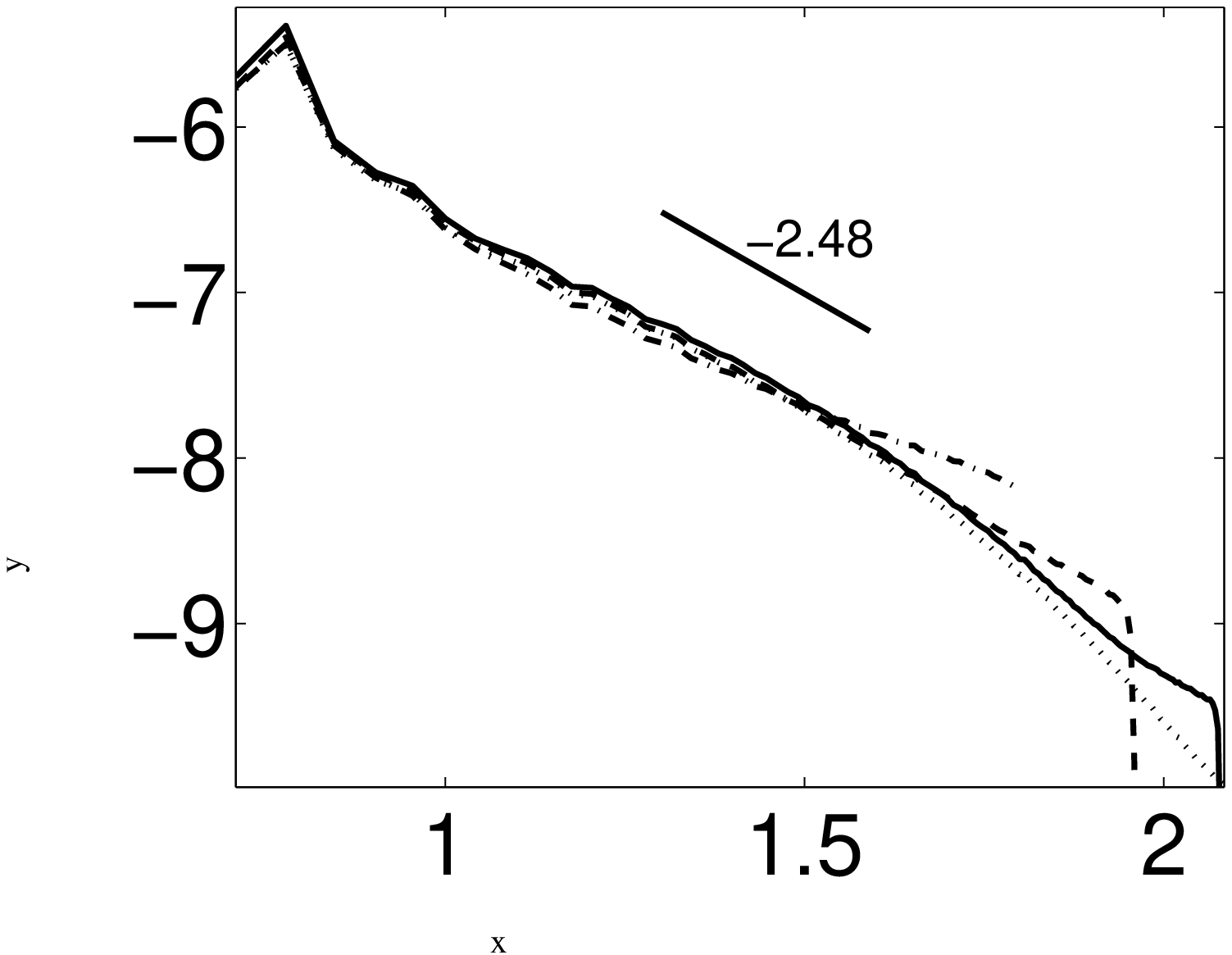}
\\
\psfrag{x}{$tS_0$}
\psfrag{y}{$E(t)/E(t_{\rm decay})$}
\includegraphics[width=6cm,height=4.5cm]{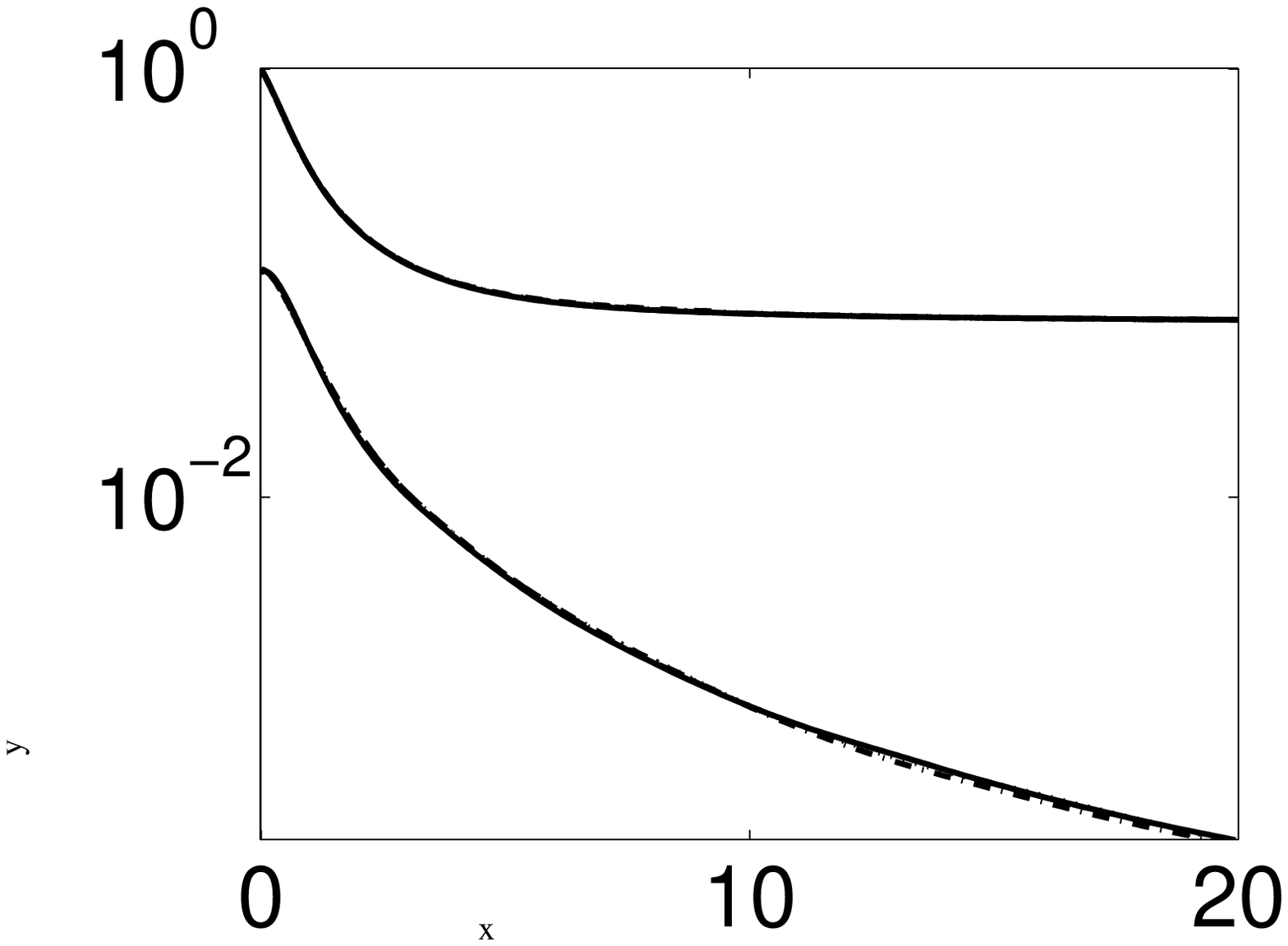}
&
\psfrag{x}{$tS_0$}
\psfrag{y}{$E(t)/E(t_{\rm decay})$}
\includegraphics[width=6cm,height=4.5cm]{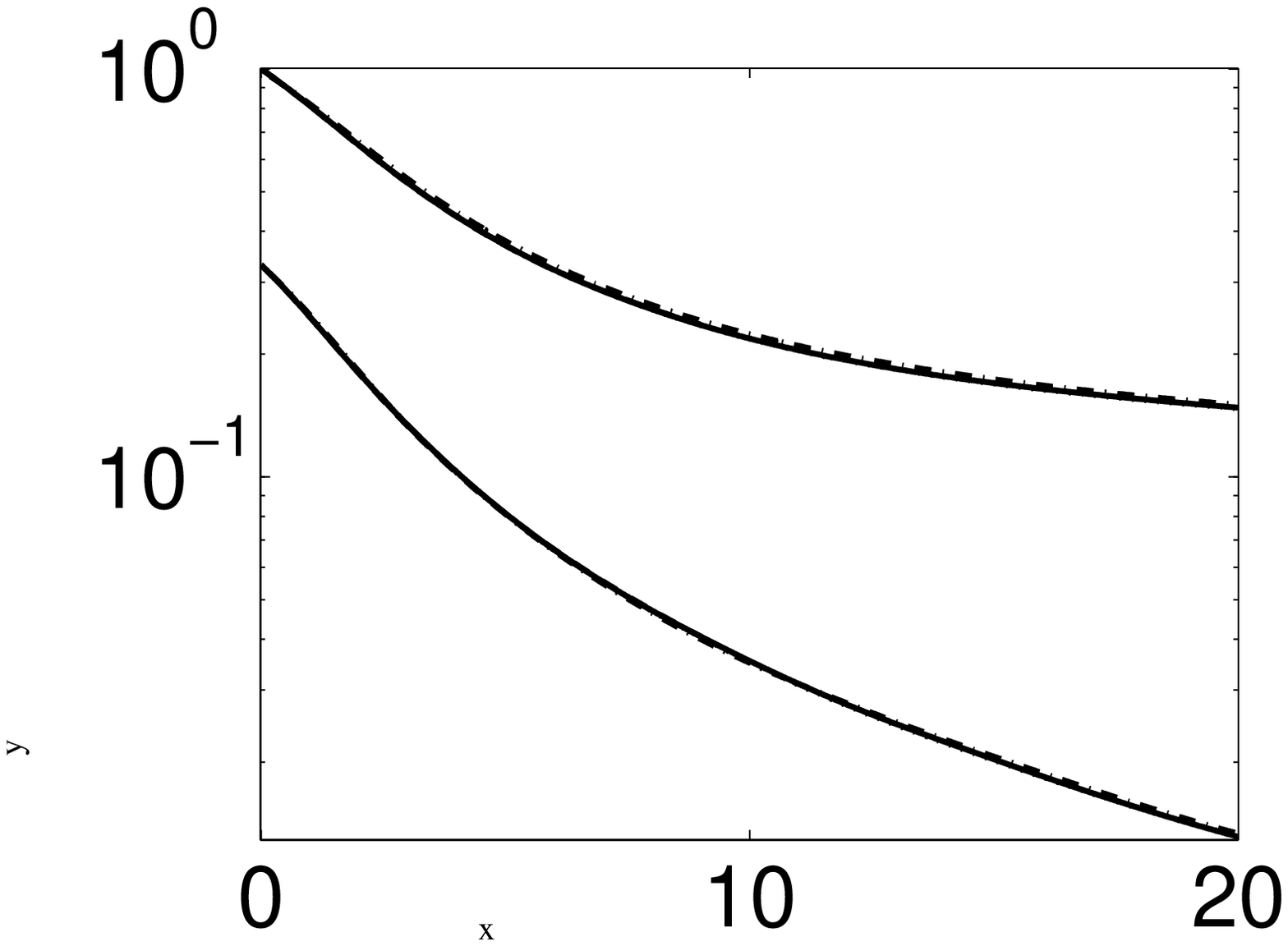}
\\
\end{tabular}

\caption{(figure \ref{LambdaEnergy} continued)}
\label{LambdaEnergy2}
\end{center}
\end{figure}
The time-averaged energy distributions 
in the $(k_\perp,k_z)$-plane ($k_\perp=\sqrt{k_x^2+k_y^2}$) 
show no visible discrepancy between the reference case and those for 
$C_\lambda\in[0.29,0.59]$. This indicates that even with the lowest resolution, 
which uses up to 64 times less modes than the reference case, the energy 
distribution, and the flow anisotropy are still qualitatively well rendered. 
An inspection of the corresponding $\lambda$-based energy and dissipation 
spectra from figure \ref{LambdaEnergy} and \ref{LambdaEnergy2} confirms and 
refines the picture: 
the small energy and dissipation pile-up that inevitably occurs at 
the high-$\lambda$ end of the spectrum certainly remains confined there 
for all $Ha$ for $C_\lambda\gtrsim0.5$. It does, however tend to slightly 
spread towards the higher end of the spectrum for lower values of $C_\lambda$, 
particularly in the dissipation spectra and in the cases at lower $Re$.
Even though it is only pronounced in the $\Ha_0=80$ case, 
this propagation of error toward larger scales is a usual symptom of 
under-resolution, and can be more easily spotted on the dissipation spectra. 
This error on the dissipation is further revealed when the flow is freely 
decaying. For each of our four reference cases, we have calculated such flows 
starting from an initial state in the established regime resolved up to the 
Kolmogorov scale. In each case, the subsequent evolution of the flow 
without forcing was calculated  several times from this same initial condition, 
for the same maximum resolutions as those used to calculate the established 
flows. The evolution was calculated over 20 Joules times, after which the flow 
had lost most of its energy. 
As for the dissipation spectra in the established state, it turns out 
that a discrepancy between reference case and cases resolved with 
$C_\lambda<0.5$ is visible in the evolution of both the total energy and of the  energy in the field direction. Cases resolved with  $C_\lambda\gtrsim0.5$, on 
the contrary, match the reference case to a great precision, both when the flow 
is established and freely decaying. As a matter of fact, the decay curves for 
$C_\lambda\gtrsim0.5$  cannot be distinguished from those of the reference case 
on the graph.\\ 
To quantify the precision reached for a given value of $C_\lambda=\sqrt{|\lambda^{\rm max}|/(4\pi^2k_f^2\Rey)}$ over a wider 
range of parameters than those of the 4 reference cases calculated above, 
we define a reduced spectral parameter normalised by scaling (\ref{eq:heur_scal_lambda}): $l=\sqrt{|\lambda|/(4\pi^2k_f^2\Rey)}$, such that for $l=C_\lambda$, 
$\lambda=\lambda^{\rm max}$.
We have calculated the variations of total energy 
$\Sigma_E(l)$ and dissipation $\Sigma_D(l)$ contained in the 
spectral subspace enclosed in the iso-$\lambda$ curve for each value of 
$l\leq C_\lambda$ for a
selection of cases resolved beyond $C_\lambda=0.5$ (summarised in table 
\ref{table:simul1}, along with their resolution expressed in terms of 
$C_\lambda$ and $C_\kappa$). 
The results are illustrated on figure \ref{fig:e_lambda}.
Firstly, it turns out that for a given value of $l$, the ratio $\alpha_E(l)$ of 
$\Sigma_E(l)$ to the total energy $\Sigma_E(C_\lambda)$ is constant for all 
calculated cases, regardless  of the values of $\Ha$, $\Gr$ and of the nature 
of the forcing (with, in particular, $\alpha_E(l=0.5)\simeq 0.99$ no matter how 
high $C_\lambda$ is, as shown in table \ref{table:simul1}).  In other 
words, the precision attained for a given value of $C_\lambda$ remains essentially constant when $\Ha$ and $\Gr$ are varied beyond their values in the four 
reference cases calculated above. This brings further confirmation of the 
validity of scaling laws (\ref{eq:heur_scal_lambda}), and of their independence 
of the nature of the forcing. Secondly, the variations of $\Sigma_E(l)$ and 
$\Sigma_D(l)$ also comfort us in the choice of $C_\lambda\simeq0.5$ as 
the minimum cutoff scale for full resolutions: this value  is indeed 
located  at the beginning of a plateau where further increase of resolution 
hardly brings any variation in the total energy and dissipation of the 
solution. Smaller values of $C_\lambda$, on the other hand, may fall outside 
this region and calculations at the corresponding resolution may thus fail to 
capture noticeable fractions of the total energy and dissipation. On these 
grounds we shall finally propose the following scaling for $\lambda^{\rm max}$:
\begin{figure}
\begin{center}
\psfrag{x}{$l=\left(\frac{|\lambda|}{4\pi^2k_f^2\Rey}\right)^{1/2}$}
\psfrag{y}{$\Sigma_E$}
\includegraphics[width=6.5cm,height=5cm]{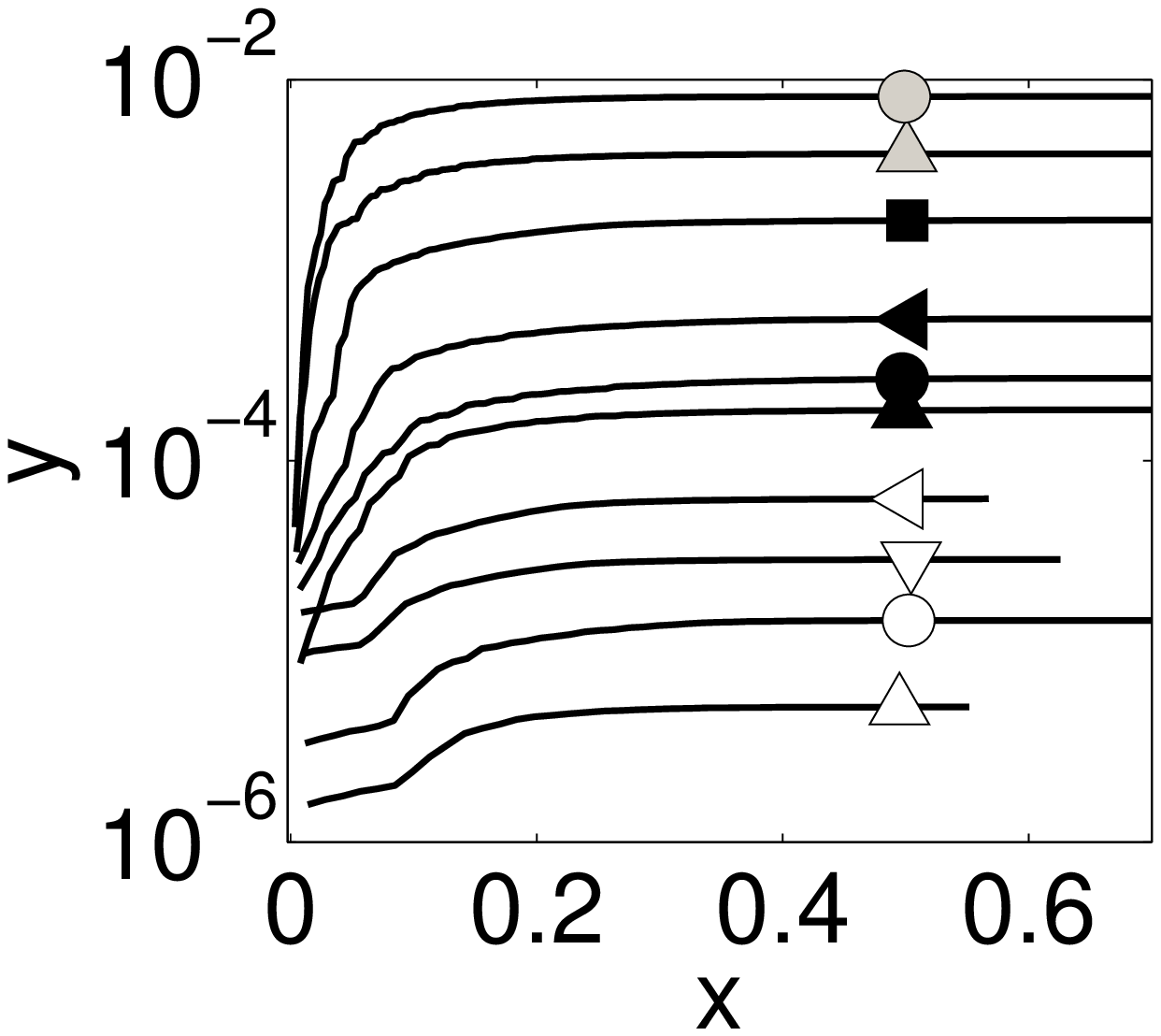}
\psfrag{x}{$l$}
\psfrag{y}{$\Sigma_E$}
\includegraphics[width=6.5cm,height=5cm]{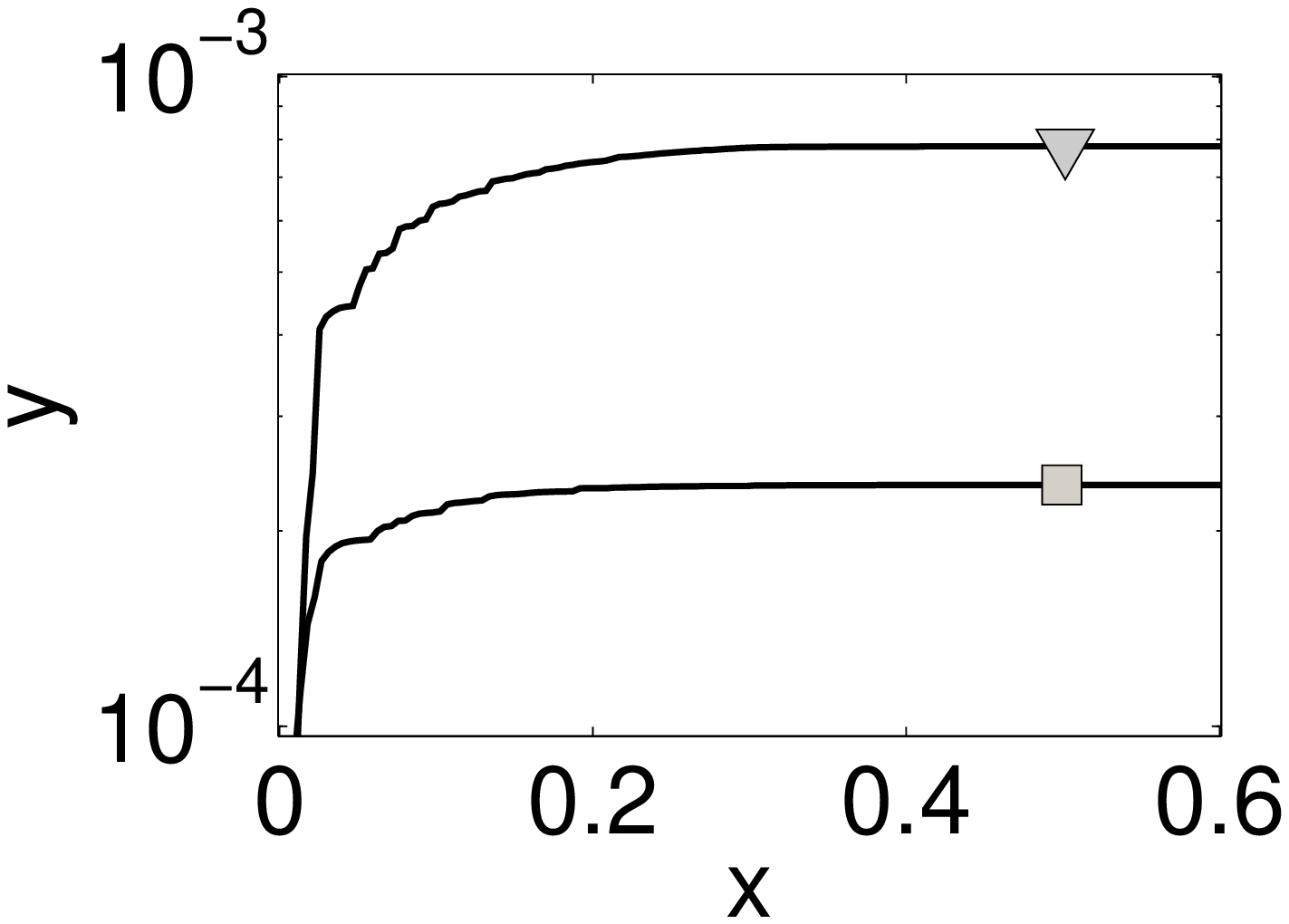}
\\
\psfrag{x}{$l$}
\psfrag{y}{$\Sigma_D$}
\includegraphics[width=6.5cm,height=5cm]{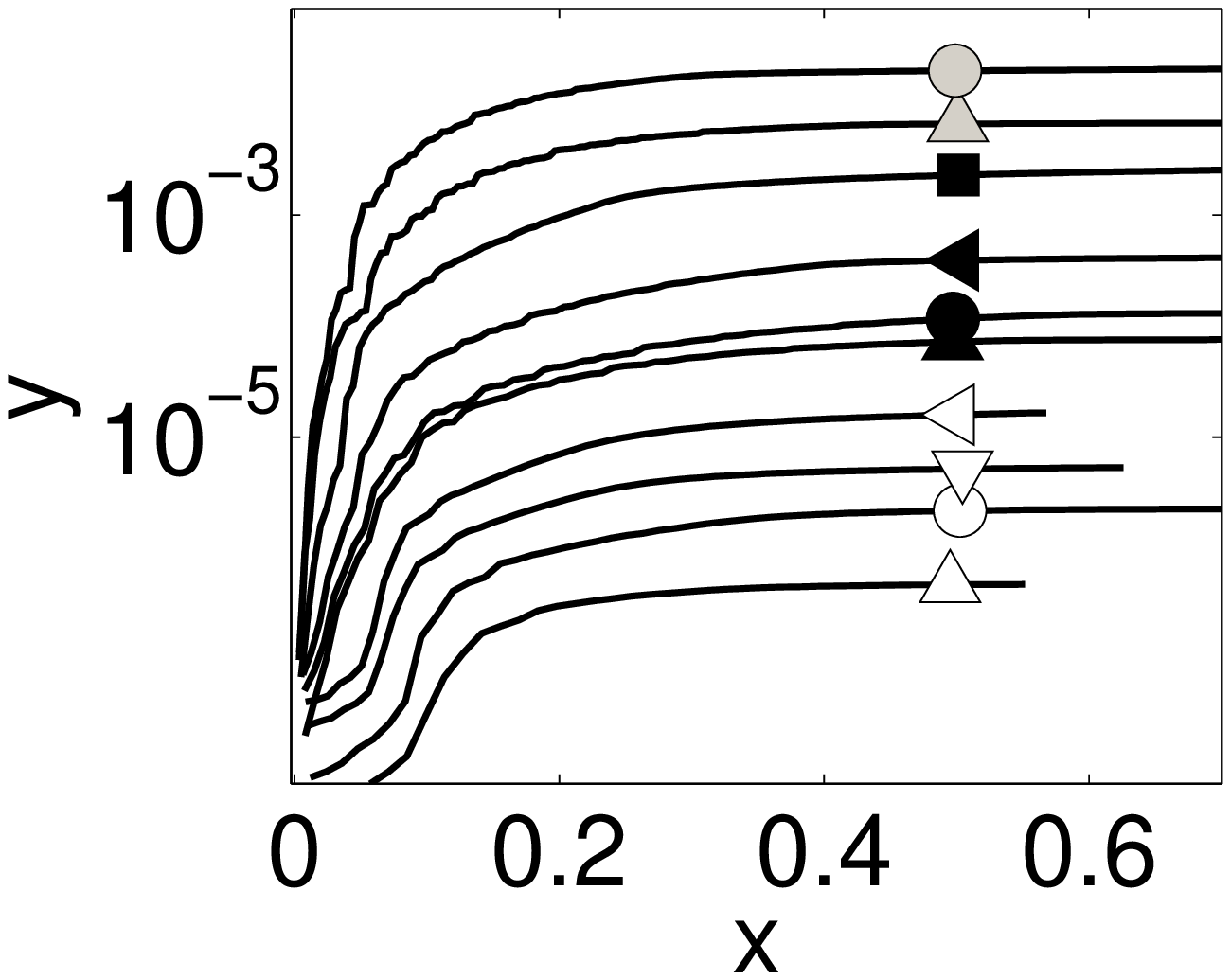}
\psfrag{x}{$l$}
\psfrag{y}{$\Sigma_D$}
\includegraphics[width=6.5cm,height=5cm]{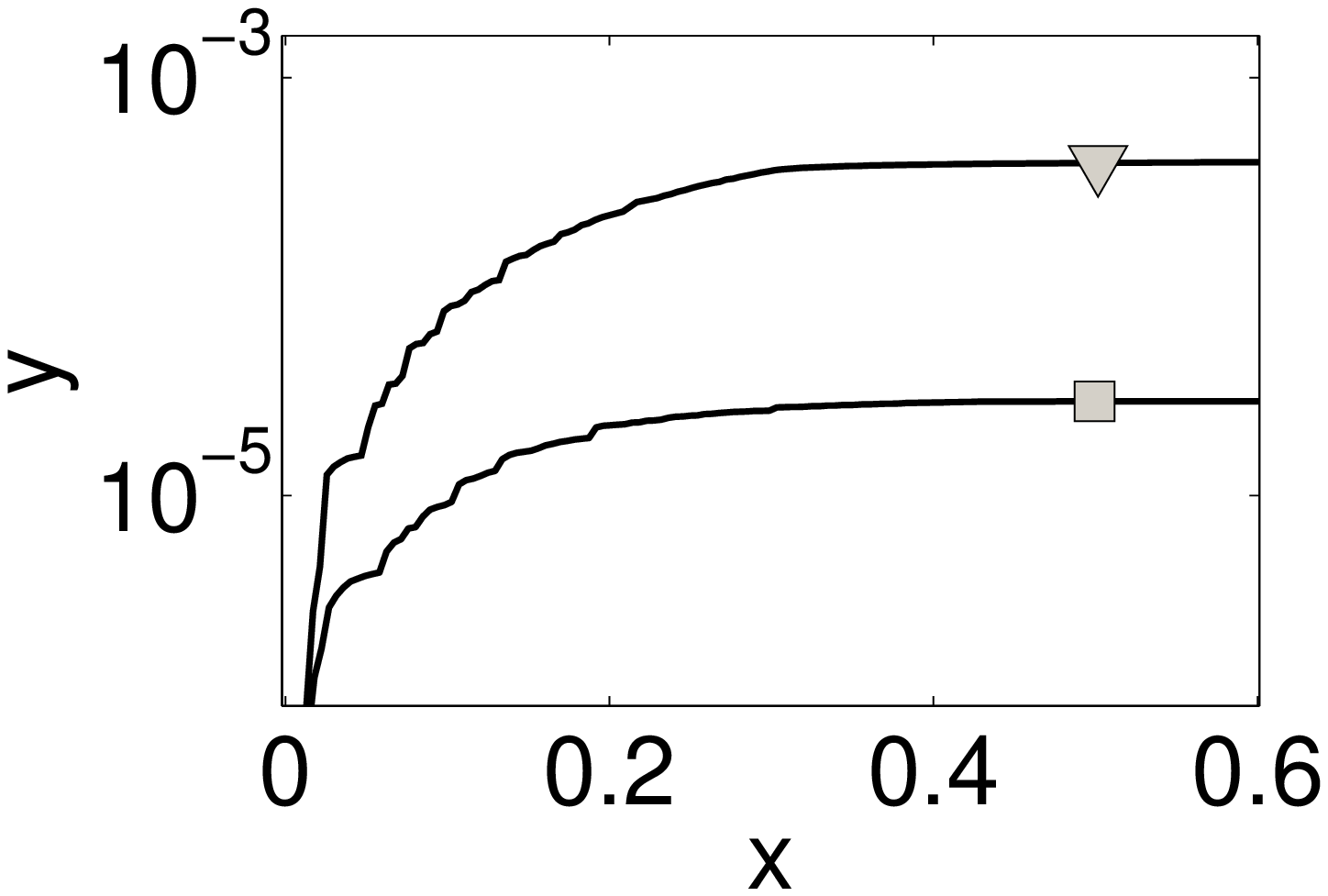}
\caption{Energy $\Sigma_E(l)$ (\textit{top}) and dissipation 
$\Sigma_D{(l)}$ 
(\textit{bottom}) contained in the subspace 
$\{\lambda, |\lambda|<4\pi^2l^2\Rey\}$ 
\textit{vs.} $l$. The symbols are those from table 1 and placed at $l=0.5$. 
\textit{Left:} two-dimensional forcing, \textit{right:} three-dimensional forcing.}.
\label{fig:e_lambda}
\end{center}
\end{figure}
\begin{figure}
\begin{center}
\psfrag{y}{$\frac12\log(\frac{|\lambda^{\rm max}|}{2\pi}(0.47\Gr_f^{0.40})^{-1})$}
\psfrag{x}{$\log|\Gr_f|$}
\includegraphics[width=8cm,height=7cm]{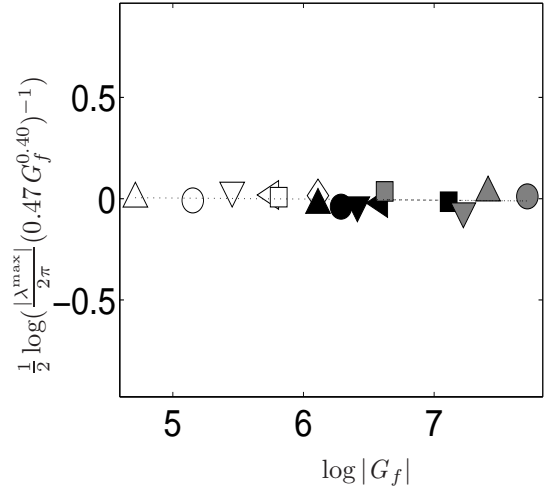}
\caption{Scaling law for the small scales expressed as a function of $\Gr$.
The symbols correspond to cases from table \ref{table:simul1}.}
\label{LambdaGrashof}
\end{center}
\end{figure}
\begin{eqnarray}
\begin{aligned}
\frac{\sqrt{|\lambda^{\rm max}|}}{2\pi k_f} & \simeq 0.5 \Rey^{1/2}
\label{eq:final_scal_lambda}
\end{aligned}
\end{eqnarray}
The values of $k_\perp^{\rm max}$  and $k_z^{\rm max}$ can be directly deduced 
from that of $\lambda^{\rm max}$ through (\ref{eq:sscale_mhd}) 
to quantify the scalings  from \cite{Potherat2003} as:
\begin{eqnarray}
\begin{aligned}
 \frac{k_z^{\rm max}}{k_f}  & \simeq 0.8 k_f \frac{\Rey}{\Ha_{\rm opt}}
&\qquad 
\frac{k_{\perp}^{\rm max}}{k_f} & \simeq 0.5 \Rey^{1/2}.
\label{eq:final_sscale_mhd}
\end{aligned}
\end{eqnarray}
%
Also, the values of $\lambda^{\rm max}$ can be expressed as a function of 
$\Gr$, which is known \textit{a priori}, unlike $\Rey$. Denoting 
$\Gr_f=\|\mathbf f\|_{2f}/(\nu^2L_f^{3/2})$, where $\|\cdot\|_{2f}$ 
represents the $L_2$ norm for a domain of volume $L_f^3=(L/k_f)^3$, the 
corresponding graph, on figure \ref{LambdaGrashof}, suggests the scaling:
\begin{eqnarray}
\begin{aligned}
\frac{\sqrt{|\lambda^{\rm max}|}}{2\pi k_f} & \simeq 0.47 \Gr_f^{0.20}.
\label{eq:final_scal_lambda_gr}
\end{aligned}
\end{eqnarray}
Finally, beyond the identification of $\lambda^{\rm max}$, 
$\lambda$-based spectra from figure \ref{LambdaEnergy} and \ref{LambdaEnergy2} 
exhibit an interesting feature, as
a remarkable steep tail is present in both energy and dissipation 
$\lambda$-spectra at high values of $\lambda$. In all cases, it 
starts when $\lambda$ reaches the value of the eigenvalue 
$\lambda_J=-\Ha_{\rm opt}^{1/2}$ 
of the first mode with a wavevector orthogonal to $\mathbf B$. Since 
$S\gtrsim1$ in all our calculations, such modes, of the form ($(0,k_\perp)$),  
are located outside the Joule cone and therefore strongly suppressed 
by Joule dissipation.  This explains why they carry very little energy.

\subsection{Practical use of the scaling laws}
The results of the present section now allow us to put forward a simple 
procedure to resolve three-dimensional MHD flows in periodic domains: 
firstly, $L_{\rm opt}$ and $L_{\rm int}$ can be  calculated at every time step
as the numerical simulation progresses (as is already usual for $L_{\rm int}$). 
When $S\gtrsim1$, (\ref{eq:final_scal_lambda}) or (\ref{eq:final_sscale_mhd}) 
then provide  criteria for the resolution necessary to represent a 
three-dimensional MHD flow completely.\\ 
When $S<1$, the flow becomes progressively more isotropic and so 
does the set of least dissipative modes. Accordingly, 
the resolution required to fully resolve the flow 
becomes higher than that predicted by scalings (\ref{eq:final_scal_lambda}) 
or (\ref{eq:final_sscale_mhd}). In this case, $L_{\rm int}$ and $L_{\rm opt}$ 
are still determined 
"on the fly" but the usual Kolmogorov criterion must be used instead of 
(\ref{eq:final_scal_lambda}) or (\ref{eq:final_sscale_mhd}).\\
When $S>>1$, the flow can be either two or three-dimensional, which poses an 
important question about the ability of the least dissipative modes to 
represent the flow accurately:
on the one hand, when $\lambda^{\rm max}$ exceeds a value that depends on 
$Ha$ only, a first three-dimensional mode  appears in the set of least 
dissipative modes, independently of the behaviour of the flow itself.
When $\Gr$ is increased from 0, on the other hand, a first three-dimensional 
physical mode appears in the flow at the actual transition between two- and 
three-dimensionality, independently of the method used to calculate it. We 
shall examine in the next section whether both coincide. This will tell us 
whether 
the least dissipative modes can be used for the simulation of MHD turbulence, 
regardless of whether it is two- or three-dimensional.\\
\section{Least dissipative modes at the transition between two-dimensional and 
three-dimensional turbulence}
\label{sec:trans}
\subsection{Two- vs. three-dimensional sets of least dissipative modes}
We now focus on the question of 
how to calculate flows using the least dissipative modes at 
the transition between two- and three-dimensional MHD turbulence. The set of 
least dissipative modes
can either contain only two-dimensional modes or both two or three-dimensional
 modes, depending on the value of $\lambda^{\rm max}$.
A transition between these two types of sets therefore occurs at a 
value $\lambda^{\rm max}=\lambda^{\rm 3D}$ 
for which the curve $\lambda=\lambda^{\rm max}$  encloses at least one mode with 
$k_z\geq 1$ (bold dashed line in Figure \ref{IsoLambda}(d)). According to our 
previous work (\cite{Potherat2003}) and in the present notations, the first three-dimensional mode in this sense is associated to the eigenvalue 
\begin{eqnarray}\label{eq:first3D} \begin{aligned} 
|\lambda^{\rm 3D}|=2\frac{\Ha_{\rm opt}}{2\pi},
\end{aligned}
\end{eqnarray}
and the modulus of the corresponding wavevector in the plane across the 
magnetic field lines is:
\begin{equation}
|{\bf k}_{\perp}^{\rm 3D}|=\sqrt{\frac{\Ha_{\rm opt}}{2\pi}-1}.
\label{eq:k3d}
\end{equation}

It is important to notice that although a flow represented by a set comprising 
three-dimensional modes is potentially three-dimensional, it  isn't 
\emph{necessarily} three-dimensional. Instead it can be either two-dimensional or in a state of 
intermittency between the two states, as in \cite{Zikanov1998}, if the 
coefficients of the three-dimensional modes in expansion 
(\ref{EigenfunctionDecomposition}) are 0 or 
intermittently become 0. This behaviour is determined by the flow dynamics, 
independently of the basis chosen to represent it (provided the 
flow is correctly resolved, obviously.). We shall now compare the first 
least dissipative three-dimensional mode to the first three-dimensional 
mode that appears in the flow.

\subsection{Numerical procedure}
We use the same numerical solver as that described in section \ref{sec:NumProc} and also 
the same type of two-dimensional forcing $\mathbf f_{2D}$ (\ref{eq:2dforce}). 
On the top of previous calculations initialised 
with the fluid at rest, we now perform two additional series of calculations, 
at $\Ha_0=80$ and $\Ha_0=400$ respectively, as follows: we start with fixed 
$\Ha_0$, low $\Gr$ and the fluid initially at rest. We look for a statistically 
steady two-dimensional solution 
and let it reach a well developed, turbulent state (after a time of the 
order of $100-200 S_0$, or dimensionally, 100-200 Joule times 
$\tau_j=\rho/(\sigma B^2)$). With
this latter state as the initial condition, we perform the next 
calculation by increasing the Grashof number by $15\%$, and 
repeat the procedure until three-dimensionality appears. 

In all simulations the numerical resolutions $n_x\times n_y\times n_z$ are 
chosen as the smallest powers of 2 such that the resolution domain encloses 
the $\lambda=1.5\lambda^{\rm max}$ curve and satisfies 
$k_\perp^{\rm max}\geq 1.2 k_K=1.2\Gr ^{1/3}(1+\log \Gr)^{1/6}$. This way, 
the flow is well resolved whether in a state of two-dimensional turbulence or 
in a state of three-dimensional MHD turbulence. 
Since $L_{\rm opt}$ cannot be determined in two-dimensional flows 
but varies little for a given $Ha$, we take the 
approximate values $L_{\rm opt}(Ha_0=80)= 24$ and 
$L_{\rm opt}(Ha_0=400)= 55$, (see figure \ref{minimize}).

\subsection{Measure of three-dimensionality}
In order to track three-dimensionality near the transition, 
we define two quantities to characterise it.
The first one 
expresses how physical quantities depend on $z$, so we shall call it 
\emph{morphological} three-dimensionality and define it as 
\begin{eqnarray}\label{eq:alpha3D}
\begin{aligned}
\alpha_{3D}=\left(\int\limits_{0}^{L}\left( f(z)-1 \right) ^2 dz\right)^{1/2},
\end{aligned}
\end{eqnarray}
where $f(z)$ expresses the ratio between the two and 
three-dimensional parts of the RMS velocity fluctuations in the plane 
$z=const$:
\begin{eqnarray}\label{eq:nmfp}
\begin{aligned} 
f(z)=\frac{L_0\int\limits_{\Omega {z'=z}}(<({\bf u}'(x,y,z'))^2>_t)^{1/2}dxdy}{\int\limits_{\Omega}(<({\bf u'(x,y,z)})^2>_t)^{1/2}dxdydz}.
\end{aligned}
\end{eqnarray}
Here, $<\cdot>_t$ denotes averaging with respect to time and  
${\bf u}'={\bf u}-<{\bf u}>_t$ is the local velocity fluctuation. 
$\alpha_{3D}$ gives a global measure of \emph{morphological} three-dimensionality as it expresses an average ratio of the three-dimensional to the 
two-dimensional part of the velocity fluctuations.

The second type of three-dimensionality is expressed as the ratio of the 
energy in the $z$ direction to that in the $x$ and $y$ direction. We shall 
therefore call it \emph{kinematic} three-dimensionality:
\begin{eqnarray}\label{eq:EzEperp}
\begin{aligned}
\beta_{3D}=\left(\frac{E_z}{E_{\perp}}\right)^{1/2}=\left(\frac{\sum\limits_{{\bf k}} w^2({\bf k})}{\sum\limits_{{\bf k}}( u^2({\bf k})+v^2({\bf k}))}\right)^{1/2}.
\end{aligned}
\end{eqnarray}
In theory, there is no reason  for the first appearance (in the sense of 
growing $\Gr$) of these 
 types  of three-dimensionality not to take place in vortices of distinct 
 wavelength, which we shall therefore name 
 $k_{\perp}^{3D\alpha}$ and $k_{\perp}^{3D\beta}$ respectively.

\subsection{First three-dimensional modes and relevance of the least dissipative modes to transitional flows}
\begin{figure}
\begin{center}
\psfrag{first3D}{$k_\perp^{3D}$}
\includegraphics[width=6.5cm,height=6.5cm]{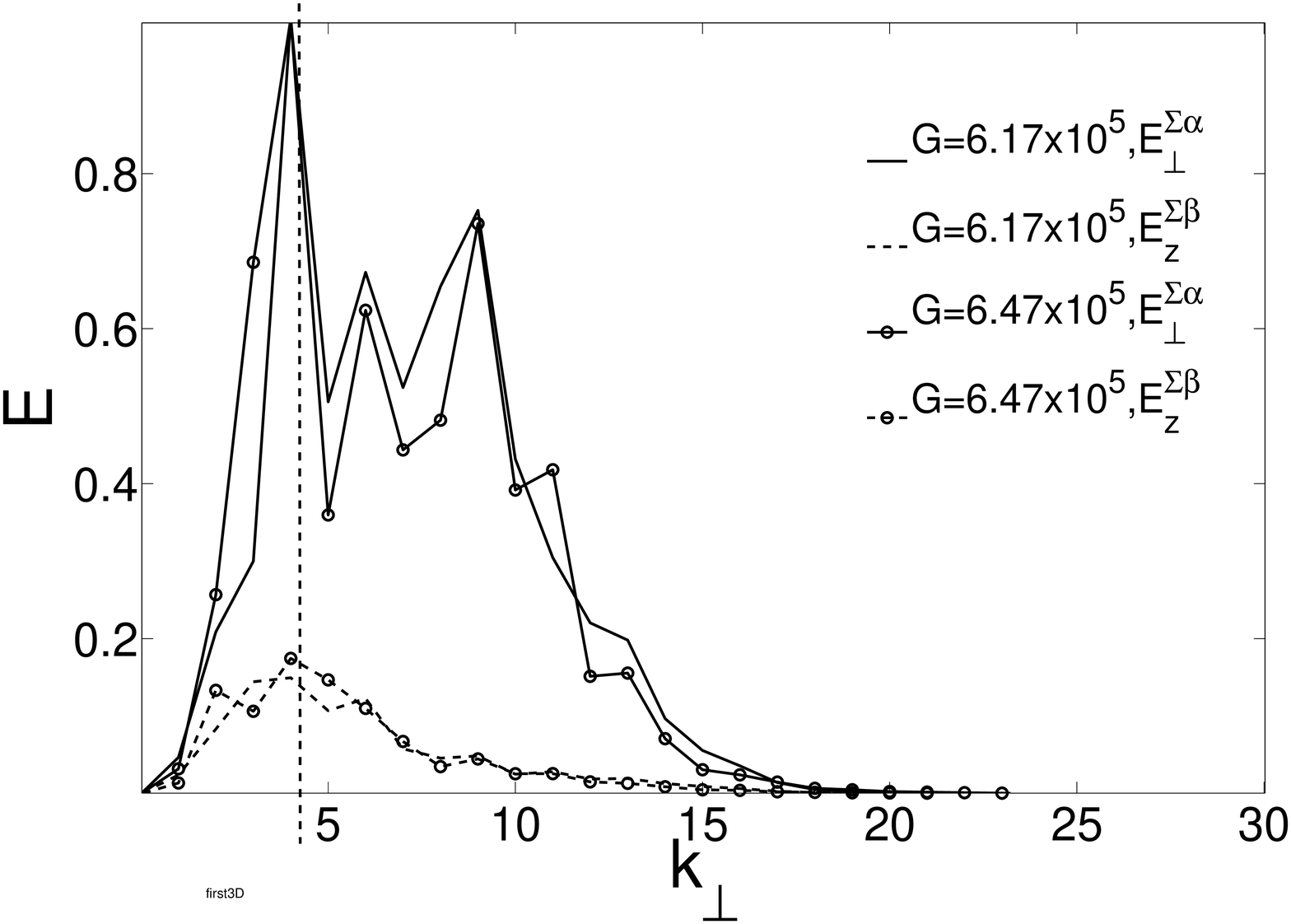}
\psfrag{first3D}{$k_\perp^{3D}$}
\includegraphics[width=6.5cm,height=6.5cm]{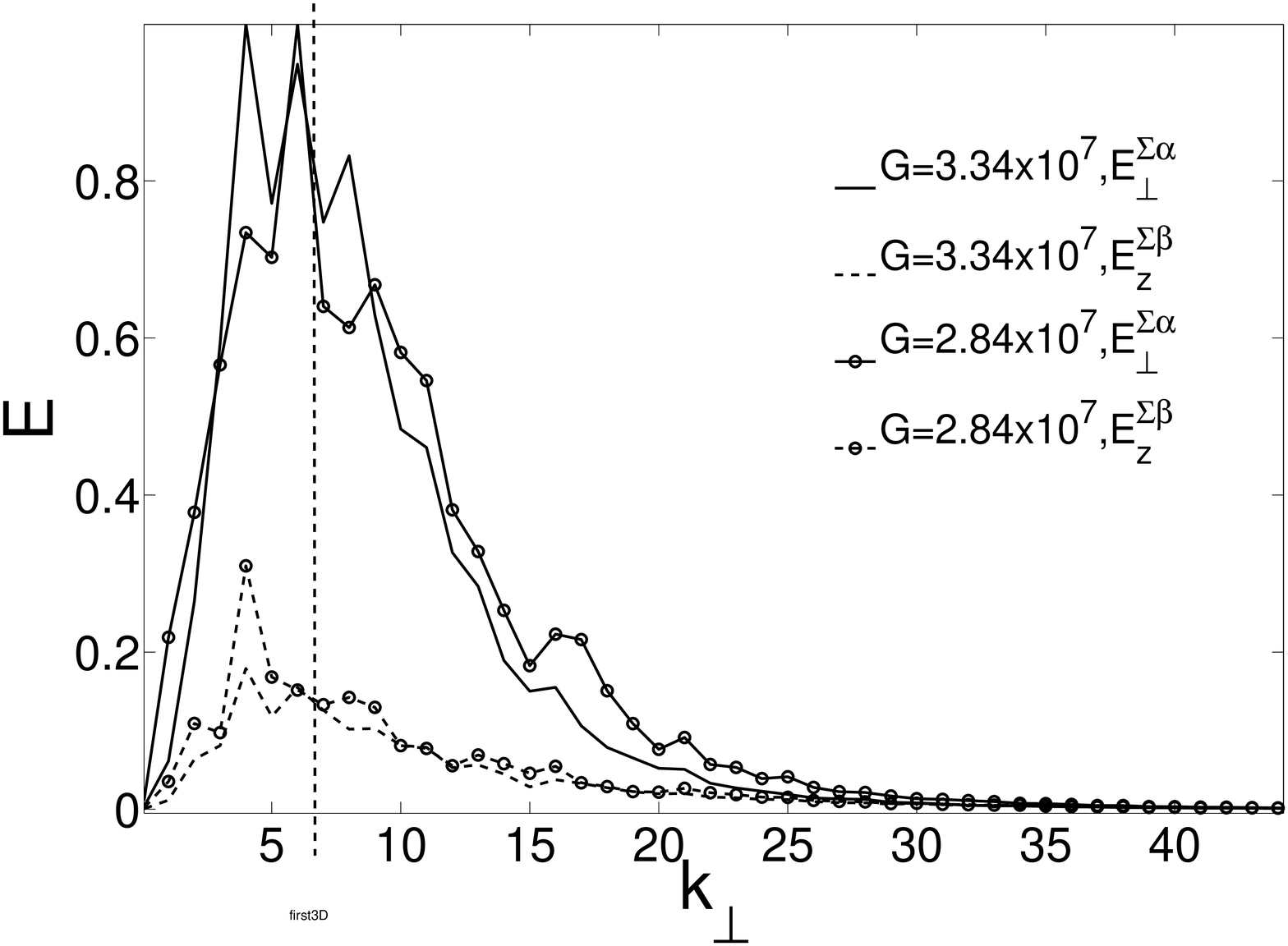}
\caption{Profiles of  $E_\perp^{\Sigma\alpha}(k_\perp)$  and  
$E_z^{\Sigma\beta}(k_\perp)$ for $\Ha_0=80$ (left) and  $\Ha_0=400$ (right): 
the corresponding flows are weakly three-dimensional. Curves marked with 'o' 
symbols indicate cases initialised in a stabilised state at slightly lower 
forcing while curves without them correspond to flows initialised at rest.
The vertical dashed lines mark the theoretical values of $k_\perp^{3D}$ given 
by (\ref{eq:k3d}).}
\label{fig:ek3d}
\end{center}
\end{figure}

On the cases initialised with the fluid at rest, we find that both 
$\alpha_{3D}$ and $\beta_{3D}$ jump to finite values at the same value of 
the forcing $\Gr^{\rm 3D}(Ha_{\rm opt})$. By contrast, when the forcing is increased 
progressively, morphological three-dimensionality appears at a lower critical 
value of $\Gr$ than dynamical three-dimensionality.
We have identified $k_\perp^{3D\alpha}$ and $k_\perp^{3D\beta}$ by calculating
 the quantities $E_\perp^{\Sigma\alpha}(k_\perp)=\sum\limits_{k_z\geq1}E_\perp(k_\perp,k_z)$ and
$E_\perp^{\Sigma\beta}(k_\perp)=\sum\limits_{k_z>0}E_z(k_\perp,k_z)$
 respectively. Both are plotted on figure \ref{fig:ek3d} for the first value of 
the forcing where three-dimensionality was observed. These quantities indeed 
remain at noise level for 
two-dimensional flows. When morphological (\textit{resp.} kinematic) 
three-dimensionality appears, several peaks rise in the 
profile $E_\perp^{\Sigma\alpha}(k_\perp)$ (\textit{resp.} 
$E_\perp^{\Sigma\beta}(k_\perp)$)  at $k_\perp=k_\perp^{3D\alpha}$ 
(\textit{resp.} $k_\perp=k_\perp^{3D\beta}$). 
Further peaks also appear around $k_\perp^{3D\alpha}$ 
and $k_\perp^{3D\beta}$. This is due to the fact that three-dimensionality can 
only be detected in slightly supercritical regime. Furthermore, since the 
maximum of the iso-$\lambda$ curve in 
$(k_\perp,k_z)$ is not only very ``flat`` but can also be located at a 
non-integer value of $k_\perp$, several peaks are expected to rise around 
the maximum. This is all the more true at high $\Ha$. 
Keeping this  in mind, one still sees that at the lowest forcings where 
either morphological or kinematic three-dimensionality were detected, 
both appeared in columnar vortices of approximately the same wavelength 
$k_\perp^{3D\alpha}\simeq k_\perp^{3D\beta}$. Importantly, this value 
is consistent with the theoretical 
estimate (\ref{eq:k3d}) for $k_\perp^{3D}$, albeit a little smaller in the 
case $\Ha_0=400$. On the top of the iso-$\lambda$ curve being very flat 
at $\Ha_0=400$, this shift towards larger scales can be explained by the fact 
that the higher $\Ha$, the higher the value of $\Gr$ at which 
three-dimensionality appears, and the higher the turbulence intensity when this 
happens. In two-dimensional turbulence, inertial transfer 
increases the energy of the large scales, that are therefore more prone to 
exhibit instabilities leading to the appearance of three-dimensionality. Among 
the least dissipative modes that dissipate energy at about the same 
rate, this favours those with the larger scales, over the strictly least 
dissipative one predicted by (\ref{eq:k3d}).\\

Importantly, one sees on figure \ref{fig:ek3d} that 
$k_\perp^{3D\alpha}\simeq k_\perp^{3D\beta}\simeq k_\perp^{3D}$ is independent of the flow's initial conditions, even though $\alpha_{3D}$ and $\beta_{3D}$ 
aren't.
 In other words, even in cases where morphological and dynamical 
three-dimensionality appear successively (in the sense of growing $\Gr$) they 
do so in vortices of the same transverse wavelength (\ref{eq:k3d}).
This implies that one can use the set of least dissipative modes together with  
scalings (\ref{eq:final_scal_lambda}) or (\ref{eq:final_sscale_mhd}) in order 
to determine \textit{a priori} the 
exact set of modes required to resolve both transitional and three-dimensional
flows completely, provided $L_{\rm opt}$ is known (It can be obtained 
from the calculation of a three-dimensional flow at the same value of $Ha_0$, 
for instance.).
For flows that lay at the transition between two- and 
three-dimensionality, a slight over-resolution is advisable that will absorb 
the peaks of three-dimensionality that appear around $k_\perp^{3D} \simeq 
k_\perp^{3D\alpha} \simeq k_\perp^{3D\beta}$.\\ 
%
It is quite remarkable that for the forcing (and the forcing scale) we have 
chosen, $k_\perp^{3D}$ follows (\ref{eq:k3d}) rather well.
Just how universal this behaviour is, however, remains to be clarified. For a 
sufficiently turbulent two-dimensional flow forced at $k_f>k_\perp^{3D}$, the 
inverse energy cascade can be expected to transfer energy back to 
$k_\perp^{3D}$ where  three-dimensional vortices would form. More generally, 
our recent experiments on MHD turbulence in cubic box have shown that the 
appearance of three-dimensionality was governed by a subtle interplay between 
inertia and the Lorentz force at the scale of each structure (\cite{kp10_prl}). 
The former is determined on the one hand by the forcing, which arbitrarily 
injects energy in the flow and, on the other hand, by the turbulent 
redistribution of energy amongst structures. Flows where turbulence is absent 
or too weak to sufficiently erase the non-universal trace of the forcing, 
therefore don't exhibit the ideal behaviour predicted by (\ref{eq:k3d}). 
This was spectacularly illustrated in our experiment where at low $\Ha$ and low $\Rey$, the destabilisation of a periodic array of columnar vortices led to 
remarkable steady three-dimensional Y-shaped vortices.
\section{Conclusions}
In this article, we have 
shown that DNS of Low-$Rm$ MHD turbulence in a three-dimensional periodic 
domain could be achieved by using the 
sequence of least dissipative eigenmodes from the dissipation operator instead
of the traditional Fourier basis. Not only is this technique far more 
cost effective at fully resolving the flow without modelling, 
but it also enlightens some of its 
properties that don't appear otherwise. Indeed, the iso-energy 
lines follow the lines of constant linear decay rate $\lambda$ well in regions 
of the spectral space that are not directly influenced by the forcing.
Furthermore, energy and dissipation spectra expressed in terms of the 
eigenvalue $\lambda$ associated to these modes instead of $k$, exhibit a 
clear cutoff that identifies modes located inside the Joule cone, and therefore 
strongly suppressed by Joule dissipation. Most importantly, analysing this 
spectra for $S\gtrsim1$ allowed us to derive laws that play the 
role of Kolmogorov laws, of determining the small scales in MHD turbulence: 
$\sqrt{|\lambda^{\rm max }|}/(2\pi k_f)\simeq 0.5 \Rey^{1/2}$ or 
$\sqrt{|\lambda^{\rm max }|}/(2\pi k_f)\simeq 0.47 \Gr^{0.20}$.
Finally, MHD flows in a periodic domain 
can be resolved as follows:  $L_{\rm opt}$ and $L_{\rm int}$
can be obtained on the fly, by minimising functional $\Sigma_{E_\lambda}$ at 
every time step (see section \ref{sec:lopt}). The 
discrete sequence of values of $\lambda$ then follows from (\ref{eq:lambda}), 
and ultimately, the small scales are obtained using our new scalings 
(\ref{eq:final_scal_lambda}) if $S_{\rm opt}\gtrsim1$, or the Kolmogorov laws 
if $S<1$.\\
In the last part of this work, we also showed that
the set of least dissipative modes 
encompassed the modes that first exhibit three-dimensionality when the forcing 
was increased from either zero or from that of a two-dimensional flow. 
This proves that the set of least dissipative 
modes is also suitable for the resolution of transitional flows, and not only 
for three-dimensional flows. On the top of this, for two-dimensional flows, 
that occur in the limit of large $S$, the 
Lorentz force vanishes so the set of least dissipative modes coincides with 
the usual set of two-dimensional Fourier modes. They can therefore be used 
in conjunction with Kraichnan's law for the size of the smallest scales 
$|\lambda^{\rm max}|^{1/2}/(2\pi)\simeq \Gr^{1/3}$. 
The least dissipative modes can therefore be used to calculate MHD flows in 
a periodic box for all values of $S$.\\

Finally, we wish to underline the large potential field of application of the
method presented in this work. The initial idea 
was to use a basis of modes that already incorporates the main constitutive 
structures of the flow, so as to save the costs of having to reconstruct them 
using elements of a less suited basis. In the present case, the basis of 
least dissipative modes readily rendered the anisotropic properties of MHD 
turbulence. Using this 
basis therefore reduced the cost of DNS by confining the spectral domain 
of resolution to that strictly relevant to the flow dynamics.
This procedure can clearly be extended to MHD and non-MHD problems with 
more complex boundary conditions. We have recently shown that the 
orthogonal set of  least dissipative modes in a channel flow with transverse 
magnetic field were exponential functions that incorporated the profile of 
the very thin Hartmann boundary layers which arise along the walls 
(\cite{dyp09}). 
Currently, channel flow DNS are limited to $\Ha$ below a few hundred because of the 
computational cost involved in meshing these layers. Using the least 
dissipative 
modes for this problem not only brings the same benefits as in the 
periodic case studied in the present work, but it also eliminates the 
difficulty posed by the Hartmann layers as they do not have to be reconstructed 
nor meshed.  As a spectacular consequence, the
computational cost of DNS based on these modes decreases with $\Ha$ 
instead of increasing as in current methods based on Tchebychev Polynomials. 
Using the least dissipative modes is therefore not only beneficial to the 
simulation of turbulent flows but also potentially to all flows where 
the reconstruction of anisotropic structures with unsuited elements 
incurs computational costs far beyond those strictly required by the dynamics.\\ 

The authors would like to express their gratitude to the Deutsche 
ForschungsGemeinschaft for their financial support under grant P01210/1-1. 
Part of the work presented here was performed during the MHD summer school 
organised by the Statistical and Plasma Physics department at the Universit\'e 
Libre de Bruxelles in 2007. The bulk of the numerical computations was 
performed on the computational facilities of the Applied Mathematics 
Research Centre at Coventry University.


\begin{thebibliography}{34}
\expandafter\ifx\csname natexlab\endcsname\relax\def\natexlab#1{#1}\fi

\bibitem[Alemany {\em et~al.\/}(1979)Alemany, Moreau, Sulem \&
  Frish]{Alemany1979}
{\sc Alemany, A., Moreau, R., Sulem, P. \& Frish, U.} 1979 Influence of an
  external magnetic field on homogeneous {MHD} turbulence. {\em J. Mec.\/} {\bf
  18:2}, 277--313.

\bibitem[Canuto {\em et~al.\/}(2006)Canuto, Hussaini, Quarteroni \&
  Zang]{canuto06_1}
{\sc Canuto, C., Hussaini, M.~Y., Quarteroni, A. \& Zang, T.~A.} 2006 {\em
  Spectral Methods: Fundamentals in Single Domains\/}. Springer-Verlag.

\bibitem[Constantin {\em et~al.\/}(1985)Constantin, Foias, Mannley \&
  Temam]{constantin85_jfm}
{\sc Constantin, P., Foias, C., Mannley, O.P. \& Temam, R.} 1985 determining
  modes and fractal dimension of turbulent flows. {\em J. Fluid. Mech.\/} {\bf
  150}, 427--440.

\bibitem[Davidson(2004)]{Davidson2004}
{\sc Davidson, P.A.} 2004 {\em Turbulence: An Introduction for Scientists and
  Engineers\/}. Oxford University Press.

\bibitem[Davidson(1997)]{dav97}
{\sc Davidson, P.~A.} 1997 The role of angular momentum in the magnetic damping
  of turbulence. {\em J. Fluid Mech.\/} {\bf 336}, 123--150.

\bibitem[Delannoy {\em et~al.\/}(1999)Delannoy, Pascal, Alboussi\`ere, Uspenski
  \& Moreau]{Delannoy1999}
{\sc Delannoy, Y., Pascal, B., Alboussi\`ere, T., Uspenski, V. \& Moreau, R.}
  1999 Quasi-two-dimensional turbulence in {MHD} shear flows: The matur
  experiment and simulations. In Transfer Phenomena and Electroconducting Flows
  (ed. A. Alemany et al.). Kluwer.

\bibitem[Doering \& Gibbons(1995)]{Doering1995}
{\sc Doering, C.R. \& Gibbons, J.D.} 1995 {\em Applied analysis of the
  Navier-Stokes equation\/}. Cambridge University Press.

\bibitem[Dymkou \& Poth\'erat(2009)]{dyp09}
{\sc Dymkou, V. \& Poth\'erat, A.} 2009 Spectral methods based on the least
  dissipative modes for wall-bounded {MHD} flows. {\em J. Theor. Comp. Fluid
  Dyn.\/} {\bf 23}~(6), 535--555.

\bibitem[Foias {\em et~al.\/}(2001)Foias, Manley, Rosa \& Temam]{Temam2001}
{\sc Foias, C., Manley, O., Rosa, R. \& Temam, R.} 2001 {\em Navier-Stokes
  Equations and Turbulence\/}. Cambridge University Press.

\bibitem[Frisch(1995)]{frisch95}
{\sc Frisch, U.} 1995 {\em Turbulence, The legacy of A.N. Kolmogorov\/}.
  Cambridge University Press.

\bibitem[Klein \& Poth\'erat(2010)]{kp10_prl}
{\sc Klein, R. \& Poth\'erat, A.} 2010 Appearance of three-dimensionality in
  wall-bounded {MHD} flows. {\em Phys. Rev. Lett\/} {\bf 104}~(3), 034502.

\bibitem[Klein {\em et~al.\/}(2009)Klein, Poth\'erat \& Alferjonok]{kpa09}
{\sc Klein, R., Poth\'erat, A. \& Alferjonok, A.} 2009 Experiment on an
  electrically driven, confined vortex pair. {\em Phys. Rev. E\/} {\bf 79}~(1),
  016304 (14 pages).

\bibitem[Knaepen \& Moin(2004)]{Knaepen2004}
{\sc Knaepen, B. \& Moin, P.} 2004 Large-eddy simulation of conductive flows at
  low magnetic {R}eynolds number. {\em Phys. Fluids\/} {\bf 16:5}, 1255--1261.

\bibitem[Kolmogorov(1941)]{k41}
{\sc Kolmogorov, A.N.} 1941 Local structure of turbulence in an incompressible
  fluid at very high {R}eynolds numbers. {\em Dokladi Akademii Nauk SSSR\/}
  {\bf 30}, 299--303.

\bibitem[Kraichman(1967)]{Kraichnan1967}
{\sc Kraichman, R.H.} 1967 Inertial ranges in two-dimensional turbulence. {\em
  Phys. Fluids\/} {\bf 10}, 1417.

\bibitem[Mininni {\em et~al.\/}(2006)Mininni, Alexakis \& Pouquet]{mininni06}
{\sc Mininni, P.~D., Alexakis, A. \& Pouquet, A.} 2006 Large-scale flow
  effects, energy transfer, and self-similarity on turbulence. {\em Phys. Rev.
  E\/} {\bf 74}, 016303.

\bibitem[Moffatt(1967)]{moffatt67}
{\sc Moffatt, H.K.} 1967 On the suppression of turbulence by a uniform magnetic
  field. {\em J. Fluid Mech.\/} {\bf 28}, 571--592.

\bibitem[Moreau(1990)]{moreau90}
{\sc Moreau, R.} 1990 Magnetohydrodynamics. Kluwer Academic Publisher.

\bibitem[Nakauchi {\em et~al.\/}(1992)Nakauchi, Oshima \& Saito]{nakauchi92_pf}
{\sc Nakauchi, N., Oshima, H. \& Saito, Y.} 1992 Two-dimensionality in
  low-magnetic {R}eynolds number magnetohydrodynamic turbulence subjected to a
  uniform external magnetic field and randomly stirred two-dimensional force.
  {\em Phys. Fluids A\/} {\bf 12}~(4), 2906--2914.

\bibitem[Ohkitani(1989)]{ohkitani89}
{\sc Ohkitani, J.} 1989 Log corrected energy spectrum and attractor dimension
  in two-dimensional turbulence. {\em Phys. Fluids A\/} {\bf 1}~(3), 451--452.

\bibitem[Orszag \& Patterson(1971)]{patterson71}
{\sc Orszag, G.~S. \& Patterson, S.~A.} 1971 Spectral calculations of isotropic
  turbulence: Efficient removal of aliasing interaction. {\em Phys. Fluids\/}
  ~(14), 2538--2541.

\bibitem[Poth\'erat \& Alboussi\`ere(2003)]{Potherat2003}
{\sc Poth\'erat, A. \& Alboussi\`ere, T.} 2003 Small scales and anisotropy in
  low-{Rm} magnetohydrodynamic turbulence. {\em Phys. Fluids\/} {\bf 15:10},
  3170--3180.

\bibitem[Poth\'erat \& Alboussi\`ere(2006)]{Potherat2006}
{\sc Poth\'erat, A. \& Alboussi\`ere, T.} 2006 Bounds on the attractor
  dimension for low-{Rm} wall bound {MHD} turbulence. {\em Phys. Fluids\/} {\bf
  18:12}, 125102.

\bibitem[Poth\'erat {\em et~al.\/}(2000)Poth\'erat, Sommeria \& Moreau]{psm00}
{\sc Poth\'erat, A., Sommeria, J. \& Moreau, R.} 2000 An effective
  two-dimensional model for {MHD} flows with transverse magnetic field. {\em J.
  Fluid Mech.\/} {\bf 424}, 75--100.

\bibitem[Roberts(1967)]{Roberts1967}
{\sc Roberts, P.H.} 1967 {\em Introduction to Magnetohydrodynamics\/}.
  Longsmans, London.

\bibitem[Rogallo(1981)]{Rogallo1981}
{\sc Rogallo, R.S.} 1981 Numerical experiments in homogeneous turbulence.
  National Aeronautics and Space Administration. Ames Research Center, Moffett
  Field, CA.

\bibitem[Schumann(1976)]{schumann76}
{\sc Schumann, U.} 1976 Numerical simulation of the transition from three- to
  two-dimensional turbulence under a uniform magnetic field. {\em J. Fluid
  Mech.\/} {\bf 35}, 31--58.

\bibitem[Sommeria(1986)]{Sommeria1986}
{\sc Sommeria, J.} 1986 Experimental study of the two-dimensional inverse
  energy cascade in a square box. {\em J. Fluid Mech.\/} {\bf 170:139}.

\bibitem[Sommeria(1988)]{Sommeria1988}
{\sc Sommeria, J.} 1988 Electrically driven vortices in a strong magnetic
  field. {\em J. Fluid Mech.\/} {\bf 189}, 553--569.

\bibitem[Sommeria \& Moreau(1982)]{Sommeria1982}
{\sc Sommeria, J. \& Moreau, R.} 1982 Why, how and when, {MHD} turbulence
  becomes two-dimensional. {\em J. Fluid Mech.\/} {\bf 118:507}.

\bibitem[Thess \& Zikanov(2007)]{thess07_jfm}
{\sc Thess, A. \& Zikanov, O.} 2007 Transition from two-dimensional to
  three-dimensional magnetohydrodynamic turbulence. {\em J. Fluid Mech.\/} {\bf
  579}, 383--412.

\bibitem[Vorobev {\em et~al.\/}(2005)Vorobev, Zikanov, Davidson \&
  Knaepen]{vorobev05}
{\sc Vorobev, A, Zikanov, O., Davidson, P.~A. \& Knaepen, B.} 2005 Anisotropy
  of magnetohydrodynamic turbulence at low magnetic {R}eynolds number. {\em
  Phys. Fluids\/} ~(17), 125105.

\bibitem[Williamson(1980)]{Williamson1980}
{\sc Williamson, J.H.} 1980 Low -storage {R}unge-{K}utta schemes. {\em J. Comp.
  Phys.\/} {\bf 35}, 48--56.

\bibitem[Zikanov \& Thess(1998)]{Zikanov1998}
{\sc Zikanov, O. \& Thess, A.} 1998 Direct numerical simulation of forced {MHD}
  turbulence at low magnetic {R}eynolds number. {\em J. Fluid Mech.\/} {\bf
  358}, 299--333.

\end{thebibliography}

%
%

\end{document}